# The Quadruplon in a Monolayer Semiconductor

Jiacheng Tang,[1,2,3,4] Hao Sun,[1,2,3] Qiyao Zhang,[1,2,3,4] Xingcan Dai,[1] Zhen Wang,[4,1,2,3] Cun-Zheng Ning[4,1,2,3]*

[1]Department of Electronic Engineering, Tsinghua University, Beijing 100084, China
[2]Frontier Science Center for Quantum Information, Beijing 100084, China
[3]Tsinghua International Center for Nano-Optoelectronics, Beijing 100084, China
[4]College of Integrated Circuits and Optoelectronic Chips, Shenzhen Technology University, Shenzhen 518118, China

* Corresponding author. Email: ningcunzheng@sztu.edu.cn

**Abstract**

**Understanding the structure of matter or materials and interaction or correlations among the constituent elementary particles are the central tasks of all branches of science, from physics, chemistry, to biology. In physics, this ultimate goal has spurred a constant search for high-order correlated entities or composite particles for nearly all states and forms of matter, from elementary particles, nuclei, cold atoms, to condensed matter. So far, such composite particles involving two or three constituent particles have been experimentally identified, such as the Cooper pairs, excitons, and trions in condensed matter physics, or diquarks and mesons in quantum chromodynamics. Although the four-body irreducible entities have long been predicted theoretically in a variety of materials systems alternatively as quadruplons[1], quadrons[2], or quartets[3], the closely related experimental observation so far seems to be restricted to the field of elementary particles (*e.g.* the recent tetraquark at CERN[4]) only. In this article, we present the first experimental evidence for the existence of a four-body irreducible entity, the quadruplon, involving two electrons and two holes in a monolayer of Molybdenum Ditelluride. Using the optical pump-probe technique, we discovered a series of new spectral features that are distinct from those of trions and bi-excitons. By solving the four-body Bethe-Salpeter equation in conjunction with the cluster expansion approach, we are able to explain these spectral features in terms of the four-body irreducible cluster or the quadruplons. In contrast to a bi-exciton which consists of two weakly bound excitons, a quadruplon consists of two electrons and two holes without the presence of an exciton. Our results provide experimental evidences of the hitherto theorized four-body entities and thus could impact the understanding of the structure of materials in a wide range of physical systems or new semiconductor technologies.**

**Main**

Whereas the search for the most "elementary" constituent particles of matter has been a never-ending pursuit in high-energy physics, the understanding of the interactions or correlations among these particles dominates studies in lower energy scales that are more relevant to our daily experience and technology. Such interactions lead to the formation of correlated entities or composite particles that determine the basic material properties, underline our fundamental understanding of almost all fields of physical and material sciences, and provide the foundation for all modern technologies. In condensed matter physics, correlated entities, such as excitonic complexes: excitons (X), trions (T), and bi-excitons (BX), are critical to our understanding of the basic material properties, and especially to the ever-richer physics of the celebrated Mott transition beyond the simple exciton-plasma picture. Recently, higher-order correlated entities have attracted much interests, including the Bose-Einstein condensation (BEC)[5], BEC-BCS crossover via different bosonizations of Fermions in strongly correlated systems, or dropletons[6-8] in semiconductors. In superconductivity theories, the charge-4$e$ configuration was proposed as an alternative to the conventional Cooper-pair mechanism[3,9]. Multiplons (including quadruplons) were also recently studied theoretically in the 1D Hubbard model[1]. It was shown theoretically that the formation of the quadron (quadruplon) was more favorable than the bi-exciton in a strongly confined parabolic quantum dot[2]. The situation is completely analogous in elementary particle physics where the bi-exciton analogue, the meson-meson molecules[10,11], and the quadruplon analogue, the genuine tetraquarks[4,11-13] (see Fig. 1 in Ref. [11]), were both predicted theoretically[10] and observed experimentally[4,12,13]. In terms of cluster expansion language, bi-exciton and meson molecules are two weakly interacting irreducible clusters of order 2, or △ △, while quadruplon and tetraquarks correspond to irreducible cluster of order 4, or △.

To date, experimental evidence for the existence of such a 4B irreducible cluster (or entity) is still lacking in other fields beyond the high-energy physics.

Two-dimensional (2D) layered semiconductors, such as monolayer transition metal dichalcogenides (ML-TMDCs), provide a unique platform for the study of high-order correlated entities or excitonic complexes. The reduced dielectric screening in ML-TMDCs leads to extremely large excitonic binding energies[14,15] and much more stable high-order excitonic complexes than those in conventional semiconductors. Indeed, trions[16,17], bi-excitons[18-22], and even charged bi-excitons[23-27] were experimentally observed in such ML-TMDCs with larger binding energies and at higher temperatures than those in bulk semiconductors. In addition, the unique spin-valley locking[28,29] leads to more varieties of the correlated entities than in conventional semiconductors. For the same reason, the correlated entities with specific spin-valley polarizations are addressable by choosing the helicity of the pump or probe light[21,22]. The combined unique features described above have never before been available for the study of these correlated entities.

To take advantages of these unprecedented opportunities, we conducted a combined theoretical and experimental investigation into the possible existence of higher-order correlated complexes in ML-TMDCs beyond the known trions and bi-excitons (including those that are reducible to excitons, trions, or bi-excitons). Using the helicity-resolved pump-probe technique, we observed a series of unexpected spectral features in the transient reflectance in a wide range of probe photon energies in gate-controlled monolayer molybdenum ditelluride (ML-MoTe$_2$) samples. Up to six spectral peaks were revealed, extending over 40 meV from below T all the way up to X. These spectral features cannot be attributed to familiar origins such as defects or phonon-related processes. To understand the new spectral features, we developed a perturbation theory based on the 4-body Bethe-Salpeter equation (4B-BSE) for the two-electron-two-hole ($2e2h$) system. By recasting the Feynman diagrams of the 4B-BSE into the cluster expansion formalism, we are able to compare the spectral contributions of the irreducible clusters of various orders up to the 4$^{th}$. Interestingly, we show that the clusters corresponding to the trions and bi-excitons cannot produce most of the new spectral features, thus excluding the trions and bi-excitons as their origin. Importantly, our theory-experiment comparison shows that the 4$^{th}$-order irreducible 2$e$2$h$-cluster, or quadruplon, is necessary and sufficient in producing all the experimental spectral features, thus providing experimental verification of the existence of the quadruplons.

**Four-body interactions and the cluster expansion**

The most natural way of visualizing correlated entities of various orders and associated many-body interactions is through the cluster expansion method[30]. This theoretical method has more recently been proven successful in describing the high-order excitonic correlations in $e$-$h$ dropletons[6,7]. Generalizing such cluster expansion of excitons to the case of individual electrons and holes with the spin degree of freedom, we can write down the complete sequence of irreducible clusters for the 2$e$2$h$ system as in Fig. 1a – 1c. The specialization of such cluster expansion for ML-MoTe$_2$ leads to the similar sequence of clusters, now expressed together with the band structures in Fig. 1d & 1e. $\triangle_n$ represents an irreducible cluster of the n$^{th}$-order with n Fermions. Clearly $\triangle_1$ is a quasi-free electron or hole. $\triangle_2$ represents a direct or indirect $e$-$h$ pair, or 2-body (2B) state, including all the excitonic Rydberg series: 1$s$-X, 2$p$-X, 2$s$-X, …. $\triangle_3$ represents an $e$-$e$-$h$ or $e$-$h$-$h$ 3-body (3B) state. In the language of multiplons proposed by Rausch et al[1], $\triangle_2$, $\triangle_3$, and $\triangle_4$ are doublon, triplon, and quadruplon, respectively. Each time we include one more cluster of higher order into a truncated expansion, it re-introduces weak interactions (indicated by the wavy lines in, e.g. Fig. 2a & Extended Data Fig. 1) among irreducible clusters of all the lower orders, as explained in more details in Methods S1. Here, we truncated the cluster expansion up to the 4$^{th}$ order. As a result, the original non-interacting cluster $\triangle_2$ $\triangle_2$ shown in Fig. 1b – 1e (also Extended Data Fig. 1d) becomes weakly interacting, i.e. $\triangle_2$ ~ $\triangle_2$ (Extended Data Fig. 1f), representing a bi-exciton (1$s$-X) ~ (1$s$-X) including all its excited states such as (1$s$-X) ~ (2$p$-X), (1$s$-X) ~ (2$s$-X) …. Obviously, cluster $\triangle_4$ or quadruplon (or quadron[2]) represents generally a distinct physical entity than the bi-exciton $\triangle$ ~ $\triangle$ . The most fundamental difference between a bi-exciton and a quadruplon is the absence of an exciton in the latter and the lack of a clear association of any one of the two electrons to a given hole. But it is easy to imagine that a certain disassociation event (*e.g.* an excitation) of a quadruplon could possibly lead to the formation of a bi-exciton. In this sense, a bi-exciton could be an excited state of a quadruplon.

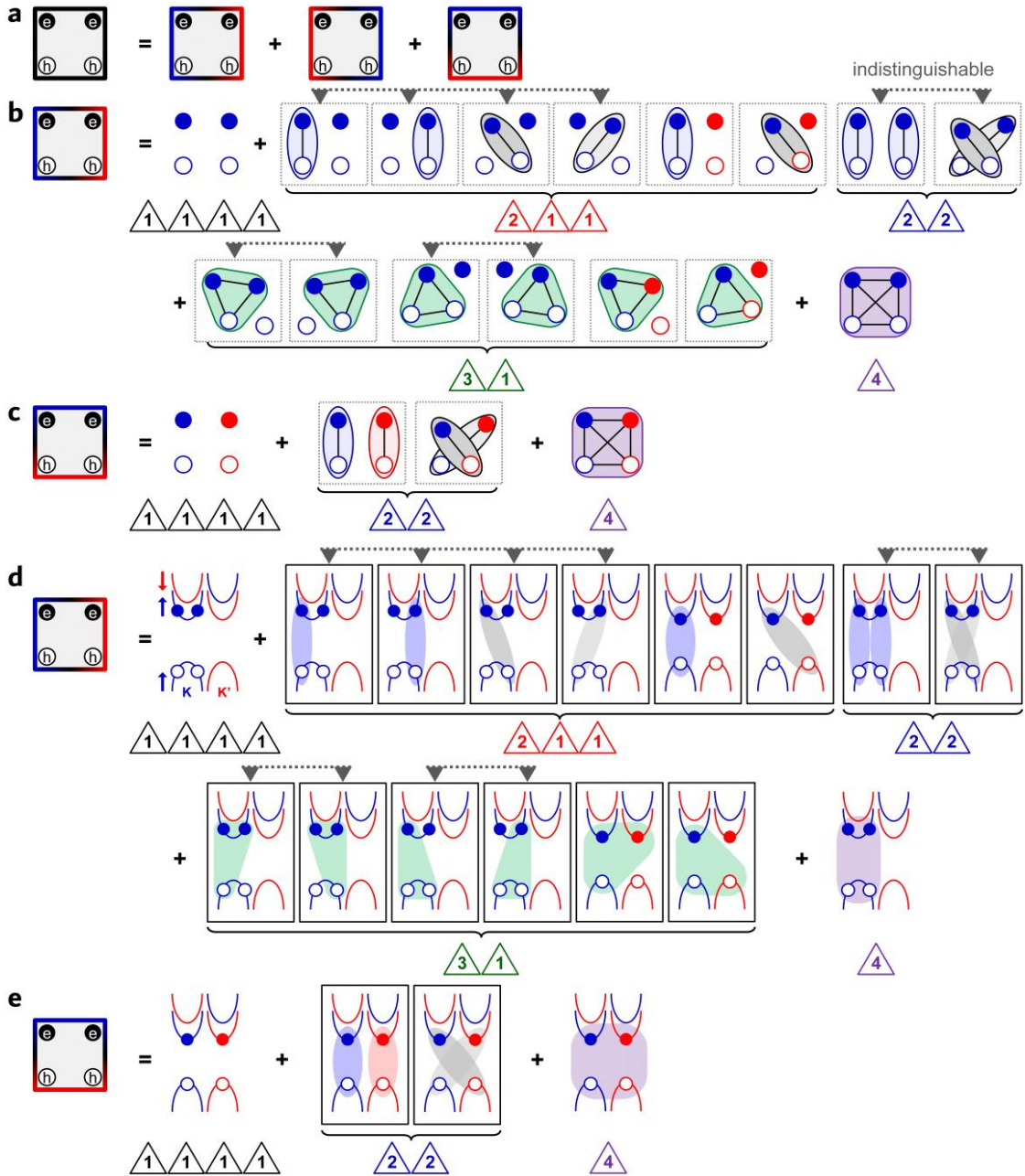

**Fig. 1 | A cluster expansion picture of a four-body system. a,** Decomposition of the 2*e*2*h* 4B system into three parts according to the time reversal (K ↔ K') symmetry, as indicated by the blue-red symmetry of the square boxes. The first two terms/boxes are mutually exchanged under the time reversal and only term is presented in detail in **b** or **d**. The third term/box is time reversal invariant, as presented in **c** or **e**. **b, c,** Cluster expansions of the 4B system into irreducible clusters (represented by the filled or unfilled circles connected by the straight lines) of various orders (sizes). **d, e,** Analogous to **b & c**, but directly in the representation of a band structure for ML-MoTe$_2$ with valleys (K and K') locked to the spin degree of freedom of electrons. The spin-up and spin-down electrons are colored in blue and red, respectively. For brevity, we exclude those 2B clusters with the same charges, such as *e-e* and *h-h*, in the figures, but these clusters are included in our theoretical calculations.

**Experimental results**

Figure 2a represents schematically the experimental situation where a strong pump produces an *e-h* "soup": a combination of various excitonic complexes or correlated entities of various orders and their corresponding excited states. Figure 2b shows our device structure of a ML-MoTe$_2$ sandwiched between two hexagonal boron nitride (h-BN) layers, with fabrication details presented in Methods S2. A back gate was used to control the background charge of the ML-MoTe$_2$. The temperature-dependent reflection contrast spectra (RCS) of Device #1 are obtained from the continuous-wave (CW) reflectance spectra with (R) and without (R') the sample, *i.e.* defined as (R – R')/R' (for the details of data processing see Methods S3.1, Extended Data Fig. 2a – 2h, and Ref. [31]). The RCS are shown under the charge-neutral condition at $V_g$ = –1 V in Fig. 2c and for the case of p-type doping at $V_g$ = –6 V in Fig. 2d. The grey dots with an additional minus sign (*i.e.* –(R – R')/R') could represent the ground-state absorption (GSA) of the material. According to the fittings for the results at 4K, the X peak is spectrally positioned at ~ 1.168 eV at the gate-compensated charge-neutral voltage, $V_g$ = –1 V (Fig. 2c), and the T peak appears at ~ 1.149 eV in the doped regime at $V_g$ = –6 V (Fig. 2d).

The GSA spectra show two simple features: **1)** The single T peak for ML-MoTe$_2$ (see also previous papers on MoTe$_2$[31-35]) corresponds to the inter-valley spin-singlet trion. This is closely related to the special band-structure of ML-MoTe$_2$ (or ML-MoSe$_2$): *i.e.* intravalley bright exciton has a smaller energy than intravalley dark exciton[27,33] (in contrast to ML-WS$_2$ and ML-WSe$_2$[23-26]), and the spin-orbit (SO) splitting of conduction band (CB), ~ 30 – 60 meV[32], is sufficiently large (in contrast to ML-MoS$_2$, ~ 3 – 15 meV[36,37]). **2)** More importantly, there are no additional absorption features below X in the charge-neutral regime without pump (see the low-energy side of X in Fig. 2c). The simple GSA spectra can be used as a base or reference for later study of complicated features in the transient differential absorption or reflection spectra (TDAS or TDRS). The fact that there is no more complicated feature below T and X makes the MoTe$_2$ one of the ideal systems among various TMDCs (for more specific reasons or advantages that we chose to study MoTe$_2$ see Methods S4).

Typically by measuring the CW RCS and PL spectra (for both of low and high excitation densities), we pre-screened the samples for the following pump-probe experiments and selected those without visible defect features. Representative CW RCS and PL spectra of ML-MoTe$_2$ could be seen with or without defect peaks below T, as presented in Methods S3.2, Extended Data Fig. 3 & 4.

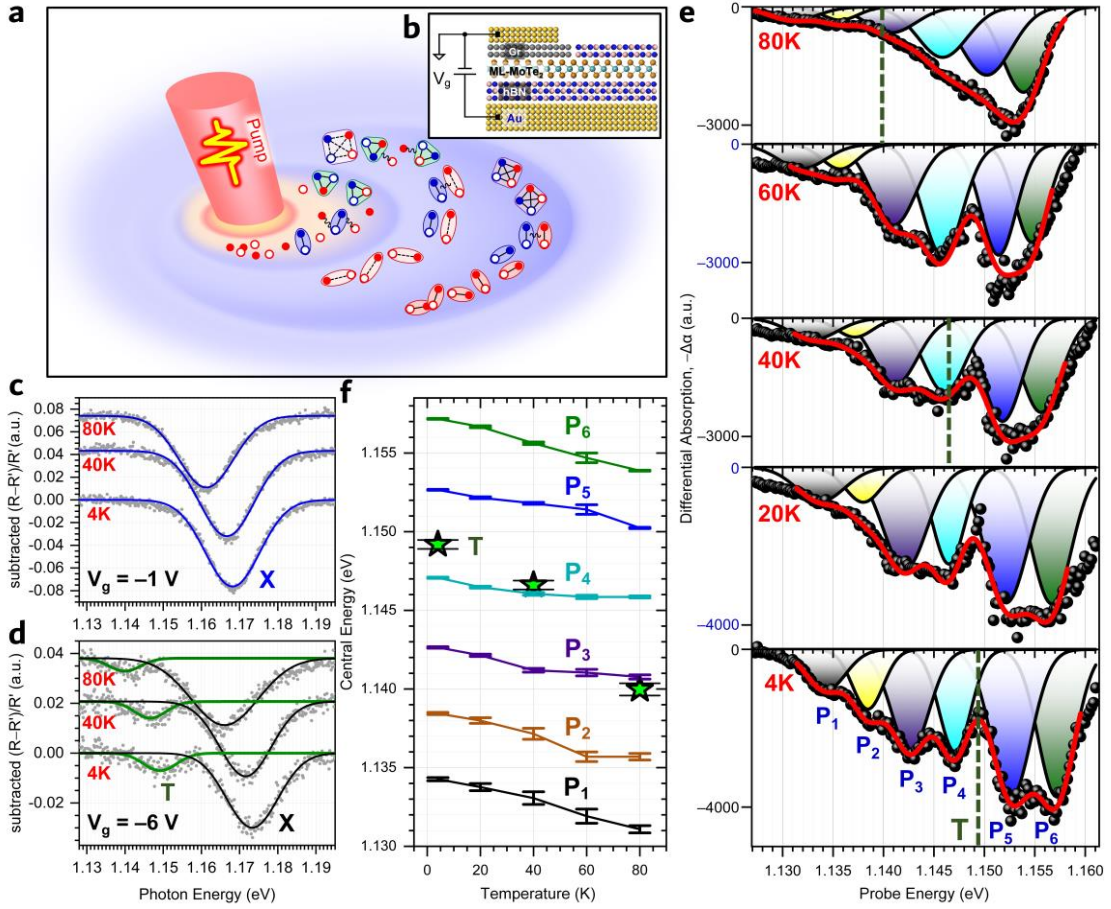

**Fig. 2 | Basics of experiment and key observations. a,** Illustration of the *e-h* "soup" generated by the intense pump pulse, showing the possible 2B and 4B states. **b,** Schematic of the charge-tunable device by the gate voltage, $V_g$. **c, d,** CW RCS (R – R')/R' (see Methods S3.1 for definition) at several temperatures under the charge-neutral condition at $V_g$ = –1 V **(c)** and under the p-doping condition at $V_g$ = –6 V **(d)** (Device #1). The solid lines are the results of the Gaussian fittings. **e,** TDAS (see the main text for definition) at different temperatures for the cross- (σ+ σ+) circularly polarized pump-probe configuration at the pump-probe delay time t ≈ 0 ps at the charge-neutral voltage, $V_g$ = –1 V. Each spectrum for a given temperature was fitted with 6 Gaussian peaks marked by $P_1$ – $P_6$. The points are experimental values, while the red solid lines are the results of the Gaussian fittings. **f,** Plots of the fitted central energies of the Gaussian peaks with respect to the temperatures for $P_1$ – $P_6$ **(e)** and T **(d)**. The spectral locations of T marked with the green dashed lines in **e** were obtained from the CW results in **d**.

The ultrafast pump-probe experiment is described in Methods S3.3, with the experimental setup illustrated in Extended Data Fig. 5. The pump energy of 1.174 eV is ~ 6 meV above X with a fluence of ~ 60 µJ·cm$^{-2}$ (corresponding to an *e-h* pair density $n_p$ estimated to be ~ 3 × 10$^{12}$ cm$^{-2}$, see Methods S3.4 for the estimate), while the probe energy was tuned from 1.127 to 1.161 eV with a resolution of ~ 0.2 meV. The gate voltage was set at −1 V to maintain charge neutrality. The temperature-dependent transient differential absorption spectra (TDAS) of Device #1 are shown in Fig. 2e. Here, the TDAS ($-\Delta\alpha$) is defined as $-\Delta\alpha = -(\alpha_p - \alpha_0) \propto R_p - R_0$, where $R_p$ and $R_0$ are the RCS from the sample with and without pump ($\alpha$ denotes the corresponding absorption. See Methods S3.3 and Ref. [34] for more details about the relation between TDAS and TDRS). It is worth noting that $-\Delta\alpha < 0$ means a pump-induced absorption increase, typically related to those of excited-state absorption (ESA) processes. In Fig. 2e, we observe rich spectral features with absorption increases as marked with $P_1 - P_6$. Based on their relative positions to T and X, these features and the associated fitted peaks can be divided into three energy intervals: $P_1$, $P_2$ (below T), $P_3$, $P_4$ (near but below T), and $P_5$ & $P_6$ (above T or between T and X).

To obtain more quantitative information about these features, we performed multi-peak Gaussian fittings on the spectra at various temperatures in Fig. 2e (The applicability of such multi-peak fittings for $P_1 - P_6$ in the TDAS will be discussed then in connection with Fig. 3l – 3n). The central energies of the obtained Gaussian peaks $P_1 - P_6$ (Fig. 2e) and the fitted peak T (Fig. 2d) are plotted in Fig. 2f versus temperature. The intervals between the neighboring peaks in $P_1 - P_6$ are in the range of 4 – 7 meV, while the total spread of these peaks is around 25 meV. The lowest peak ($P_1$, ~ 1.134 eV at 4K) is about 35 meV below the original X peak (~ 1.168 eV at 4K) and ~ 15 meV below the T peak (~ 1.149 eV at 4K). When temperature increases from 4K to 80K, $P_1 - P_6$ in the ultrafast spectroscopy (Fig. 2e) show a redshift of ~ 1 – 3 meV, while peaks T and X extracted from the CW results (Fig. 2c & 2d) show a redshift of ~ 7 meV. As can be seen in Fig. 2e, $P_1 - P_6$ are well resolved at 4K, but merged more together with increase in temperature due to increased broadening with temperature. To show the reproducibility of the results (the 4K case in Fig. 2e), we measured the TDAS with a finer spectral resolution of 0.1 meV. The similar six peaks can be seen quite clearly even without the multi-Gaussian fitting, as presented in Extended Data Fig. 6, Methods S5.

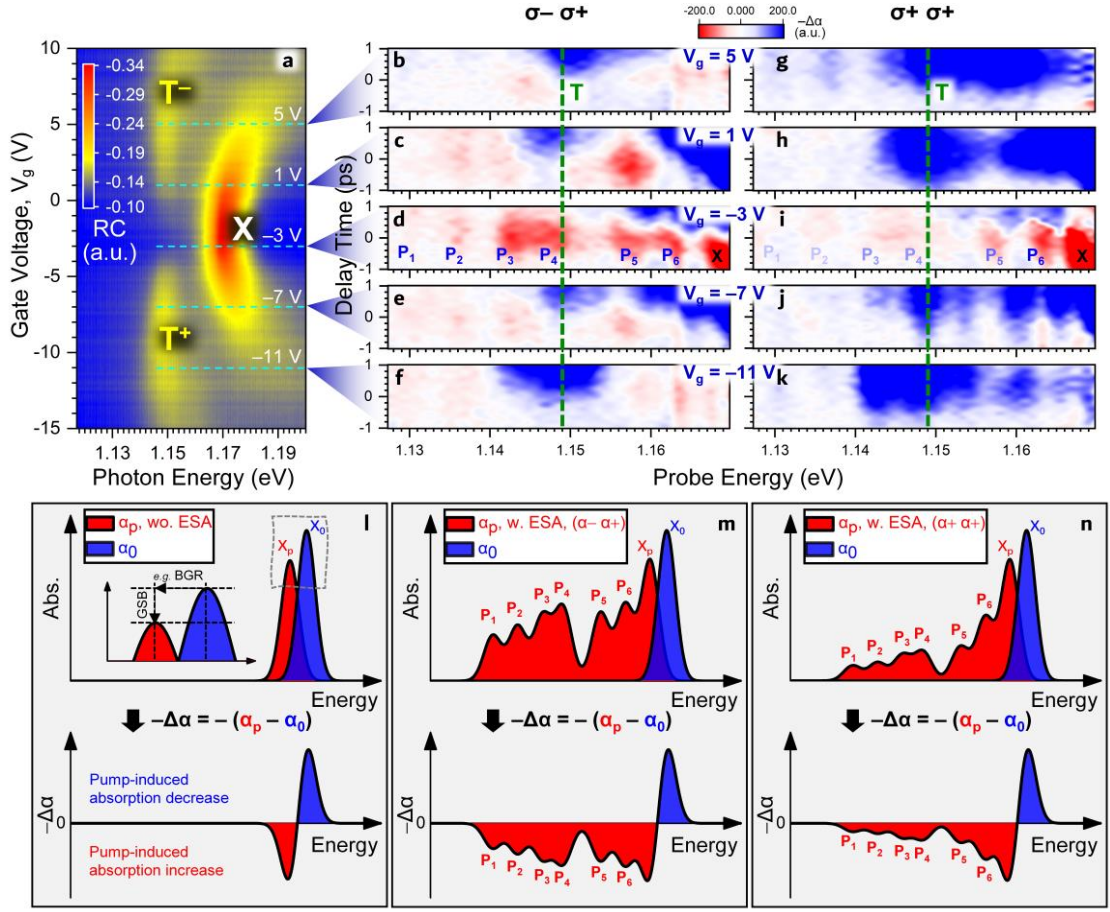

**Fig. 3 | Gate-voltage and polarization dependence of the TDAS. a,** CW RC contour in the plane of photon energy and gate voltage (Device #2 at 4K). **b – k,** TDA contours in the plane of probe energy and delay time for the cross- (σ– σ+) **(b – f)** and co- (σ+ σ+) **(g – k)** circularly polarized pump-probe configurations. The voltages (5 V, 1 V, –3 V (charge-neutral), –7 V, –11 V) applied for observing the TDAS in **b – k** are marked with the five cyan dashed lines in **a**. In **d & i**, the similar new spectral features ($P_1 – P_6$) to those in Fig. 2e are marked accordingly. **l – n,** Schematics of the relationship between the TAS (upper panels) and the TDAS (lower panels) under the charge-neutral condition. In the upper panels, the curves filled with red and blue denote the absorption spectra with ($\alpha_p$) and without ($\alpha_0$) pump, respectively. The zoomed-in inset in l shows the X with ($X_p$) and without ($X_0$) pump to illustrate GSB and BGR. **l – n** show the cases with **(m & n)** and without **(l)** ESA caused by the potential many-body effect of high-order correlation, thus corresponding to the presence and absence of $P_1 – P_6$ in **m & n** and **l**, respectively.

We notice that previous studies have also observed a peak below T attributed to charged bi-excitons[23-27]. To examine the charge dependence of our new spectral peaks, we performed a gate-dependent experiment on a device of the same design (Device #2) at 4K. The gate-dependent CW RC (or absorption) map is shown in Fig. 3a. Similar to the case of Device #1 (Fig. 2c), there are no observable features below X in the charge-neutral regime ($V_g$ = –3 V) without pump. A few selected TDAS are shown in Fig. 3b – 3k. From top to bottom, the system was gated into the charge-negative, neutral, and positive regimes corresponding to the five voltages as marked by the cyan dashed lines in Fig. 3a. In Fig. 3b – 3k, the spectral features of $P_1 – P_6$ similar to those in Fig. 2e are visible, and are the strongest in the charge-neutral situations of $V_g$ = –3 V (Fig. 3d & 3i). The features fade away as the system deviates from charge neutrality. Such gate-dependent behavior was also observed for another device of the same design (Device #5) (see Methods S7, Extended Data Fig. 10 for details). Contrary to the previous observations of the PL of charged bi-excitons[23-26] that are the weakest in the charge-neutral regimes, our new peaks $P_1 – P_6$ are the strongest in these regimes, pointing to the existence of charge-neutral entities.

As can be seen from the polarization-dependent results, the spectral features are stronger for the case of (σ– σ+) (Fig. 3d) than for the case of (σ+ σ+) (Fig. 3i) and this is also true for two other devices we measured (see Methods S5, Extended Data Fig. 6 for Device #1, and Methods S6, Extended Data Fig. 7 for Device #4). Compared to $P_1 – P_6$, the signals around X show less such a polarization contrast. The polarization contrast for $P_1 – P_6$ was also observed for bi-exciton signals in previous experiments[21,22,27]. It means that the inter-valley (σ– σ+) configuration is always more favorable than the intra-valley (σ+ σ+) one[21,22,27], which is also similar to the case of bounding and anti-bounding states of a Hydrogen molecule. Such a polarization contrast is also consistent with our theoretical results (see Methods S13.1, Extended Data Fig. 15 for details). In contrast to the six peaks observed here, previous studies have shown one peak for BX[18-21,23-27,38-40] and no more than three peaks for fine structure of BX (BXFS)[22,41]. The spectral positions (or the corresponding binding energies) of BX and BXFS have been calculated for ML-TMDCs previously[22,38-41], and widely accepted to be between T and X (14.4 meV below X for ML-MoTe$_2$[38]). Therefore, the sequence of our experimental spectral features, *i.e.* $P_1 – P_6$, which extend ~ 40 meV from below T all the way up to X, could not be explained by BXFS.

To better understand the new spectral features $P_1 - P_6$ & X in Fig. 3d & 3i (also Extended Data Fig. 7), we schematically illustrate the relationship between the TAS ($α_p$ & $α_0$) and the TDAS ($-Δα$), as shown in Fig. 3l – 3n. Under the charge-neutral condition without pump, $α_0$ has only a single X peak, $X_0$ (the corresponding RCS can be seen in Fig. 2c). Under optical pumping, the X peak in $α_p$ ($X_p$) shows a reduced oscillation strength due to ground-state bleaching (GSB) and a redshifted resonance energy due to bandgap renormalization (BGR), as illustrated in the inset in Fig. 3l. To obtained TDAS, a subtraction of $α_0$ from $α_p$ with a redshifted X peak leads to the typical anti-symmetric feature shown in the lower panel of Fig. 3l (also Ref. [42]) schematically or in actual experimental data (see Methods S6, Extended Data Fig. 7). In addition to GSB and BGR, a strong pump could lead to the excited-state absorption (ESA) corresponding to transitions from existing 2B states to 4B entities such as bi-exciton[21,22,39,40], *etc*. Figure 3m & 3n and 3l compare cases with and without ESA. With ESA, new peaks in $α_p$ emerge with pump in addition to X. A subtraction of $α_0$ from $α_p$ produces the corresponding pump-induced peaks in the TDAS, *i.e.* $P_1 - P_6$ as shown in the lower panel of Fig. 3m & 3n (see also Ref. [21,22,39,40] for the BX peak in TDAS). In other word, peaks ($P_1 - P_6$) resulting from ESA remain in TDAS as in TAS, in contrast to the anti-symmetric feature that corresponds to GSA processes.

To fit the TDAS, we assign six negative Gaussian peaks for $P_1 - P_6$ and a combination of negative (for $X_p$) and positive (for $X_0$) peaks for the antisymmetric line-shape around X (see Methods S6 for the fitting details, the fitting method can be seen also in Ref. [22]). We notice that $X_p$ is red-shifted from $X_0$ by ~ 5 – 7 meV, while peaks $P_1 - P_6$ are red-shifted by from $X_p$ ~ 6 – 40 meV. Since peaks $P_1 - P_6$ are sufficiently narrow and well-separated from $X_p$, they can be independently fitted.

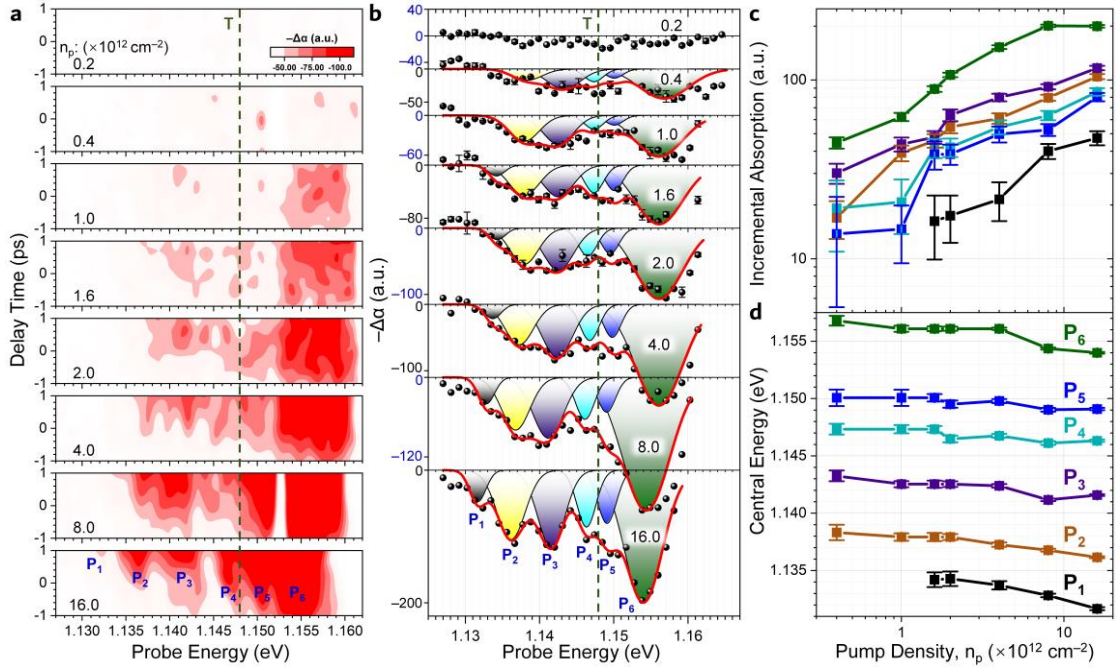

**Fig. 4 | Pump-density dependence of the TDAS. a,** TDA contours in the plane of probe energy and delay time for the cross- (σ– σ+) circularly polarized pump-probe configuration (Device #3, measured at 4K in the charge-neutral regime). **b,** Gaussian fittings (red solid lines) of the TDA extracted from **a** (dots) at a delay time of ~ 0.9 ps. Each spectrum is fitted with 6 Gaussian peaks marked by $P_1 - P_6$ as in Fig. 2e. The spectral location of T is marked with the green dashed lines in **a & b**. The total *e-h* pair density, $n_p$, generated by the pump was marked in each panel of **a & b**. **c, d,** The total areas (corresponding to absorption increment) **(c)** and central energies **(d)** of $P_1 - P_6$ with respect to the above pump densities.

To study the pump-density dependence of $P_1 - P_6$, we performed a series of pump-probe experiments with a varying pump fluence on a device of the same design (Device #3) at 4K. Based on the results of Device #2 presented in Fig. 3, the system was gated into the charge-neutral regime to have the maximum effects of the spectral features. The pump-fluence-dependent TDAS are shown in Fig. 4a. From top to bottom, the pump fluence was varied from ~ 4 μJ·cm$^{-2}$ to ~ 320 μJ·cm$^{-2}$ ($n_p$ estimated to be 2.0× 10$^{11}$ cm$^{-2}$ to 1.6×10$^{13}$ cm$^{-2}$, see Methods S3.4 for the estimates). Figure 4b show the spectra in x-y plot (corresponding to those of Fig. 4a at the delay time of ~ 0.9 ps, where the signals are the strongest) with the multi-Gaussian fittings similar to those shown in Fig. 2e. The integrated intensities and central energies of the fitted Gaussians are plotted in Fig. 4c & 4d, respectively (We re-plotted Fig. 4c in a linear scale, as can be seen in Methods S8). As can be seen in Fig. 4a & 4b, the spectrum is nearly featureless when the pump density is below $4.0\times10^{11}$ cm$^{-2}$. As the pump density is above $4.0\times10^{11}$ cm$^{-2}$, $P_6$ (between T and X) is the first to appear in the spectrum, followed by $P_2 - P_4$ (below or near T) & $P_5$ (between T and X) with blurred features. With the further increase of the pump density beyond $1.6\times10^{12}$ cm$^{-2}$, $P_2 - P_4$ & $P_5$ become increasingly visible and better distinguishable, and $P_1$ (below T) starts to appear (better visible in Fig. 4b than in Fig. 4a). When the pump density reaches $8.0\times10^{12}$ cm$^{-2}$, all the peaks of $P_1 - P_6$ are visible and can be fitted within relatively small errors (Fig. 4c & 4d). At any pump level as shown in Fig. 4c, $P_1$ is always the weakest among the 6 peaks while $P_6$ is the strongest. The pump dependence described here will be further explained in connection with the discussions of Fig. 5. With the increase of the pump density, $P_1$, $P_2$, & $P_6$ exhibit larger redshifts of ~ 2.5, 2.2, and 2.8 meV than $P_3 - P_5$ of ~ 1.5 meV or smaller (Fig. 4d).

Spectral features represented by $P_1 - P_6$ as we characterized so far have not been seen before in other semiconductors. It is important to rule out other effects as potential origins of these new features such as defects[18-20,23,26], phonons[43-46] (*e.g.* exciton phonon replicas, *etc*), any emission signals, second-harmonic generation, and other non-linear mixing effects typically related to *e.g.* $\chi^{(3)}$, $\chi^{(5)}$, … contributions and those charge-neutral entities involving more than two excitons (tri-excitons △ ~ △ ~ △, quad-excitons △ ~ △ ~ △ ~ △, … dropletons △ ~ △ ~ △ ~ … ~ △, as we discuss in more detail in Methods S15. The significant difference between the cross- and co-polarized pump-probe configurations is also a strong indication that many of the above origins can be excluded. In addition, the good agreement between experiment and theory as presented in the following also favors the intrinsic origins of 4B states, since the following theory does not include any of the above effects.

To understand the origin of $P_1 - P_6$, we will develop a many-body theory based on the Bethe-Salpeter equation for the 4B case in the following.

**Theory for four-body systems of two electrons and two holes**

The standard 2-body Bethe-Salpeter equation (2B-BSE) has been extensively applied to excitonic systems[47,48] to obtain the Hydrogen-like Rydberg series[49] (for details about the 2B-BSE see Methods S10 – S12.1). The 3B-BSE has also been applied to explain the trion-related features in 2D materials[37,50,51]. Figure 5a presents the absorption spectrum calculated using the 3B-BSE for ML-MoTe$_2$, where we see clearly only the T peak in its neighborhood. Obviously, such 3B-BSE does not provide an explanation for the features related to $P_1 - P_6$. The BX-related spectral features have been calculated theoretically[22,38-41] to be between T and X. Specifically, the BX resonance of 14.4 meV below X for ML-MoTe$_2$ was calculated to be between T and X by combining the density functional theory with the path-integral Monte Carlo method[38]. But so far, there has been no experimental observation of the BXs in MoTe$_2$. In addition, our spectral features $P_1 - P_6$ contain more spectral peaks in a much larger spectral range than those BX related peaks[22,39-41]. To explain these new features, we developed the 4B-BSE theory that describes the 2$e$2$h$ system by including all the 4B correlations (as illustrated in Fig. 1). Details of the theory and the calculation of absorption spectra are presented in Method S12.2 – S13.2. Here we show only the key steps. The wavefunctions of the 4B and 2B states, $|e_1 h_1 e_2 h_2\rangle$ and $|e_3 h_3\rangle$, are expressed in terms of creation and annihilation operators,

$$|e_1 h_1 e_2 h_2\rangle = \sum_{(v_1, c_1, v_2, c_2)} B^{v_1, c_1, v_2, c_2} \hat{a}^\dagger_{c_1} \hat{a}^\dagger_{c_2} \hat{a}_{v_1} \hat{a}_{v_2} |0\rangle, \quad (1)$$

$$|e_3 h_3\rangle = \sum_{(v_3, c_3)} A^{v_3, c_3} \hat{a}^\dagger_{c_3} \hat{a}_{v_3} |0\rangle, \quad (2)$$

where $v_i$'s and $c_i$'s index the single-particle states including the valence ($v$) and conduction ($c$) band indices with the momenta. Coefficients $A^{v_3, c_3}$ and $B^{v_1, c_1, v_2, c_2}$ are determined by the solutions of the 2B-BSE and 4B-BSE, respectively. The corresponding eigen-energies of the 2B and 4B states can also be solved and denoted by $\varepsilon_{e_3h_3}$ and $\varepsilon_{e_1h_1e_2h_2}$, respectively (see Methods S12.1 & S12.2, respectively; for the numerical techniques see Methods S14). In this way, dielectric function $\epsilon$ for each given $|e_3h_3\rangle$ can be calculated via

$$\epsilon \propto \frac{2\pi}{\hbar} \sum_{\alpha} \left| \sum_{(v_i,c_i,i=1,2,3)} \langle e_3h_3|\mathbf{e}\cdot\hat{\mathbf{p}}|e_1h_1e_2h_2\rangle_\alpha \right|^2 \Gamma\left(\varepsilon_{e_1h_1e_2h_2}^\alpha - \varepsilon_{e_3h_3} - \hbar\omega - i\gamma\right), \qquad (3)$$

where, $\hbar\omega$ is the photon energy, $\mathbf{e}$ is the unit vector of the in-plane electric field component for a normal incident field, $\hat{\mathbf{p}}$ is the dipole momentum operator defined via $\hat{\mathbf{p}} = \sum_{(v_4,c_4)} \mathbf{p}^{v_4,c_4} \hat{a}_{c_4} \hat{a}_{v_4}^\dagger$ with the independent-particle inter-band dipole momentum $\mathbf{p}^{v_4,c_4}$ given in Methods S10, eq. (S8) & (S9), $\alpha$ is the serial number of all the 4B eigen-states in the sequence of the eigen-energies from low to high, and $\Gamma$ represents the Gaussian broadening function with a phenomenological broadening parameter, $\gamma$. Finally, the absorption coefficient is given by the imaginary part of the dielectric function for both (σ– σ+) and (σ+ σ+) pump-probe configurations.

Clearly, one has to calculate the dipole matrix elements of all the 2B-4B transitions to obtain the dielectric function (eq. (3)). Even more challenging is the knowledge of the distributions and weightings of all the possible species in the "soup" at a given set of parameters (temperatures and the total *e* or *h* densities, *etc.*). In the case of *e-h* 2B systems such as excitons and plasmas, a self-consistent description of the coupling between distributions and polarizations is given by the Semiconductor Bloch equation[52]. Since a complete theory involving self-consistent coupling between distributions of all the species in a 2*e*2*h* 4B system and the possible polarizations is lacking, we will only consider possible spectral contributions from all the 2B-4B transitions, without considering the exact distributions of various entities and the weighting of each transition.

Another important notice to make of is the relationship between the cluster expansion approach and direct solutions of the 4B-BSE using the Feynman diagrammatic technique. The relations between the terms in these two expansion techniques are given in Methods S12.3. Due to the intuitive pictures and transparent physics of the cluster expansion approach, we will use the language of cluster expansion to describe the results. For the 2B states, we have a total of 15 discrete states from 1*s*-X, 2*p*-X, 2*s*-X, ... plasma for ML-MoTe$_2$ (see Methods S12.1). For the 4B states, we consider following three cases of truncation (in connection with Methods S1, Extended Data Fig. 1), in order to extract and identify the effects of each: Case 1, up to △△ ; Case 2, up to △△ ; Case 3, up to △ (for details see Methods S12.2 – S13.2). For each case, a series of 4B states are solved from the corresponding truncated 4B-BSE. The dipole matrix elements of all the possible 2B-4B transitions are then calculated as follows,

$$\langle e_3 h_3 | \hat{\mathbf{p}} | e_1 h_1 e_2 h_2 \rangle = \sum_{(v_i, c_i, i=1,2,3,4)} \left( A^{v_3,c_3} \right)^* B^{v_1,c_1,v_2,c_2} \mathbf{p}^{v_4,c_4} \times \langle 0 | \hat{a}^{\dagger}_{v_3} \hat{a}_{c_3} \hat{a}_{c_4} \hat{a}^{\dagger}_{v_4} \hat{a}^{\dagger}_{c_1} \hat{a}^{\dagger}_{c_2} \hat{a}_{v_1} \hat{a}_{v_2} | 0 \rangle. \quad (4)$$

An expansive form of eq. (4) can be seen in Methods S13, eq. (S19). The dipole matrix elements thus calculated are inserted into eq. (3) (also eq. (S20)) to obtain the spectra corresponding to photon absorption or emission shown in Fig. 5b – 5e (also Extended Data Fig. 15 & 17).

To obtain a correct and complete picture of the possible 2B-4B transitions, especially the relative positions of spectral lines and their origins, we emphasize here the difference between the typical optical spectrum and the total-energy spectrum. Such total-energy spectrum of the 2*e*2*h* 4B system is shown in Fig. 5g (see Extended Data Fig. 18 for the 4B system truncated to △△ , △△ and △ ). It is important to realize that the optical spectrum measured in an absorption or emission experiment does not reflect the absolute energies on the total energy scale. As shown in Fig. 5g (also Extended Data Fig. 18), a series of transitions occur between different 2B states and the corresponding 4B states on the total energy scale, as marked by the vertical double-arrowed lines with ① – ⑥ (6 examples, selected out of 15 2B states shown in Extended Data Fig. 17). The actual spectrum of the transitions corresponding to the superposition (Fig. 5j) of all possible individual transitions at different total energies are shown schematically in Fig. 5g. Unfortunately, such total-energy spectra are not easy to obtain in an optical experiment, where only the energy difference (photon energy) between the initial and final states of these transitions is measured. Or equivalently, all these total energy spectra are shifted relative to a common reference (*e.g.* the 1*s*-X energy), as shown in Fig. 5i. The actual optical spectrum is the superposition (Fig. 5j) of 15 of these spectra shown in Fig. 5i, or 15 vertically "collapsed" spectra shown in Fig. 5h (the same layout can be seen for each figure in Extended Data Fig. 18).

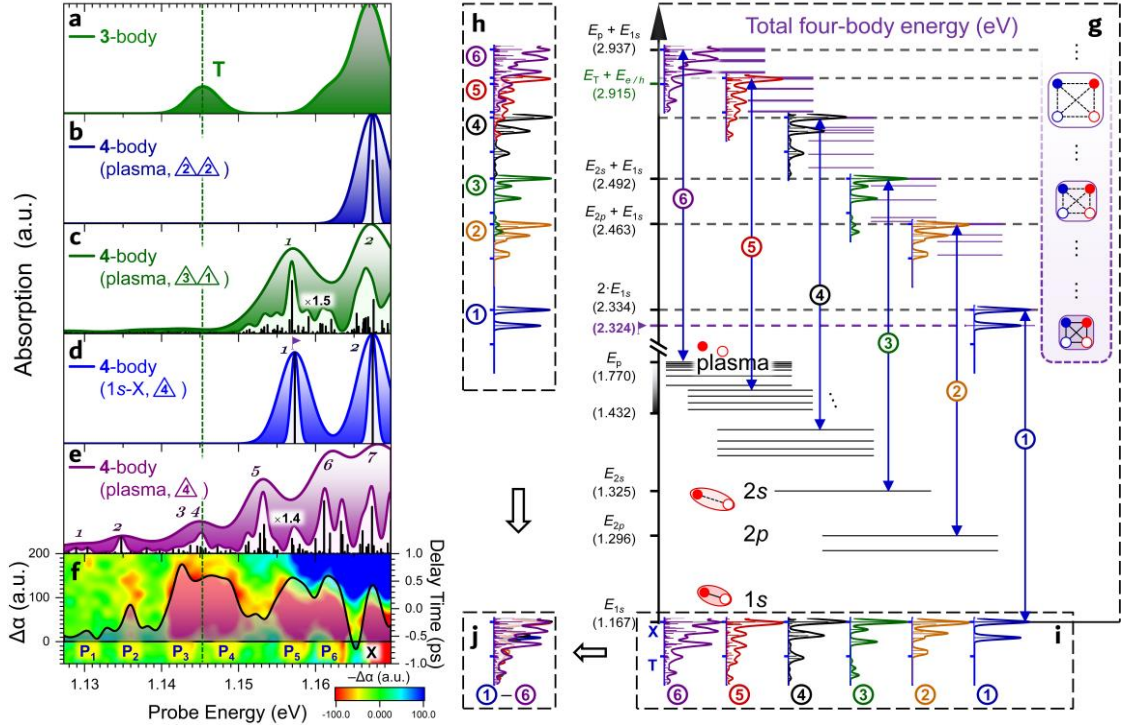

**Fig. 5 | Theory-experiment comparison of absorption spectra. a – e,** Absorption spectra calculated based on the 3B-BSE **(a)** and the 4B-BSE **(b – e)** for ML-MoTe$_2$. **d** shows the spectra of transition between the 2B state (1$s$-X) and the 4B states solved from the full 4B-BSE. **b, c, & e** show the cases with the same 2B state (plasma), but the 4B states solved from the 4B-BSE truncated up to $\triangle_2 \triangle_2$ **(b)**, $\triangle_3 \triangle_1$ **(c)**, and $\triangle_4$ **(e)**. The vertical dashed line is the calculated trion (T) energy. The vertical black lines underneath the spectral profiles mark the calculated spectral positions with the height representing the strength of the dipole transitions calculated from eq. (3) (also eq. (S19) in Methods S13). The spectral function, Γ, in eq. (3) is a Gaussian with a broadening parameter, $\gamma$, chosen to be 0.5 meV (unfilled profiles) or 2 meV (shaded profiles). We label the Gaussian-broadened peaks in each spectrum with italic numerals from low to high energy. The solid black profile in **f** represents Δα at the zero delay for Device #2 at 4K in the charge-neutral regime, where we also overlay the contour of −Δα in the plane of probe energy and delay time (see also Fig. 3d). The features are labelled with $P_1 – P_6$ (and X), consistent with those in Fig. 3d. The relationship between the total-energy spectra and the optical spectra is shown in **g – j** (for the explanation in more details see the main text). The calculated total energies are shown beside the upward axis in **g** for those typical 2B and 4B states (not to the scale, for such a total-energy spectrum and the optical spectra plotted strictly to energy scale see Extended Data Fig. 16 in Methods S13.1). The energy range of 1.432 – 1.770 represents the continuous absorption band (see Methods S12.1). Transition ① and ⑥ correspond to **d & e**, respectively. The flag notation in **g** marks the 4B ground state, whose spectral feature corresponds to peak *1* in **d**.

Figure 5b & 5c show the optical spectra corresponding to $\langle e_3 h_3 | \hat{\mathbf{p}} | e_1 h_1 e_2 h_2 \rangle$ in eq. (4) with the 4B states calculated from the 4B-BSE truncated up to △₂△₂ (Fig. 5b) and △₃△₁ (Fig. 5c) and the 2B state given the plasma state (see transition ⑥ in Extended Data Fig. 18a & 18b, respectively). We notice a peak between T and X in the case of △₃△₁ (peak *1* in Fig. 5c) versus the featureless background in △₂△₂ (Fig. 5b). Strictly speaking, the only peak in Fig. 5b represents a 4B state of non-interacting (1s-X)(plasma) (without wavy lines). In the case of Fig. 5c, △₂△₂ introduces interaction (wavy lines) between the low-order clusters, leading to the formation of △₂~△₂△₂, or (1s-X) ~ (plasma) (in connection with Extended Data Fig. 1e, Methods S1). Therefore, peak *1* and the tiny splitting peaks below X in Fig. 5c should correspond to those of △₃△₁ or △₂~△₂△₂. Clearly, we do not see any spectral features corresponding to P₁ – P₄, especially the features below T. This means that P₁ – P₄ do not originate from cluster △₂△₂ or △₃△₁.

Our next approximation is to include all the 4B states calculated from the full 4B-BSE (truncated up to △₄). The relationship between the total-energy spectra and the optical spectra in this case is shown in Fig. 5g – 5j (also Extended Data Fig. 18c). The optical spectra calculated with the 2B states given the 1s-X (transition ①) and plasma (transition ⑥) are shown in Fig. 5d & 5e, respectively (see Methods S14, Extended Data Fig. 19 & 20, for the convergence test for the spectra in Fig. 5c – 5e). A direct comparison is shown in Fig. 5f and Extended Data Fig. 15c & 15f (It is very important to point out that the experimental result should not be compared with only one of these calculated spectra, because the actual spectrum is a superposition of all possible 2B-4B transition spectra as we illustrated in Fig. 5g – 5j and Extended Data Fig. 18c). We see that the spectral features (peak *1 – 6* in Fig. 5e, and those in Extended Data Fig. 17c & 18c) in the entire spectral region resemble closely to those of P₁ – P₆. In other words, the spectral features of $P_1 - P_4$ originate from cluster ④. By comparing various 2B-4B transitions in Fig. 5g – 5j for cluster ④ (also Extended Data Fig. 17c & 18c), clusters ③④ (Extended Data Fig. 17b & 18b), and ②② (Extended Data Fig. 17a & 18a) both on the total-energy scale and optical spectrum, we could make the following conclusion: The states from cluster ④ for each of the same 2B states have lower energies and are thus more stable than those of ②~② (BX) and ③~④ (T$^+$~e and T$^-$~h), indicating the most stable existence of cluster ④ or quadruplon. Importantly, we notice from the transitions in the total-energy scale that the lower-frequency features (below T) in the optical spectrum such as $P_1 - P_3$ correspond to the transitions (such as ④, ⑤, and ⑥ in Fig. 5g) between the more-excited states in the 4B manifold. This explains why $P_1 - P_3$ decays faster as we mentioned in connection with the discussions of Extended Data Fig. 7a & 7d for Device #4. Nearly all the 2B-4B transitions shown in the total-energy spectrum in Fig. 5g, *e.g.* from ① (low energy) to ⑥ (high energy), can contribute to the spectral peaks between T and X. Especially, the spectral peak related to the 4B ground state is calculated to be between T and X (close to $P_5$ or $P_6$, see the flag notation in Fig. 5d). This explains why $P_6$ has the largest intensity and appears the earliest with increasing pump (Fig. 4). However, only the transitions between the highly excited 2B and 4B states such as those of ③ – ⑥ in Fig. 5g, can contribute to the peaks well below T. The fact that $P_1$ only emerges at high pump levels signifies the existence of highly excited states of both the 2B & 4B entities. Through such 2B-4B spectroscopy, we observe a much more complex many-body system with more refined interplays of these 2B, 3B, and 4B complexes than the simple picture of the Mott transition. The similar spectra (to *e.g.* Fig. 5d & 5e) were also calculated for the inter- (σ– σ+) and intra- (σ+ σ+) valley 4B states (Extended Data Fig. 15), and extensively discussed in connection with such continuous Mott transition (see Methods S13).

One more experimental fact to support the picture of 2B-4B transitions is that, the relative position of $P_1 - P_6$ are less affected by the temperature than the T and X peaks (Fig. 2e & 2f). We notice that the transition energies of T and X follow the same sensitive temperature dependence as bandgap, given by the Varshni formula. The total energy of a 4B system also follows a similar temperature dependence. But since the transition energies of $P_1 - P_6$ are determined by the differences between the total energies of the 4B system and the 2B system, the difference becomes much less sensitive to temperature change. This explains the much weaker temperature dependence of the $P_1 - P_6$.

It is important to note that $\triangle_4$ in the 4B cluster expansion represents a $2e2h$ 4B irreducible cluster, where none of the 2 electrons or 2 holes belongs to a specific exciton, or there is no a well-defined exciton in such a 4B system. The entity is therefore not a bi-exciton or an excited-state bi-exciton anymore, rather a quadruplon, as schematically shown as the last cluster in Fig. 1b – 1e. We notice that exactly such a system was recently calculated for a quantum dot system and was called a "quadron"[2]. Therefore, our consistent theoretical and experimental results show that the spectral features corresponding to $P_1 - P_4$ indeed originate from the quadruplons. Similar quadruplon consisting of 4 identical fermions was recently theoretically studied in a 1D Hubbard model[1]. Recent study showed that the $e$-$h$/$e$-$e$ exchange interaction[22,41] can result in fine structure of bi-exciton, which would correspond to $\triangle_2 \sim \triangle_2$. Therefore, quadruplon features from irreducible cluster $\triangle_4$ should not have been called BXFS[53] (see Reference List for an explanation in more detail).

**Conclusion**

The focus of this paper is the experimental observation of several new spectral peaks that appeared below the trion features and the development of a microscopic theory based on the 4B-BSE to explain the origin of these peaks. The observation of these peaks covering a large energy range of ~ 40 meV was made experimentally possible through a helicity-resolved pump-probe spectroscopy in gate-controlled ML-TMDC samples. It was at the charge-neutral point that we observed the new rich spectral features. This might be one of possible reasons that the they were not observed in previous ultrafast-spectroscopic experiments on ML-TMDCs (for more reasons in detail see Methods S9). Through the establishment of a systematic relationship between the Feynman diagrams of the 4B-BSE and cluster expansions of various orders, we showed that these new spectral features could not be explained by clusters associated with trions or bi-excitons, while the irreducible cluster associated with 2$e$2$h$ was necessary and sufficient in producing these new spectral features. This agreement between our new theory and experiment established the existence of a new correlated 4B entity, the quadruplon. The existence of the 4B entity is further corroborated by the pump dependent experiment where systematic changes of the absorption spectra are consistent with the possible occupation of higher 2B states and the existence of the final 4B states upon the absorption of the probe photons. Interestingly, these 4B states can stably exist even at a relative high temperature of ~ 60 K.

We would like to make a few comments about other multi-particle states that are products of lower order clusters. Examples include the Suris tetron[54-56] that contains a conventional exciton and another $e$-$h$ pair formed near the Fermi surface, or △ ~ △ bi-exciton-like, or trion-hole-like[57] entity in the language of this paper. Other examples include the quarternion[58] (an exciton bound to two free charges of the same sign), charged bi-exciton or trion-exciton[23-27], and even 6B and 8B ones: the hexciton and oxciton[59] (an exciton plus multiple fermi-sea e-h pair, also known as the sub-terms of the exciton polaron together with the Suris tetron), or tri-exciton[60] △ ~ △ ~ △ and quad-exciton △ ~ △ ~ △ ~ △ and dropleton △ ~ △ ~ △ ~ … ~ △ as exciton ladders, *etc*. All these many-body complexes are reducible to product states of excitons, trions, and charges. Not association with higher irreducible clusters than 3 has been made. Thus the possibility of 4$^{th}$ order irreducible clusters has not been considered so far[23-27,54-56,58-60]. The difference between product of lower order irreducible clusters and the corresponding higher order irreducible clusters are also important and analogous in the field of elementary particles, where meson molecules and genuine tetra-quarks represent separate milestone discoveries[4,10-13]. In the context of ultracold gas, the 4B states (related to the Efimov effects) were theoretically discussed[61] in terms of all irreducible clusters such as $B_4$, $B_3+B$, $B_2+B+B$, $B_2+B_2$, or $B+B+B+B$. Attempts were also made to explain an earlier experiment[62] in terms of the above theory[61]. Also, the quadruplons consisting of 4 holes were theoretically revealed in the sequence of Q (quadruplon), T (triplon), D (Doublon), and B (Band-like part) in the calculated two-hole excitation spectrum at different filling factors in a 1D Hubbard model[1].

To conclude, the observation of quadruplons would also contribute in a very important way to the fundamental understanding of the Mott transition, one of the most celebrated aspects of condensed matter physics. The existence of a new 4B entity adds to the complexity of the Mott physics and would definitely lead to more new physical phenomena. For example, the co-existence and mutual conversion of various known species have been the subjects of extensive studies for their interesting optical transitions, leading to bi-exciton gain[63] or trion gain[31]. Bi-exciton emission processes are also related to the generation of entangled photons. It would be of great interests to study the co-existence and mutual conversion of quadruplons with other known species. The existence of quadruplons opens many new exciting opportunities to study its consequences on optical gain, generation of quantum states, and nonlinear optics, beyond the Mott transitions.

Finally, our study may stimulate more similar studies to search for clusters or many-body complexes of even higher orders beyond the quadruplons and their excited states. Another interesting issue that arises from this study is the experimental study of the total energy spectrum of a 2*e*2*h* 4B system. As shown in Fig. 5g, the total energy of the 2*e*2*h* system extending over ~ 0.3 – 0.6 eV shows very rich spectral features that cannot be determined through optical spectroscopy. A comparison of our 4B-BSE theory with the total energy spectrum would allow us to study many different states of the 4B irreducible clusters in a more unique fashion. More importantly, it would allow the determination of the relative energetic stabilities of various states. Among all of these states, the experimental verification of the relative stability of bi-excitons versus quadruplons would be of great special interest.

**Data availability**

The data that support the findings of this study are available from the corresponding author upon request.

**Acknowledgements**

The authors thank Alan H. Chin for his critical reading of the manuscript and for his comments. The authors acknowledge the following financial supports: National Natural Science Foundation of China (Grant No. 91750206, No. 61861136006, and No. 61705118); National Key R&D Program of China (Grant No. 2021YFA1400700); Beijing Natural Science Foundation (Z180012); Beijing National Center for Information Science and Technology; Tsinghua University Initiative Scientific Research Program.

**Author Contributions**

J.T. and Q.Z. prepared the materials and fabricated the devices with help of Z.W. for the metal electrodes.   X.D. and H.S. built the pump-probe optical setup.   H.S., X.D., J.T., and Q.Z. optimized the optical system and automated the data acquisition.   J.T. and Q.Z. conducted the optical experiments and measurements.   J.T. processed the data under guidance of C.Z.N..   J.T. developed the many-body theory and performed the theoretical modelling and calculations under guidance of C.Z.N..   C.Z.N. made the initial connections with the irreducible cluster expansion model and the quadruplons, while J.T. and C.Z.N. eventually worked to establish these connections.   J.T. and C.Z.N. performed the data analysis and wrote the manuscript.   C.Z.N. supervised the overall project.

Corresponding authors
Correspondence to Cun-Zheng Ning, ningcunzheng@sztu.edu.cn

**Ethics declarations**
Competing interests
The authors declare no competing interests.

# SI: The Quadruplon in a Monolayer Semiconductor

Jiacheng Tang,[1,2,3,4] Hao Sun,[1,2,3] Qiyao Zhang,[1,2,3,4] Xingcan Dai,[1] Zhen Wang,[4,1,2,3]
Cun-Zheng Ning[4,1,2,3]*

[1]Department of Electronic Engineering, Tsinghua University, Beijing 100084, China

[2]Frontier Science Center for Quantum Information, Beijing 100084, China

[3]Tsinghua International Center for Nano-Optoelectronics, Beijing 100084, China

[4]College of Integrated Circuits and Optoelectronic Chips, Shenzhen Technology University, Shenzhen 518118, China

* Corresponding author. Email: ningcunzheng@sztu.edu.cn

## List of Methods

**Methods**

## S1. The cluster expansion model with different orders of truncation

The cluster expansion model is one of the most common and intuitive approaches to many-body interactions, in which an interacting N-body system is expressed as the sum of all the possible combinations involving irreducible clusters of the N$^{th}$ and all the lower orders.

In our special case of the two electrons and two holes (2$e$2$h$) with the inclusion of the spin degree of freedom, the 4-body (4B) cluster expansion is expressed in terms of the following 5 classes, *i.e.* ①①①①, ②①①, ②②, ③① and ④, as shown in Fig. 1 in the main text. To better recapitulate the essence of the 4B interaction, we consider a simplified model with the Hamiltonian truncated up to different orders, *i.e.* up to cluster ②②, ③① and ④, as illustrated in Extended Data Fig. 1a, 1b, & 1c, respectively. $H_{ii}$'s ($i$ = 1, … 5) stand for the diagonal matrices of the Hamiltonians for the 5 corresponding classes, with their dimensions given by the numbers of the combinations of the irreducible clusters in Fig. 1 in the main text. The base vectors are illustrated accordingly on the right side in Extended Data Fig. 1a – 1c. The full 5 basis in Extended Data Fig. 1c correspond to the above 5 classes in the 4B cluster expansion.

To illustrate appearance of the weak interactions (the wavy lines in Fig. 2a in the main text) among the irreducible clusters, we recall the basic process of successive truncations in the cluster expansion approach. Starting from $H_{11}$ (①①①①) for the case without any correlation or interaction, the inclusion of the next order, $H_{22}$ (②①①), not only increases the dimension of the new total Hamiltonian, but also introduces the coupling ($H_{12}$ and $H_{21}$) between the two orders. Such a process continues until a final cut off of the cluster expansion is made. Each time we include one more cluster of higher order into a truncated expansion, mainly two changes occur to the Hamiltonian (see the variances in Extended Data Fig. 1a – 1c): First, new diagonal elements are introduced (*i.e.* $H_{ii}$, $i = 3, 4, 5$), whose appearance expands the dimension (size) of the Hamiltonian matrix and would lead to new many-body states; Second, the original off-diagonal elements (blocks) are updated to new ones (*i.e.* $H_{ij} \to H'_{ij} \to H''_{ij}$). Such off-diagonal elements in Extended Data Fig. 1a – 1c introduce the weak interactions among the irreducible clusters of lower orders indicated by wavy lines, as depicted in Extended Data Fig. 1d – 1f, respectively. In the above 3 cases with different orders of truncation, the introduced weak interactions are accordingly different as well (see the wavy lines indicated in the different grey levels in Extended Data Fig. 1d – 1f). Once a final cut-off is made, we obtain the total Hamiltonian, whose off-diagonal elements determine the final interactions among the elements of the various clusters. In our specific case, the diagonalization of the final Hamiltonian in Extended Data Fig. 1c allows the determination of all the states of the 2*e*2*h* 4B system, or of all the possible 2*e*2*h* entities.

If we truncate the Hamiltonian up to cluster △₂△₂ (Extended Data Fig. 1a), cluster △₂△₂ would produce the entities such as non-interacting (1*s*-X)(1*s*-X), (1*s*-X)(2*s*-X), … (1*s*-X)(plasma) and so on (△₂△₂ without the wavy line, see the △₂△₂ entities in Extended Data Fig. 1d). Once we introduce one more cluster of higher order, *i.e.* △₄, into the Hamiltonian, cluster △₂△₂ would produce the entities such as weak-interacting (1*s*-X)~(1*s*-X), (1*s*-X)~(2*s*-X), … (1*s*-X)~(plasma) and so on (△₂~△₂ with the wavy line, see the △₂~△₂ entities in Extended Data Fig. 1f), or namely, bi-excitons and excited-state bi-excitons. Similarly, cluster △₃△₁ would produce the entities of △₃~△₁ (see the △₃~△₁ entities in Extended Data Fig. 1f). Last but not least, we are more interested in cluster △₄, and the resulting 4B irreducible entities, quadruplons. Based on the above model, it is clear that the quadruplons are completely new and different entities beyond those conventional bi-excitons.

We point out that the simplified model is only presented here as an intuitive picture for the understanding of the essence of the 4B physics involved, but not employed for any specific calculation in this paper. Instead, our numerical results in this paper are obtained completely using the microscopic many-body Hamiltonian, as will be discussed in Methods S12.2 – S13.2.

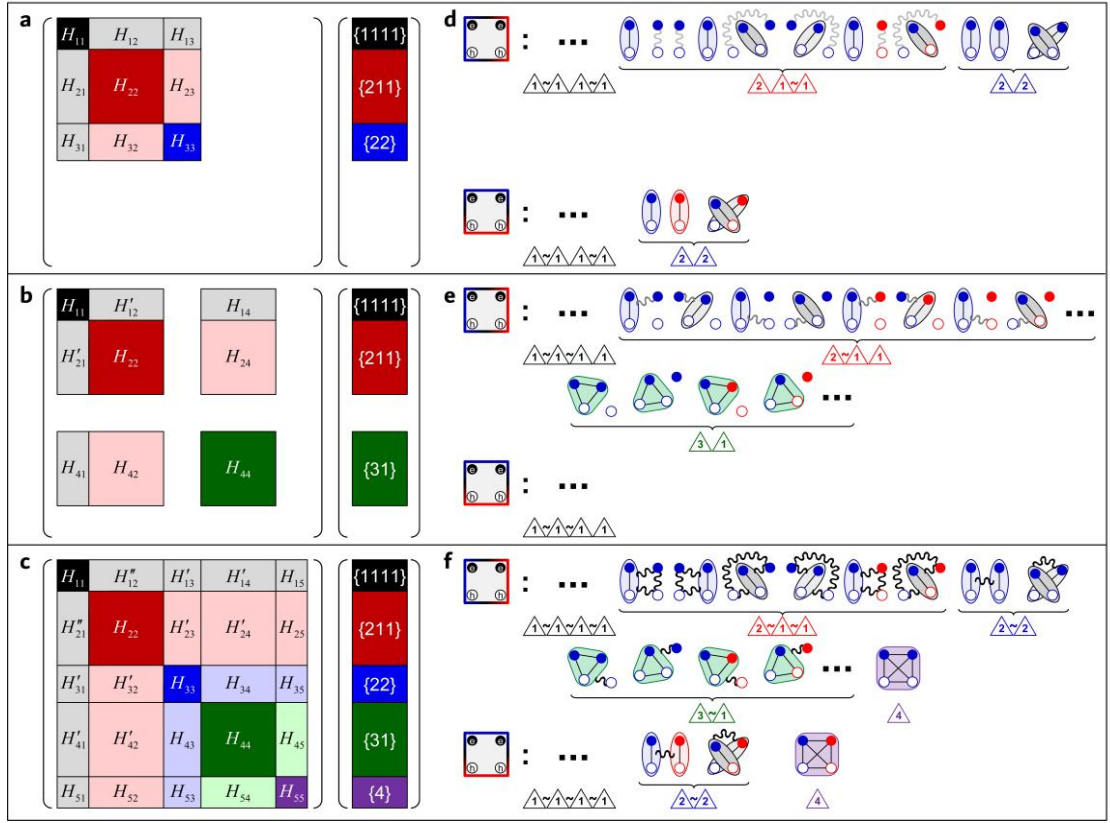

**Extended Data Fig. 1 | a – c,** A simplified model with the Hamiltonian truncated up to cluster △₂ (a), △₂△₂ (b) and △₄ (c). The full basis on the right side in **c** correspond to the clusters of different orders, while the basis in **a & b** are incomplete due to the corresponding truncation. **d – f,** Representative 2$e$2$h$ 4B entities solved from the eigen-energy equations with the Hamiltonian truncated up to cluster △₂△₂ (d), △₃△₁ (e) and △₄ (f). The wavy lines in the different grey levels in **d – f** indicate the different weak interactions introduced by the joining of the different clusters of higher orders.

## S2. Device fabrication

To fabricate an electrically-gated device (Fig. 2b in the main text), the metal electrodes were predefined by photolithography on the SiO$_2$/Si substrate and then 50/30 nm

Au/Ti were deposited through electron beam evaporation. The monolayer of $MoTe_2$ and the films of hexagonal boron nitride (h-BN) and graphite were mechanically exfoliated from those commercial bulk crystals (2D semiconductors or HQ graphene Inc.) and then transferred onto the polydimethylsiloxanes (PDMS). The ML-$MoTe_2$ samples were first identified through the contrast of the optical microscope images, with their layer thickness finally identified by the photoluminescence (PL) emission. More importantly, we screen the samples to select the best ones for the following optical experiments. The screening methods include the measurements of continuous-wave (CW) reflection contrast spectra (RCS) and PL spectra (such methods will be discussed in more details in Methods S3.2), by which one can know the crystal quality of a sample or the level of defects that the sample contains. All the transfer processes were performed using a home-build transfer stage integrated with an optical microscope at a moderate heating temperature. The h-BN film (~ 50 nm) and ML-$MoTe_2$ were transferred subsequently onto one of the electrodes. The graphite stripe (~ 10 nm) was transferred as a top contact bridging between the ML-$MoTe_2$ and one of the other metal electrodes. Another h-BN film (~ 10 nm) was transferred on top of the device for a protection. Finally, the device was annealed at ~ 200 ℃ for 3 hours to reduce the bubbles between the layers. Similar device structure and the fabrication method can be seen in Ref. [1,2].

**S3. Optical spectroscopy**

The fabricated ML-$MoTe_2$ devices were characterized in a home-build micro-PL system at a liquid helium temperature of ~ 4K. For the measurement of the CW RCS, a tungsten halogen lamp (Thorlabs SLS201L) was used as the white light source to detect the sample through a 100x NIR-optimized objective (Mitutoyo NA = 0.7) with the spot diameter of ~ 3 μm. The reflection signals were collected by the same objective and finally delivered to a Princeton Instruments spectrometer (SpectraPro HRS-550) equipped with a LN-cooled InGaAs CCD (PyLoN-IR). The gate-dependent measurement was carried out by using a semiconductor parameter analyzer (Keysight B1500A).

**S3.1. Identification of the spectral energies of T and X in the CW optical spectroscopy**

As mentioned in the main text, the RCS is defined as (R – R') / R', where R and R' are the reflectance spectra from the regions with and without the ML-MoTe$_2$ sample. Here, we take the results of Fig. 3a in the main text (for Device #2 at 4K) as an example to explain the data processing. The results of other devices were obtained in the same way. First, the unprocessed signals of R and R' are shown in Extended Data Fig. 2a & 2e for the charge-neutral case (Extended Data Fig. 2a) and for the doping case (Extended Data Fig. 2e). The corresponding results of (R – R') / R' are given in Extended Data Fig. 2b & 2f. Then, we utilized the approach of adjacent-averaging smoothing to obtain the baselines (see the red lines in Extended Data Fig. 2b & 2f), with the results of subtracting the baselines shown in Extended Data Fig. 2c & 2g, respectively. Similar data processing can be seen typically in Ref. [3] and *etc*. In addition, we also performed the second derivatives (SD) for (R – R') / R' (the black lines in Extended Data Fig. 2b & 2f) with respect to the spectral energy. The corresponding results are shown in Extended Data Fig. 2d & 2h. As we see in Extended Data Fig. 2a – 2h, the obtained spectral energies of T and X are less dependent on the methods of the data processing, thus can be well identified. All these results can well describe the CW absorption features of the ML-MoTe$_2$ sample. A simplified effective dielectric multi-layer system (air/MoTe$_2$/h-BN/metal) was modelled in Ref. [1] to determine the absorption and to discuss the influences of the substrates (see SI text S1 & S2 in Ref. [1]).

The gate-dependent CW PL results are shown in Extended Data Fig. 2i & 2j. The peaks of T and X are spectrally located at ~ 1.146 – 1.148 eV and ~ 1.168 – 1.170 eV, respectively, which are consistent with the CW absorption results in Extended Data Fig. 2a – 2h.

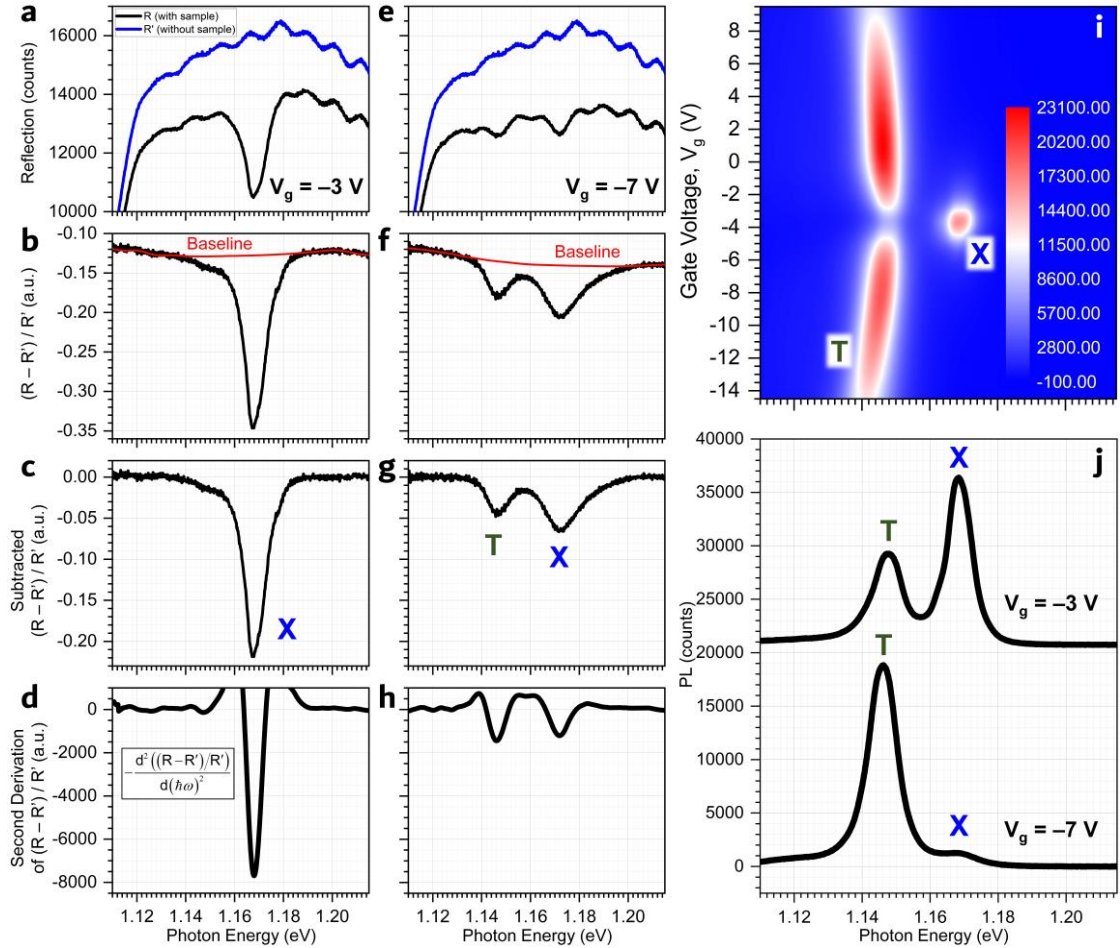

**Extended Data Fig. 2 | a, e,** Reflectance spectra from the regions with (R) and without (R') the ML-MoTe$_2$ sample (Device #2 at 4K) for the charge-neutral case at $V_g$ = –3 V **(a)** and for the doping case at $V_g$ = –7 V **(e)**. **b, f,** (R – R') / R' and the corresponding baselines for the cases of $V_g$ = –3 V **(b)** and $V_g$ = –7 V **(f)**. **c, g,** (R – R') / R' with the baselines subtracted for the cases of $V_g$ = –3 V **(c)** and $V_g$ = –7 V **(g)**. **d, h,** SD performed versus photon energy for (R – R') / R' shown in **b** & **f**, respectively. **i,** CW PL contour in the plane of photon energy and gate voltage. **j,** CW PL spectra selected from **i** at $V_g$ = –3 V and $V_g$ = –7 V.

### S3.2. Representative CW RCS and PL spectra for ML-MoTe$_2$ samples with or without visible defect features

Typically for ML-TMDC samples, the defect-related PL peaks (or so-called localized states) can be observed below T with CW laser excitation[4-8]. Extended Data Fig. 3 & 4 show representative CW RCS and PL spectra for ML-MoTe$_2$ samples with or without such visible defect features. In our PL tests, we used a 633-nm CW laser (weak excitation), and additionally a 650-nm femtosecond (fs) pulsed laser (strong excitation), to excite ML-MoTe$_2$ samples. The PL results under these two excitation conditions are shown in Extended Data Fig. 4. The levels of the weak- and strong-excitation densities were estimated to be ~ $10^{8-9}$ and ~ $10^{13}$ cm$^{-2}$, respectively. We note that the strong-excitation condition was similar to that used in our pump-probe experiments, and our weak excitation conditions (*e.g.* 200 μW 633-nm CW laser excitation, used for Extended Data Fig. 3) are the typical ones that were similarly used in Ref. [4-8] to observe defect PL as well in other ML-TMDCs. For a good sample we selected, we do not see any defect emission under both of the strong- and weak-excitation conditions (*e.g.* Extended Data Fig. 2i, 2j, 3b, and the red and black lines in Extended Data Fig. 4), or any defect absorption (*e.g.* Extended Data Fig. 3a). Typically for a bad sample, we expect to see defect emission at even a low pump level (*e.g.* Extended Data Fig. 3d, or the blue lines in Extended Data Fig. 4). Sometimes, one can also see the defect-related absorption feature below the T peak (*e.g.* Extended Data Fig. 3c) for a bad sample in the CW RCS.

In addition, we notice the emission peaks of exciton (or trion) phonon replicas were also observed spectrally below T for W-based materials in recent reports[9-12]. However, it is clear that for our high-quality ML-MoTe$_2$ samples, there are no such additional emission or absorption features existing below T, or in the same spectral range as P$_1$ – P$_6$ (*e.g.* Extended Data Fig. 3a, 3b, and the red and black lines in Extended Data Fig. 4).

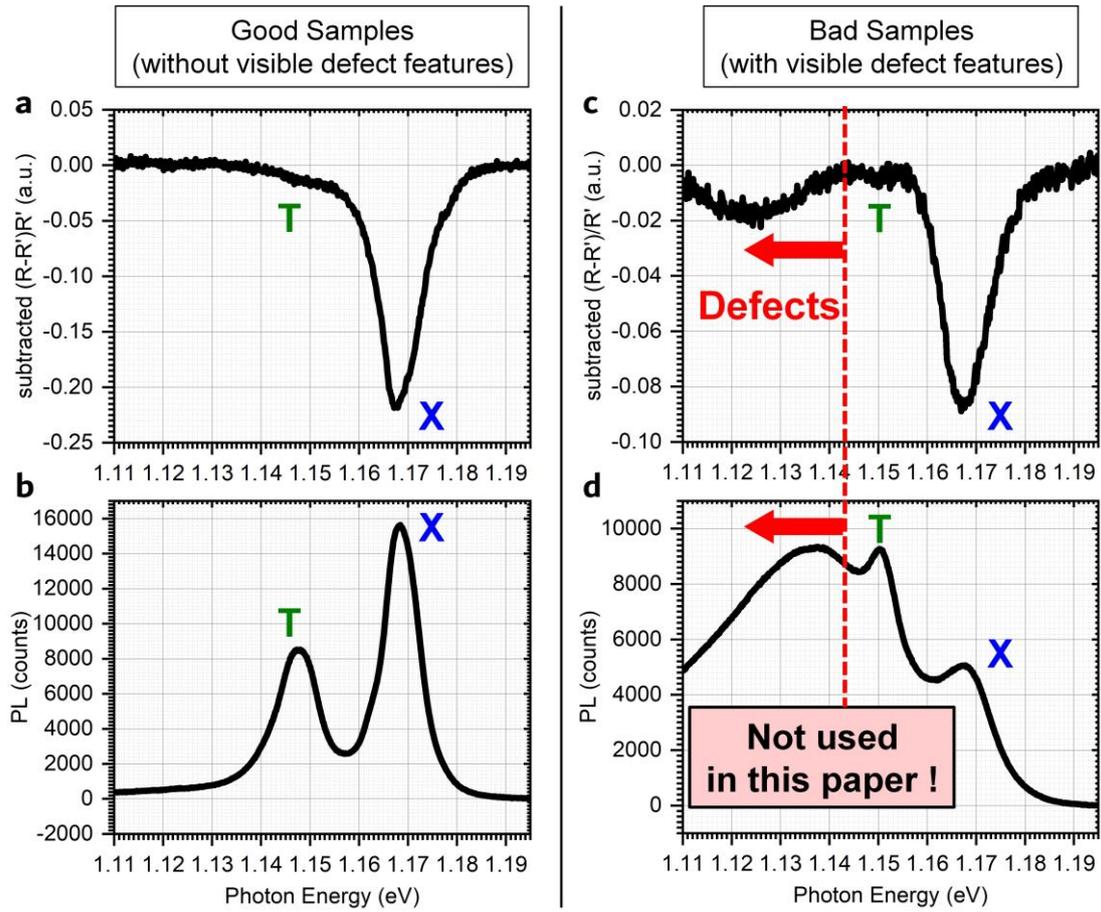

**Extended Data Fig. 3 |** Representative CW RCS **(a & c)** and PL **(b & d)** spectra for ML-MoTe$_2$ samples with **(c & d)** or without **(a & b)** visible defect features. All of these spectra were measured at 4K under the charge-neutral condition.

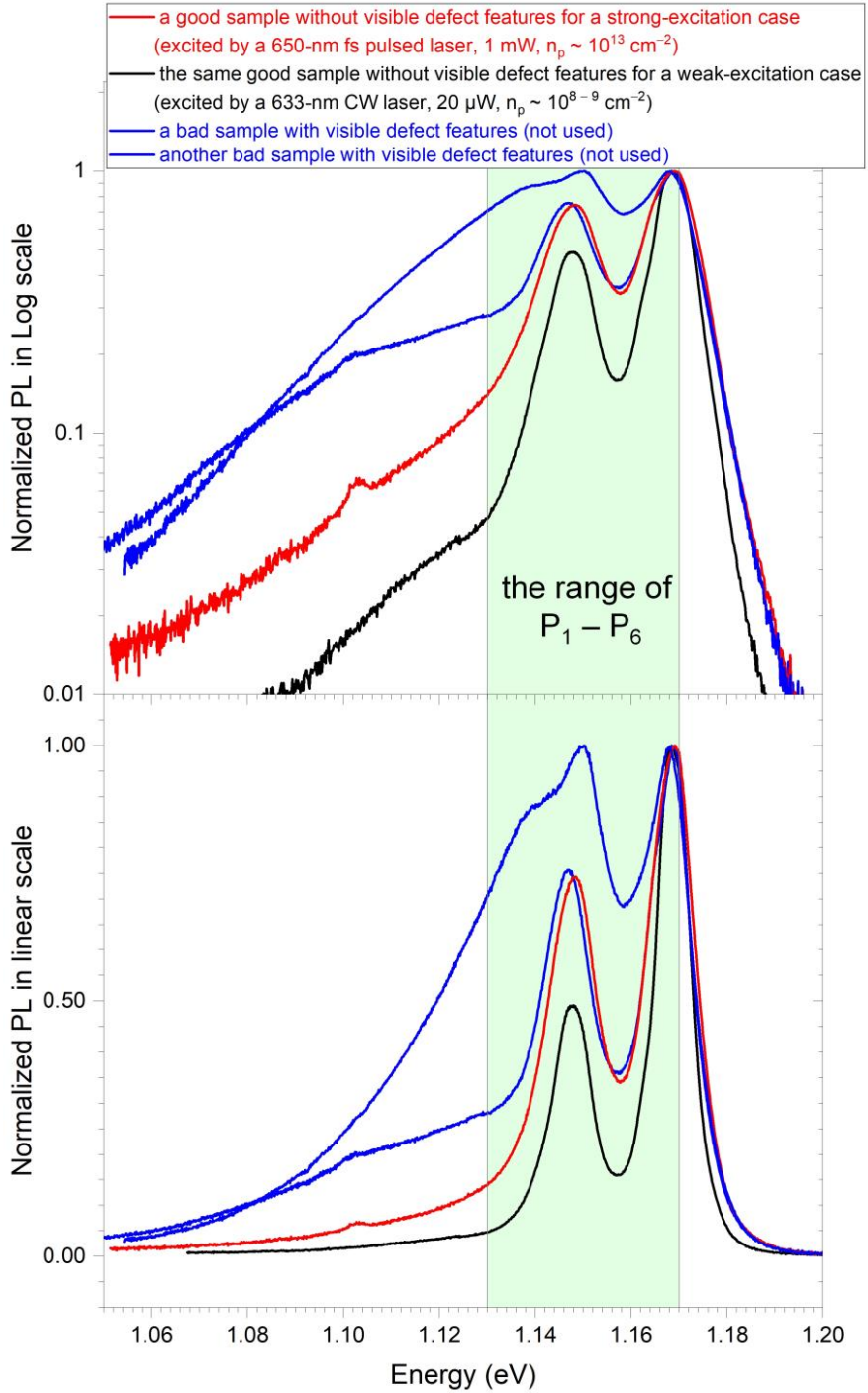

**Extended Data Fig. 4 |** Representative PL spectra for ML-MoTe$_2$ good samples for strong- (red) and weak-excitation (black) cases, as plotted in Log (upper panel) and linear scales (lower panel). All of these spectra were measured at 4K under the charge-neutral condition. For better comparison, we also showed the PL spectra for those ML-MoTe$_2$ bad samples (blue) with visible defect features below T, which were not used in the pump-probe experiments.

### S3.3. Helicity-resolved TDAS (or TDRS)

For the helicity-resolved transient differential absorption (or reflection) spectroscopy (TDAS or TDRS) (Extended Data Fig. 5), a 1040 nm femtosecond laser (Spirit from Spectra Physics with the pulse width of ~ 500 fs, and the repetition rate of 400 kHz) was divided into two beams: pump and probe. The pump beam was sent to an OPA system (Light Conversion TOPAS) to achieve the tunability of its central wavelength and then modulated by a mechanical chopper. The typical pump-excited *e-h* pair density ($n_p$) was estimated to be at $n_p$ ~ $10^{12}$ /cm$^2$ (see Methods S3.4). The probe pulse was spectrally broadened by a sapphire crystal and then filtered by a grating-based pulse shaper to achieve the narrow FWHM of ~ 1 – 2 meV. The intensity of the probe pulse was typically one order smaller than that of the pump pulse. A combination of a λ/2 waveplate, two linear polarizers, and a λ/4 waveplate was used to control the circular polarizations. Both the pump and probe beams were first polarized by the λ/2 waveplate and the linear polarizers (LP), then combined by a non-polarized beam splitter, then delivered to the λ/4 waveplate, and finally delivered to a 50x NIR-optimized objective (Mitutoyo NA = 0.4). The strong pump signal was greatly filtered out via setting the band-pass center of the grating in the spectrometer to the desired probe energies, which are always off-resonant to the central energy of the pump pulse. The TDR signals at a series of discrete probe wavelengths were detected by a high-gain InGaAs detector using a lock-in technique. As mentioned in the main text, the TDR (ΔR) is defined as the variance between the transient reflection (TR) signals with ($R_p$) and without ($R_0$) the pump: $\Delta R = R_p - R_0$. The TDR (ΔR) is proportional to $-\Delta\alpha$, the negative value of the TDA (Δα), thus can well describe the transient absorption (TA) features. Their relation can be also written as $-\Delta\alpha = -(\alpha_p - \alpha_0) \propto \Delta R = R_p - R_0$. The similar data processing for the TDAS and TDRS in the pump-probe experiments were described in Ref. [2].

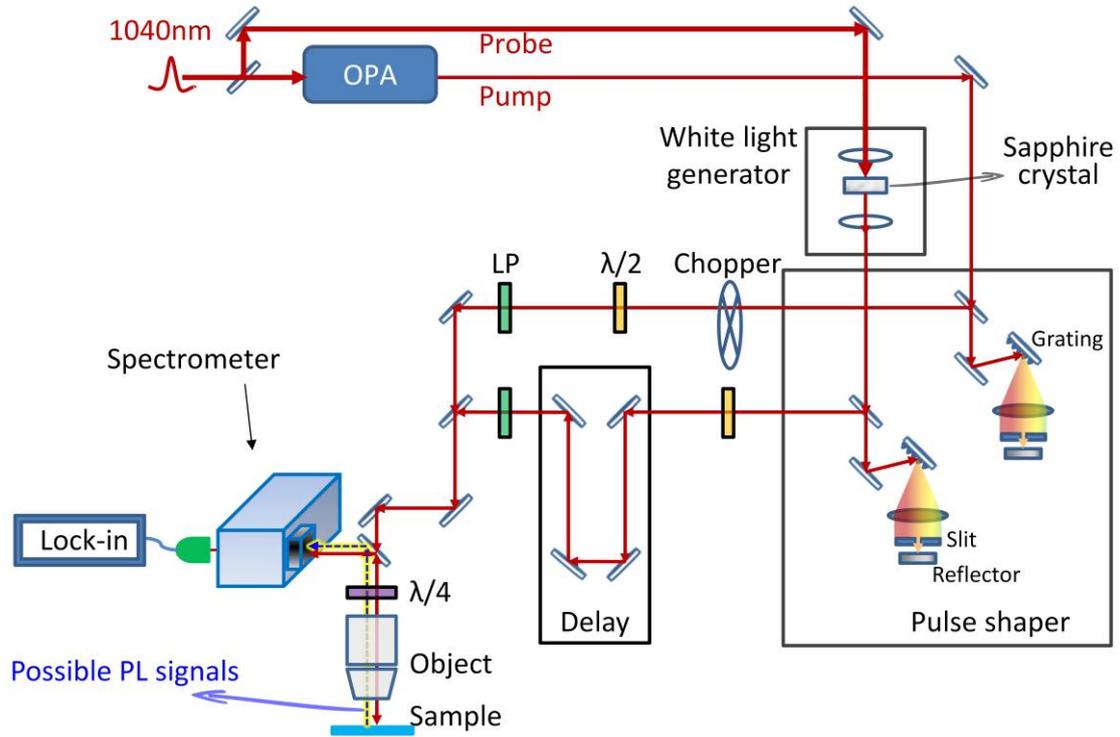

**Extended Data Fig. 5 |** Schematic of the time-resolved pump-probe setup.

### S3.4. Estimates of the *e-h* pair density

The optically generated total *e-h* pair density ($n_p$) is estimated via $n_p = \dfrac{2\lambda\alpha_{\text{eff}}}{\pi^2 d^2 \hbar c f} \cdot P$, where $P$ is the pump power, $\lambda$ is the central wavelength of the pump photon, $\alpha_{\text{eff}}$ is the effective absorption coefficient of the sample, $d$ is the diameter of the pump spot, and $f$ is the pulse repetition frequency. In our specific case, $\lambda \approx 1050$ nm, $\alpha_{\text{eff}} \approx 1\%$ (Ref. [1]), $d \approx 4$ μm, $f = 400$ kHz.

### S4. The reasons and advantages for choosing to study MoTe$_2$ among TMDCs

**1)** Among the family of ML-TMDCs (*i.e.* WS$_2$, WSe$_2$, MoS$_2$, MoSe$_2$, and MoTe$_2$), MoTe$_2$ is the only one system that the optical emission (or absorption) is located in the spectra range of near-infrared wavelength[1,13-15]. Such spectral range is transparent for the Silicon absorption, thus providing great opportunities for Silicon-based opto-electronic integration, such as the on-Silicon MoTe$_2$ nanolaser[15]. MoTe$_2$ is also known for the relatively high carrier mobility, which will be of benefit to nano electronic applications, such as MoTe$_2$-based field effect transistors[16,17]. Recently, longer valley polarization lifetime[2] was discovered for ML-MoTe$_2$ than other ML-TMDCs, which enables it to be an ideal platform for potential quantum information applications. Therefore, one of the reasons that we choose to study MoTe$_2$ is the advantages for the potential development of the ever new opto-electronic devices.

**2)** We believe that the existence of quadruplon is a more generic phenomenon for semiconductors. The question is which system is more favorable for experimental observation. In W-based TMDCs, there are more complicated features in the similar spectral range due to the lower energy of dark excitons and the related excitonic species[4-12,18-23]. These features include fine structures due to different spin-valley configurations (including bright/dark states)[5-7,18-22], phonon replicas[9-12], *etc*. The poor spectral resolution and linewidth broadening add to the difficulty for a unique identification. In the Mo-based TMDCs, there is much less spectral complexity in the spectral range of interest due to the lower energy of the bright exciton than the dark exciton. Among the three Mo-based TMDCs, MoS$_2$ has two conduction bands separated by only ~6 meV, making the situation more similar to the W-based materials. The conduction band splitting in the MoSe$_2$ and MoTe$_2$ is around 30 – 60 meV. Therefore the quadruplon features are well separated from those of other excitonic complexes mentioned above. Therefore, we believe that one should be able to observe similar quadruplon effects in MoSe$_2$ as well.

For the above reasons, as we concerned, possible quadruplon signals would highly overlap with those of fine structures (bright/dark state), *etc*, if we chose to study WS$_2$, WSe$_2$, or MoS$_2$, which would make the quadruplon signals difficult to extract. To avoid such potential confusion for the quadruplon features, therefore, we consider MoTe$_2$ and MoSe$_2$ are more favorable for the type of studies of the present paper than other TMDCs.

**3)** In fact, there have been growing number of papers on MoTe$_2$. For various experiments, ML-MoTe$_2$ samples have been fabricated with high quality and demonstrated to be a stable system[1,2,13-15,24] with excellent device performance[15]. There are many papers in the literature that reported the optical characterization[1,2,13-15,24] or low-temperature tests[1,2,14,15,24] or ever advanced condensed-matter experiments[25-27] for MoTe$_2$. From theoretical side, different *ab-initio* approaches[28,29] produced consistent parameters about band structure, Keldysh potential, *etc* for ML-MoTe$_2$. Those values of parameters were also verified by our theoretical calculation (see Methods S11).

**S5. Helicity-resolved TDAS with a finer spectral resolution of 0.1 meV (Device #1)**

Extended Data Fig. 6a shows the helicity-resolved TDAS for Device #1 at 4K under the charge-neutral condition with a finer spectral resolution of 0.1 meV (see Fig. 2e in the main text for the reproducible results with the resolution of 0.2 meV). For both of the (σ– σ+) and (σ+ σ+) cases, the spectral lines were fitted by the similar six peaks of $P_1$ – $P_6$ to those in Fig. 2e in the main text. The absorption increase with pump (negative signals, –Δα < 0) for the (σ– σ+) case is stronger than for the (σ+ σ+) case. The contrast between the cross- and co-polarized cases can be seen as well in Fig. 3d & 3i in the main text for Device #2. Such a polarization contrast is also reflected in the theoretical results (see Methods S13.1, Extended Data Fig. 15). As we mentioned in Methods S3.3, the wavelength of the probe beam was tuned to a series of discrete values and swept over a wide spectral range. During the test, we monitored the laser powers of both the pump (modulated) and probe pulses. The real-time values of the laser powers are shown in Extended Data Fig. 6b. The laser power of the pump (modulated) and probe pulses are controlled to be 3.0 ± 0.1 μW and 0.15 ± 0.01 μW, respectively.

The power fluctuations of the pump and probe would lead to the corresponding fluctuations in the TDAS. According to the pump-dependent results shown in Fig. 4c in the main text and Extended Data Fig. 11 in Methods S8, such data fluctuations are less than 1.3%. For example, the signal intensity of $P_6$ in Extended Data Fig. 6a is ~ 6000, with the fluctuation evaluated to be ± 80. It is clear that such fluctuations cannot produce the spectral features of $P_1 - P_6$. Therefore, the laser-power fluctuation can be excluded as the possible source of the spectral peaks of $P_1 - P_6$.

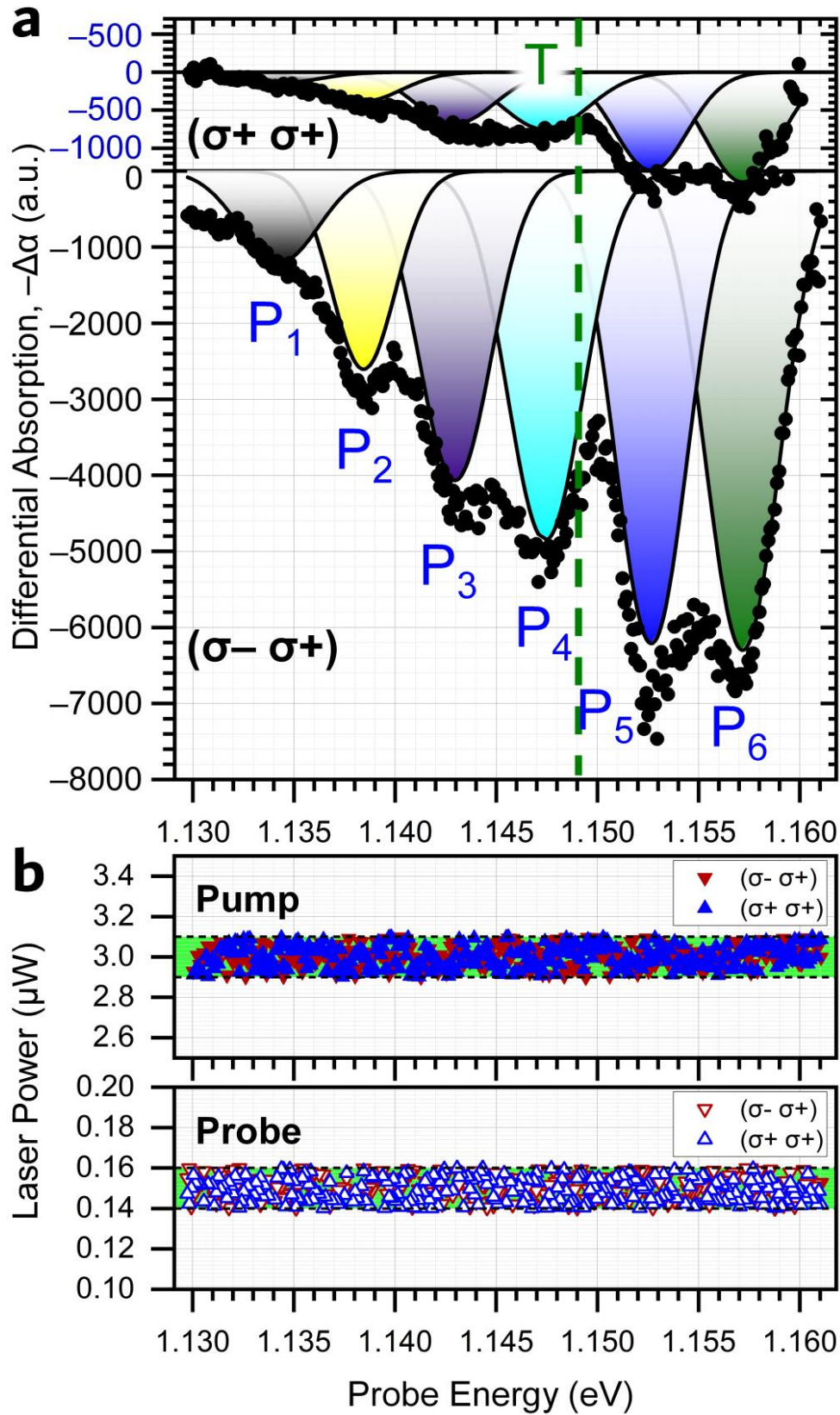

**Extended Data Fig. 6 | a,** TDAS for the cross- (σ– σ+) and co- (σ+ σ+) circularly polarized pump-probe configurations (Device #1, at 4K, under the charge-neutral condition). The fitted Gaussian peaks are marked by $P_1 – P_6$. **b,** Laser powers of the pump (modulated) and probe pulses monitored during the test.

## S6. Additional experimental results (Device #4)

The gate-dependent CW RC map of Device #4 at 4K is shown in Extended Data Fig. 7c. The strongest X peak (~ 1.170 eV) marks the gate-compensated charge-neutral voltage, $V_g = -12$ V. Both T$^-$ and T$^+$ appear at ~ 1.145 eV in the $e$- and $h$-doped regimes, respectively. We can see clearly there are no additional absorption features below X in the charge-neutral regime (white dashed box) without pump.

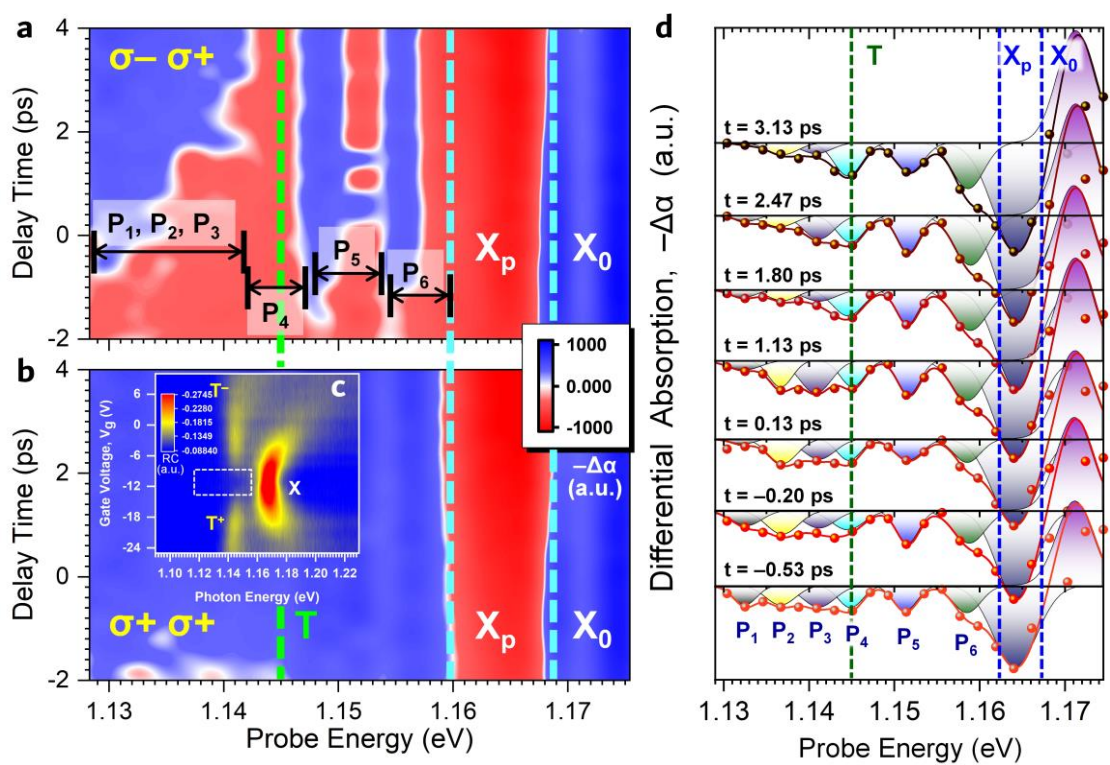

**Extended Data Fig. 7 | a, b,** TDA contours in the plane of probe energy and delay time for the cross- (σ− σ+) **(a)** and co- (σ+ σ+) **(b)** circularly polarized pump-probe configurations (Device #4, at 4K, under the charge-neutral condition, $V_g = -12$ V). **c,** CW RC contour in the plane of photon energy and gate voltage. **d,** Gaussian fittings of the TDAS measured at the different delay times extracted from **a**. The spectrum for each given time delay is fitted with six Gaussian peaks marked by $P_1 - P_6$, together with the other two Gaussians related to X (X$_0$: without pump; X$_p$: with pump). The points are experimental values from **a**, while the solid lines are the results of the Gaussian fittings. The spectral locations of X$_0$, X$_p$, and T marked with the blue and green dashed lines in **a, b, & d** were obtained from the CW RC result in **c**.

Extended Data Fig. 7a, 7b, & 7d show the results of the helicity-resolved pump-probe experiment. The pump energy of 1.182 eV is ~ 10 meV above X with a fluence of ~ 80 μJ·cm$^{-2}$ (corresponding to the $e$-$h$ pair density $n_p$ estimated to be ~ 4 × 10$^{12}$ cm$^{-2}$, see

Methods S3.4 for the estimate), while the probe energy was tuned from 1.130 to 1.175 eV. The gate voltage was switched to –12 V to maintain charge neutrality. The TDAS of Device #4 at 4K are shown in Extended Data Fig. 7a & 7b for the cross- (σ– σ+) and co- (σ+ σ+) circularly polarized pump-probe configurations, respectively. Compared to the spectrum for the case of (σ+ σ+) (Extended Data Fig. 7b), we observe rich spectral features for the case of (σ– σ+) (Extended Data Fig. 7a), as marked with $P_1 – P_6$. Based on their relative positions to T and X, these features and the associated fitted peaks in Extended Data Fig. 7d can be similarly divided into the three energy intervals: $P_1$, $P_2$ (below T), $P_3$, $P_4$ (near T), and $P_5$ & $P_6$ (above T or between T and X). In addition to the spectral features well below X, we see also a change of the TDA around X, as marked by the negative ($–\Delta\alpha < 0$, red regions) and positive ($–\Delta\alpha > 0$, blue regions) bands in both Extended Data Fig. 7a & 7b. These bands are related to the pump-induced bandgap renormalization (BGR), as illustrated in Fig. 3l – 3n in the main text and Ref. [30].

To obtain more quantitative information about the TA features, we performed the multi-peak Gaussian fittings on the TDAS selected from Extended Data Fig. 7a at seven representative delay times. The results are plotted in Extended Data Fig. 7d. Such multi-peak fittings for other devices were performed in the similar way. Here, we take the results of Device #4 at 4K (Extended Data Fig. 7d) as an example to describe the fitting methods.

➢ **<u>Fitting method for the TDAS</u>**

The multi-Gaussian function (used for Extended Data Fig. 7d) is written explicitly as follows,

$$F(\varepsilon^*) = -\sum_{i=1}^{6}\left[G(\varepsilon^*; A_{P_i}, \varepsilon_{P_i}, \gamma_{P_i})\right] - G(\varepsilon^*; A_{X_p}, \varepsilon_{X_p}, \gamma_{X_p}) + G(\varepsilon^*; A_{X_0}, \varepsilon_{X_0}, \gamma_{X_0}) + \Delta_b$$

. (S1)

Here, the Gaussian function $G(\varepsilon^*; A, \varepsilon, \gamma)$ in eq. (S1) is defined via

$$G\left(\varepsilon^*; A, \varepsilon, \gamma\right) := A \cdot \exp\left[-\ln 2\left(\frac{\varepsilon^* - \varepsilon}{\gamma}\right)^2\right], \qquad (S2)$$

where, $\varepsilon^*$ is the argument (probe energy), $A$, $\varepsilon$, and $\gamma$ denote the oscillator strength, the energy position and the line width of the Gaussian peak, respectively. As is shown in eq. (S1), up to six negative Gaussian peaks (the 1st term) are used to fit features $P_1$ – $P_6$ (excited-state absorption, ESA, *e.g.* the quadruplon effect in this paper, the conventional bi-exciton effect[31-34], *etc*). The additional negative (the 2nd term) and positive (the 3rd term) peaks are designated to account for the exciton resonances with ($X_p$) and without ($X_0$) pump, which result in the antisymmetric line-shape of the pump-induced BGR (see Extended Data Fig. 7a, 7b, & 7d in connection with Fig. 3l – 3n in the main text; See Ref. [32-34] for the illustration of the absorption increase with pump below X (ESA, −Δα < 0) caused by the bi-exciton effect; See also Ref. [30] for the illustration of the antisymmetric line-shape around X caused by BGR). Since the spectral range of sweeping the probe energies did not cover the vicinity of X when we tested Device #1 & #3, peaks $X_p$ and $X_0$ were not included in the multi-Gaussian fitting function. However, such BGR signals were well recorded around X for Device #2 & #4 (see Fig. 3 in the main text for Device #2 and see Extended Data Fig. 7 for Device #4). The last term $\Delta_b$ is used to account for an overall (frequency independent) ground-state bleaching (GSB). In contrast to the ESA effect leading to the absorption increase with pump (−Δα < 0), the GSB effect leads to the absorption decrease with pump (−Δα > 0).

The fitted values of the parameters in eq. (S1) are listed in Extended Data Tab. 1. The central energies of obtained Gaussian peaks $P_1$ – $P_6$ (as marked in Extended Data Fig. 7a & 7d) are largely independent of the delay time and given by 1.1325 eV, 1.1369 eV, 1.1411 eV, 1.1448 eV, 1.1516 eV, 1.1588 eV, respectively. The variations of the fitted peak positions in the same columns (for the same oscillators) are all within 0.2 meV. The intervals between these peaks are in the range of 4 – 7 meV, while the total spread of these peaks is around 26 meV. The lowest peak (1.1325 eV) is about 40 meV below the original position of X and ~ 13 meV below the position of T. The fitted line widths (≤ 3 meV) are all smaller than the typical values of T and X extracted from the CW absorption spectra (5 – 10 meV). It is also evident from Extended Data Fig. 7a & 7d that the first three peaks $P_1 – P_3$ decay faster than the other peaks. This phenomenon is explained when we discuss Fig. 5 in the main text.

**Extended Data Tab. 1. Fitted values of the parameters for the different pump-probe delays**

| Delay | | $P_1$ | $P_2$ | $P_3$ | $P_4$ | $P_5$ | $P_6$ | $X_p$ | $X_0$ | $\Delta_b$ |
|---|---|---|---|---|---|---|---|---|---|---|
| 3.47 ps | ε (eV) | – | 1.13695 | 1.14083 | 1.14467 | 1.15167 | 1.15886 | 1.16455 | 1.17132 | |
| | A | – | 51.09 | 95.83 | 285.46 | 392.96 | 520.27 | 1184.98 | 1217.34 | 186.77 |
| | γ (eV) | – | 0.00216 | 0.00220 | 0.00216 | 0.00157 | 0.00226 | 0.00302 | 0.00251 | |
| 3.13 ps | ε (eV) | 1.13274 | 1.13697 | 1.14087 | 1.14464 | 1.15183 | 1.15875 | 1.16442 | 1.17138 | |
| | A | 23.42 | 134.17 | 123.20 | 370.28 | 331.93 | 509.79 | 1208.98 | 1237.68 | 228.19 |
| | γ (eV) | 0.00152 | 0.00214 | 0.00216 | 0.00192 | 0.00197 | 0.00219 | 0.00324 | 0.00256 | |
| 2.47 ps | ε (eV) | 1.13275 | 1.13657 | 1.14123 | 1.14496 | 1.15167 | 1.15904 | 1.16448 | 1.17139 | |
| | A | 69.08 | 129.43 | 266.00 | 301.84 | 416.77 | 554.62 | 1178.21 | 1229.65 | 225.72 |
| | γ (eV) | 0.00153 | 0.00221 | 0.00212 | 0.00190 | 0.00159 | 0.00235 | 0.00282 | 0.00254 | |
| 1.80 ps | ε (eV) | 1.13235 | 1.13695 | 1.14128 | 1.14464 | 1.15160 | 1.15902 | 1.16444 | 1.17118 | |
| | A | 91.08 | 124.27 | 130.11 | 292.73 | 388.06 | 454.35 | 1103.08 | 1147.72 | 204.61 |
| | γ (eV) | 0.00176 | 0.00206 | 0.00233 | 0.00211 | 0.00155 | 0.00241 | 0.00283 | 0.00258 | |
| 1.13 ps | ε (eV) | 1.13248 | 1.13672 | 1.14084 | 1.14517 | 1.15174 | 1.15866 | 1.16429 | 1.17137 | |
| | A | 107.61 | 296.77 | 365.09 | 293.73 | 386.04 | 445.56 | 1079.68 | 1130.18 | 221.10 |
| | γ (eV) | 0.00225 | 0.00163 | 0.00216 | 0.00167 | 0.00155 | 0.00223 | 0.00290 | 0.00247 | |
| 0.13 ps | ε (eV) | 1.13230 | 1.13694 | 1.14125 | 1.14476 | 1.15172 | 1.15863 | 1.16421 | 1.17133 | |
| | A | 60.28 | 275.62 | 227.56 | 240.86 | 238.17 | 262.74 | 951.88 | 972.88 | 154.14 |
| | γ (eV) | 0.00157 | 0.00202 | 0.00221 | 0.00164 | 0.00195 | 0.00188 | 0.00308 | 0.00250 | |
| −0.20 ps | ε (eV) | 1.13273 | 1.13692 | 1.14124 | 1.14460 | 1.15161 | 1.15888 | 1.16435 | 1.17123 | |
| | A | 170.64 | 264.71 | 197.24 | 216.70 | 370.13 | 365.03 | 977.61 | 1025.56 | 163.36 |
| | γ (eV) | 0.00213 | 0.00224 | 0.00227 | 0.00226 | 0.00156 | 0.00225 | 0.00273 | 0.00250 | |
| −0.53 ps | ε (eV) | 1.13255 | 1.13673 | 1.14121 | 1.14493 | 1.15135 | 1.15861 | 1.16425 | 1.17129 | |
| | A | 215.58 | 216.58 | 240.89 | 246.01 | 282.78 | 301.22 | 900.88 | 933.20 | 131.54 |
| | γ (eV) | 0.00161 | 0.00207 | 0.00223 | 0.00169 | 0.00163 | 0.00194 | 0.00303 | 0.00250 | |
| −0.87 ps | ε (eV) | 1.13231 | 1.13693 | 1.14122 | 1.14495 | 1.15134 | 1.15866 | 1.16423 | 1.17125 | |
| | A | 166.31 | 221.86 | 215.06 | 243.16 | 292.05 | 305.32 | 885.36 | 952.07 | 111.60 |
| | γ (eV) | 0.00225 | 0.00211 | 0.00225 | 0.00169 | 0.00173 | 0.00234 | 0.00296 | 0.00227 | |

We note that in Extended Data Fig. 7b the spectra for Device #4 are almost featureless below X, while in Fig. 3i in the main text for Device #2 features $P_1 – P_6$ are visible. There are several possible reasons for this. First, the co-polarized signal is typically weaker than the cross-polarized one due to the more favorable many-body configuration of the cross-polarized case; Second, there are always unavoidable quality variations from sample to sample; Third, those two spectra were measured at different pumping levels: For example, Extended Data Fig. 7b was for pumping level of ~ 80 μJ·cm$^{-2}$, while Fig. 3i in the main text was for ~ 160 μJ·cm$^{-2}$. The pump pulse with different pumping levels might produce different 2B and 4B states, thus lead to different 2B-4B transition spectra (the relevant discussion can be seen in Methods S13 & S13.1, in connect with the theoretical results).

**S7. Additional experimental results (Device #5)**

We performed the similar gate-dependent CW RCS and pump-probe tests on another device (Device #5) at 4K with the same gate-tunable structure. For the gate-dependent CW RCS tests, the results are shown in Extended Data Fig. 8. For the pump-probe experiments, the pump energy was fixed at 1.180 eV (near-resonant to X). The typical signal evolutions with time in the charge-neutral regime ($V_g$ = −4 V referring to Extended Data Fig. 8) are shown in Extended Data Fig. 9. As has been described in the preceding text, the negative signals around the zero delay (−Δα < 0, the red curve in Extended Data Fig. 9a) signify the TA enhancements at the specific probe energies (*e.g.* 1.145 eV), which are relevant to the ultrafast generations of the high order correlated entities. At the other probe energies (*e.g.* 1.153 eV), there are no such negative signals as shown in Extended Data Fig. 9b. These two representative probe energies are correspondingly marked with the white arrows in Extended Data Fig. 8. According to the gate-dependent CW RCS tests, three voltages (the cyan dashed lines in Extended Data Fig. 8) were selected to perform the pump-probe tests, with the results shown in Extended Data Fig. 10.

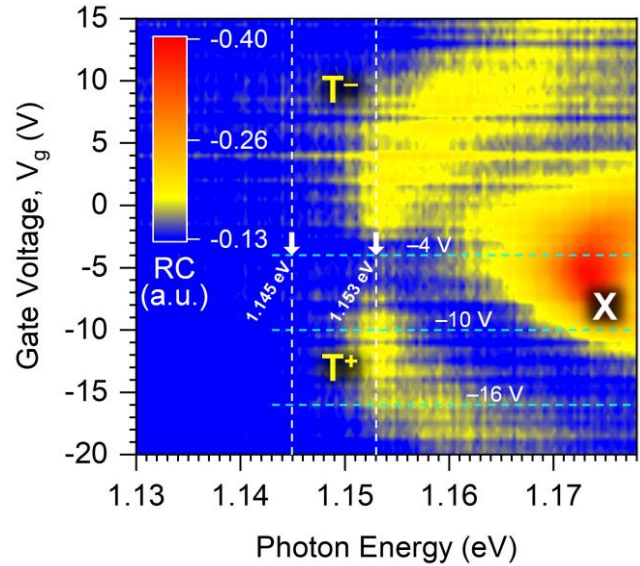

**Extended Data Fig. 8 |** CW RC contour in the plane of photon energy and gate voltage for Device #5 at 4K. The white arrows of 1.145 eV and 1.153 eV correspond to the probe energies that are used for the pump-probe tests in Extended Data Fig. 9a & 9b, respectively. The cyan dashed lines of –4 V (charge-neutral), –10 V, and –16 V correspond to the gate voltages at which the TDAS were measured and shown in Extended Data Fig. 10.

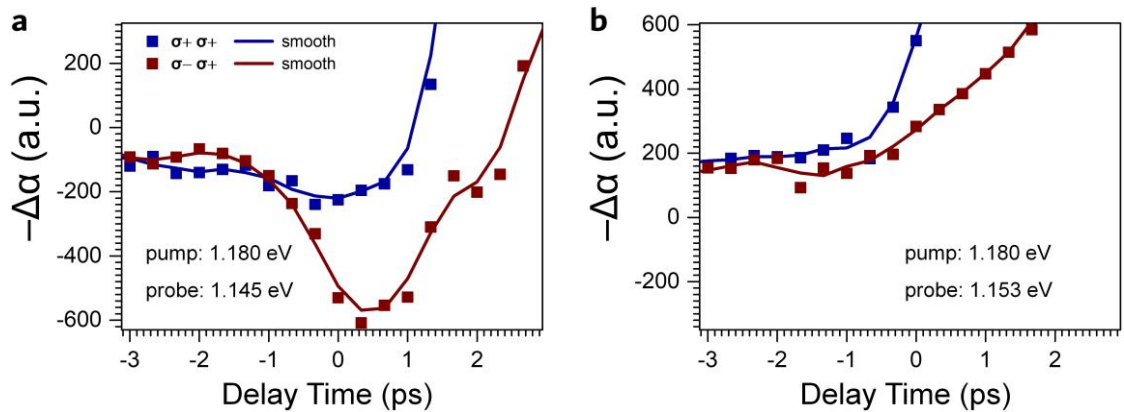

**Extended Data Fig. 9 | a, b,** Time evolutions of the TDA signals (Device #5, at 4K, under the charge-neutral condition) around the pump-probe delay t ≈ 0 ps with the cross- (σ+ σ+) and co- (σ+ σ+) circularly polarized pump-probe configurations at the different probe energies of 1.145 eV **(a)** and 1.153 eV **(b)**.

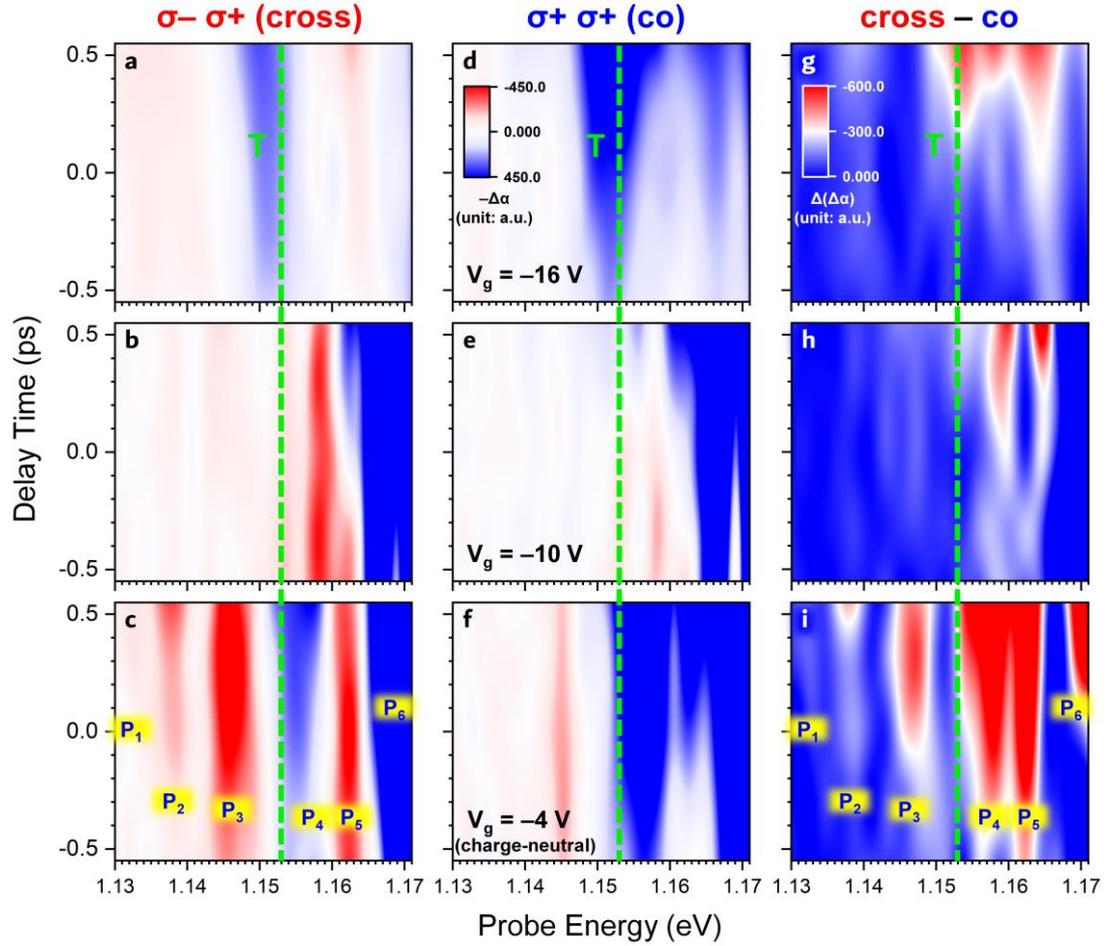

**Extended Data Fig. 10 | a – f,** TDA contours in the plane of probe energy and delay time at the three gate voltages (as given in the middle one of the three panels at the same row) with the cross- (σ– σ+) **(a – c)** and co- (σ+ σ+) **(d – f)** circularly polarized pump-probe configurations. **g – i,** Second-order TDA (defined as: cross – co) contours in the plane of probe energy and delay time at the corresponding gate voltages. Features $P_1 - P_6$ are marked in the charge-neutral cases ($V_g = -4$ V) in both **c & i**. The vertical green dashed lines (at 1.153 eV) mark the spectral position of T as seen from the CW RC results in Extended Data Fig. 8.

For Device #5, the features corresponding to $P_4$ and $P_6$ are not visible even in the charge-neutral regime, as shown in Extended Data Fig. 10c. The expected features are likely buried in the absorption reduction bands caused by other nonlinear effects such as GSB (see Methods S6). For a more defined demonstration of those many-body signals, we deducted the inessential contributions from these unconcerned nonlinear effects (*e.g.* GSB) by analyzing the second-order TDAS (defined as: $\Delta(\Delta\alpha) = \Delta\alpha_{cross} - \Delta\alpha_{co}$) for Device #5 (Extended Data Fig. 10g – 10i). The similar data processing was performed in Ref. [31] to extract the bi-exciton signals. In this way, all the features of $P_1$

– $P_6$ become visible, especially in the charge-neutral regime, as shown in Extended Data Fig. 10i. The gate-dependent behavior of $\Delta(\Delta\alpha)$ around the zero delay is consistent with that of $-\Delta\alpha$ as we see in Fig. 3b – 3k in the main text for Device #2. All these features show the dependency on the charge-neutrality, which provides the evidence for the charge-neutral entities.

## S8. Linear plot of the absorption of $P_1 - P_6$ and X vs. pump density

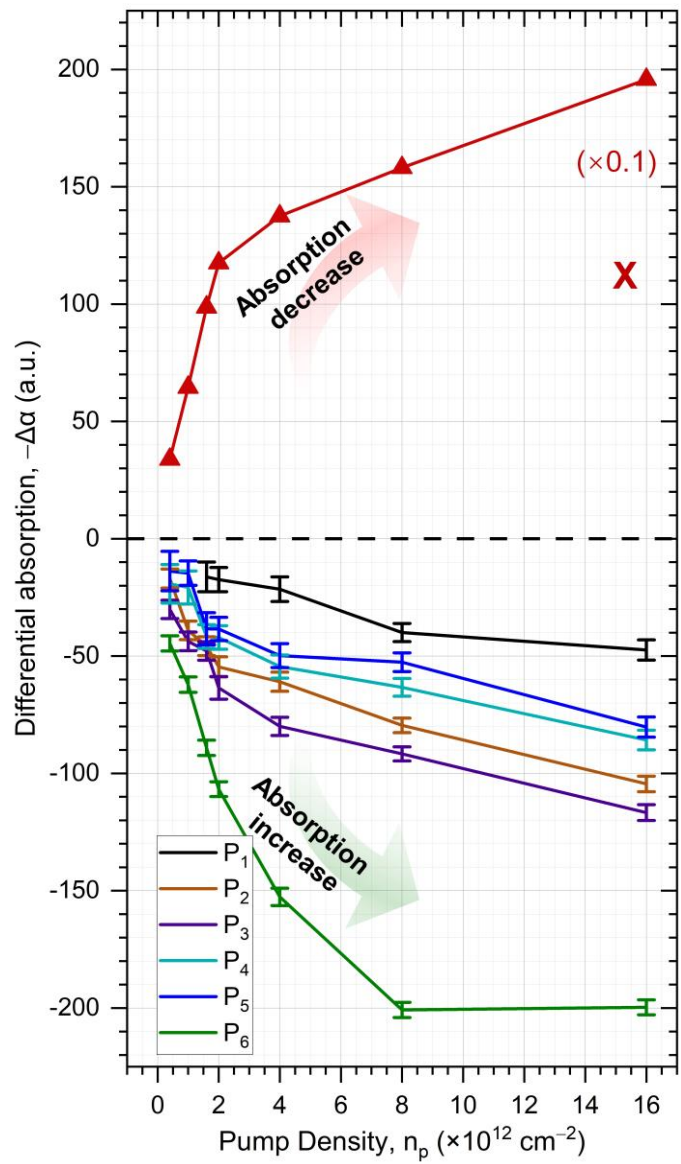

**Extended Data Fig. 11 |** TDA with respect to the pump densities (Device #3). The signal of $-\Delta\alpha$ for X were obtained at ~ 1.171 eV (near X). The same results for the 6 fitted peaks were shown in a logarithmic scale in Fig. 4c in the main text. Here, all of these results are plotted in a linear scale.

**S9. Possible reasons for the absence of P$_1$ – P$_6$ features for other ML-TMDCs**

**1)** In many of the papers for studying ML-TMDCs through ultrafast spectroscopy (*e.g.* TDAS or TDRS[31,32], 2D coherent Fourier-transform spectroscopy[35], *etc*), their samples were not gated and the samples are most likely not charge neutral. As can be seen in Fig. 3b – 3k in the main text, our features of P$_1$ – P$_6$ are the strongest at the charge-neutral point, and become weaker with gate-induced doping.

**2)** The second reason might be the spectral resolution. In our experiment, the spectral resolution is fine enough to identify the narrow peaks. While in other relevant papers[32], their spectral resolution is 1 – 2 orders coarser than ours.

**S10. Two-band tight-binding k·p model**

The 2H-phase ML-TMDC is a direct-bandgap semiconductor with the valence band maxima (VBM) and the conduction band minima (CBM) located at the two groups of inequivalent corners of the hexagonal first Brillouin zone (FBZ). The gapped massive Dirac model proposed by Xiao *et al*[36] has been proven successful in describing the band structures in the vicinity of the $\hat{C}_3$ high-symmetry points K and K'. This model was also extensively used by many of previous theories, *e.g.* the spin-valley dynamics accounting for the inter- and intra-valley *e-h* exchange effects[37] and the exciton-trion quantum coherent beating[38]. For the same k-mesh density, the dimension of a 2*e*2*h* 4B Hilbert space is estimated to be four orders of magnitude larger than that of a conventional *e-h* 2B case. To avoid such an expensive computation, we adopted the low-budget tight-binding **k·p** model instead of a fully first-principle calculation.

The two-band Hamiltonian including the SOC term can be expressed as

$$H_0(\mathbf{k}) = \begin{pmatrix} \left(-\varepsilon_g + \lambda_v \tau_\mathbf{k} S_z\right)/2 & at\left(\tau_\mathbf{k} q_x + iq_y\right) \\ at\left(\tau_\mathbf{k} q_x - iq_y\right) & \left(\varepsilon_g + \lambda_c \tau_\mathbf{k} S_z\right)/2 \end{pmatrix}. \tag{S3}$$

Here, **k** is the wavevector relative to the closest K or K' point, $\varepsilon_g$ is given by defined as $\varepsilon_g = \frac{1}{2}\left(\left(\varepsilon_{CBM,\uparrow} + \varepsilon_{CBM,\downarrow}\right) - \left(\varepsilon_{VBM,\uparrow} + \varepsilon_{VBM,\downarrow}\right)\right)$, $\lambda_v$ and $\lambda_c$ correspond to the spin-orbit splittings of the valence and conduction bands, $\tau_\mathbf{k}$ represents the valley pseudo-spin index equal to 1 (for K valley) or −1 (for K' valley), $S_z$ stands for the electronic spin index equal to 1 (for spin up) or −1 (for spin down). Accordingly, the product of $\tau_\mathbf{k}$ and $S_z$ can well describe the spin-valley locking effect. $a$ is the in-plane lattice constant, and $t$ is the effective nearest neighbor hopping integral.

The electronic dispersions around the K or K' point at the single-particle level can be written as

$$\varepsilon_{v\mathbf{k}} = \frac{1}{2}\left[\frac{1}{2}(\lambda_v + \lambda_c)\tau_\mathbf{k} S_z - \sqrt{\left(\varepsilon_g - \frac{1}{2}(\lambda_v - \lambda_c)\tau_\mathbf{k} S_z\right)^2 + (2atk)^2}\right], \quad (S4)$$

$$\varepsilon_{c\mathbf{k}} = \frac{1}{2}\left[\frac{1}{2}(\lambda_v + \lambda_c)\tau_\mathbf{k} S_z + \sqrt{\left(\varepsilon_g - \frac{1}{2}(\lambda_v - \lambda_c)\tau_\mathbf{k} S_z\right)^2 + (2atk)^2}\right]. \quad (S5)$$

The corresponding wavefunction envelopes $|\tilde{\psi}_\mathbf{k}^v\rangle$ and $|\tilde{\psi}_\mathbf{k}^c\rangle$ are the linear combinations of the Mo-$d$ orbitals, which read,

$$|\tilde{\psi}_\mathbf{k}^v\rangle = \exp(i\tau_\mathbf{k}\varphi_\mathbf{k})\cos(\theta_\mathbf{k}/2)|\tilde{\phi}_{\tau_\mathbf{k}}^v\rangle - \sin(\theta_\mathbf{k}/2)|\tilde{\phi}_{\tau_\mathbf{k}}^c\rangle, \quad (S6)$$

$$|\tilde{\psi}_\mathbf{k}^c\rangle = \sin(\theta_\mathbf{k}/2)|\tilde{\phi}_{\tau_\mathbf{k}}^v\rangle + \exp(-i\tau_\mathbf{k}\varphi_\mathbf{k})\cos(\theta_\mathbf{k}/2)|\tilde{\phi}_{\tau_\mathbf{k}}^c\rangle, \quad (S7)$$

where the spinor basis is made up of $|\tilde{\phi}_{\tau_\mathbf{k}}^v\rangle := |d_{x^2-y^2}\rangle + i\tau_\mathbf{k}|d_{xy}\rangle$ and $|\tilde{\phi}_{\tau_\mathbf{k}}^c\rangle := |d_{z^2}\rangle$, and $\theta_\mathbf{k}$ and $\varphi_\mathbf{k}$ are defined as $\tan(\theta_\mathbf{k}) = \dfrac{2atk}{\varepsilon_g - (\lambda_v - \lambda_c)\tau_\mathbf{k} S_z/2}$ and $\tan(\varphi_\mathbf{k}) = q_y/q_x$. To further study the light-matter interactions, first it is important to calculate the inter-band transition matrix elements within the independent particle approximation (IPA). Here, we use the generalized Feynman-Hellman theorem to write the matrix elements,

$$\mathbf{p}_{\mathbf{k},\mathbf{k}}^{v,c} = \frac{m}{\hbar}\langle\tilde{\psi}_\mathbf{k}^v|\nabla_\mathbf{k} H_0(\mathbf{k})|\tilde{\psi}_\mathbf{k}^c\rangle, \quad (S8)$$

where $\nabla_{\mathbf{k}} H_0(\mathbf{k})$ explicitly expressed in the $\mathbf{k_x}$-$\mathbf{k_y}$ coordinate as follows,

$$\nabla_{\mathbf{k}} H_0(\mathbf{k}) = \begin{pmatrix} 0 & at(\tau_k \mathbf{e}_x + i\mathbf{e}_y) \\ at(\tau_k \mathbf{e}_x - i\mathbf{e}_y) & 0 \end{pmatrix}, \quad (S9)$$

with $\mathbf{e_x}$ and $\mathbf{e_y}$ denoting the unit vector of the $\mathbf{k_x}$- and $\mathbf{k_y}$-axis, respectively.

**S11. *Ab-initio* calculation for the electronic structure of the ML-MoTe$_2$**

The *ab-initio* calculations for the electronic structure of the ML-MoTe$_2$ were performed using the density functional theory (DFT) within the local density approximation (LDA), as implemented in the Quantum ESPRESSO package[39]. In addition, the effect of non-collinear SOC was included with the spinor wavefunctions using fully relativistic norm-conserving pseudopotentials. The pseudopotentials were generated with the ONCVPSP code[40], explicitly including the Mo 4*s* and 4*p* semicore states. We assumed the in-plane lattice constant of 3.53 Å for the ML-MoTe$_2$ as the starting point of our calculations. A slab model with a vacuum layer thickness of ~ 30 Å along the out-of-plane direction was adopted to avoid the adjacent interactions between the periodic images. The geometric structure was fully relaxed in a non-spin polarized case with a uniform k-grid of 15 × 15 × 1, until the force on each atom was less than 0.002 eV·Å$^{-1}$. The electronic density converged in a non-collinear SOC case with an energy cutoff of 120 Ry and a k-grid of 30 × 30 × 1.

Part of the parameters obtained from the DFT calculations are shown in Extended Data Fig. 12, with $a \approx 3.53$ Å, $\varepsilon_g \approx 1.1$ eV (converted from the DFT bandgap ~ 0.95 eV), $\lambda_v \approx 222$ meV, and $\lambda_c \approx 36$ meV. $t$ is evaluated to be ~ 0.66 eV under the parabolic approximation. These results are consistent with the previous studies[29], which have been proven numerically reliable as the starting point of a many-body perturbation calculation.

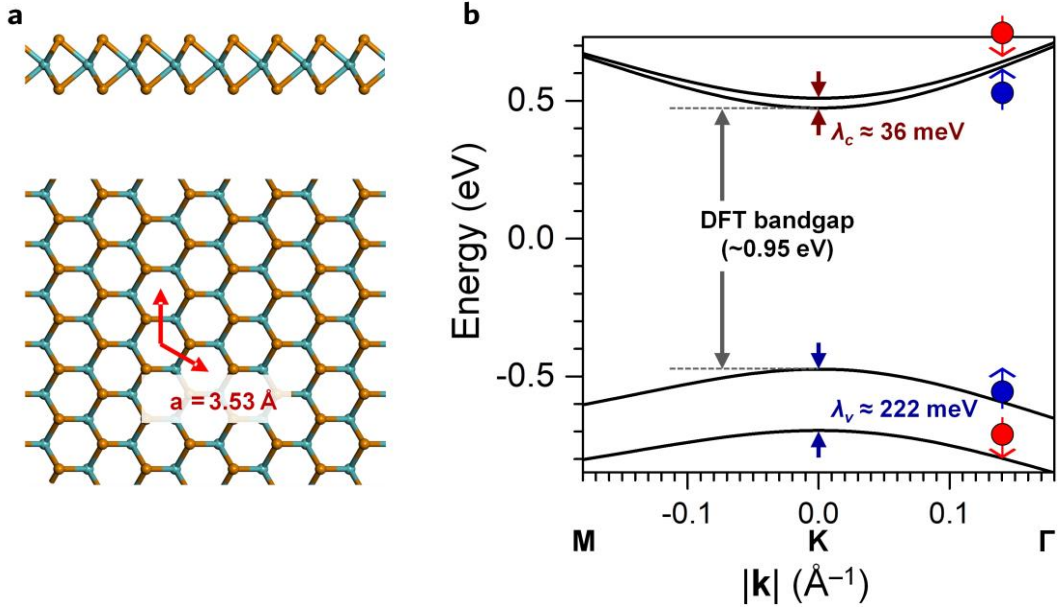

**Extended Data Fig. 12 | a,** Side (top) and top (bottom) view of the lattice structure of the 2H-phase ML-MoTe$_2$. **b,** Calculated electronic dispersions near the K point accounting for the effect of SOC.

**S12. Many-body perturbation theory**

**S12.1. Two-body (2B) perturbation theory**

The correlated 2B wavefunctions ($|e_3 h_3\rangle$, see eq. (2) in the main text) can be expanded by using the above-discussed single-particle basis (eq. (S6) & (S7)), written explicitly within the Tamm-Dancoff approximation (TDA) as

$$|e_3 h_3\rangle = \sum_{(v_3, c_3, \mathbf{k}_3)} A^{v_3, c_3}_{\mathbf{k}_3-\mathbf{Q}, \mathbf{k}_3} \psi^{v_3}_{\mathbf{k}_3-\mathbf{Q}}(\mathbf{r}_{h_3})^* \psi^{c_3}_{\mathbf{k}_3}(\mathbf{r}_{e_3}), \tag{S10}$$

where $\mathbf{r}_{e_3}$ and $\mathbf{r}_{h_3}$ are the real-space coordinates of electron $e_3$ and hole $h_3$, $A^{v_3, c_3}_{\mathbf{k}_3-\mathbf{Q}, \mathbf{k}_3}$ is the coefficient of the 2B wavefunction expanded to the corresponding microscopic basis, and the $v\mathbf{k}$ and $c\mathbf{k}$-like labels stand for the single-particle states associating the band indices of VB ($v$) and CB ($c$) with momentum $\mathbf{k}$. $A^{v_3, c_3}_{\mathbf{k}_3-\mathbf{Q}, \mathbf{k}_3}$ can be obtained by solving the standard 2-body Bethe-Salpeter equation (2B-BSE),

$$i\hbar \partial_t A^{v,c}_{\mathbf{k}-\mathbf{Q}, \mathbf{k}}(\mathbf{0}; \mathbf{Q}) = \left( \tilde{\varepsilon}^{QP}_{c\mathbf{k}} - \tilde{\varepsilon}^{QP}_{v\mathbf{k}-\mathbf{Q}} - i\gamma \right) A^{v,c}_{\mathbf{k}-\mathbf{Q}, \mathbf{k}}(\mathbf{0}; \mathbf{Q})$$
$$- \sum_{(v,c,\mathbf{q},\mathbf{k})} \left( W^{v',c,v,c'}_{\mathbf{k}+\mathbf{q}-\mathbf{Q}, \mathbf{k}, \mathbf{k}-\mathbf{Q}, \mathbf{k}+\mathbf{q}} - V^{v',c,c',v}_{\mathbf{k}+\mathbf{q}-\mathbf{Q}, \mathbf{k}, \mathbf{k}+\mathbf{q}, \mathbf{k}-\mathbf{Q}} \right) A^{v',c'}_{\mathbf{k}+\mathbf{q}-\mathbf{Q}, \mathbf{k}+\mathbf{q}}(\mathbf{q}; \mathbf{Q}) \tag{S11}$$

To correct the underestimated bandgap obtained from the DFT calculations, we use the $G_0W_0$ quasi-particle bandgap of the ML-MoTe$_2$ of 1.77 eV[29] and then applied a ~ 0.82 eV scissor operator to our result. Unlike the GW calculation, the scissor operator leads to a simple shift of the bands without changing the band curvature. This could in principle be an additional source of error. It is noteworthy that $\tilde{\varepsilon}_{c\mathbf{k}}^{QP}$ and $\tilde{\varepsilon}_{v\mathbf{k}-\mathbf{Q}}^{QP}$ in eq. (S11) stand for the electronic energies after the artificial alignment rather than restricted to the mean-field level. Additionally, a homogeneous broadening factor, $\gamma$, is used to mimic the finite excited-state lifetime. $W_{\bar{\mathbf{k}}',\mathbf{k},\bar{\mathbf{k}},\mathbf{k}'}^{v',c,v,c'}$ and $V_{\bar{\mathbf{k}}',\mathbf{k},\mathbf{k}',\bar{\mathbf{k}}}^{v',c,c',v}$ represent the screened and unscreened (bare) Coulomb interaction kernels and are written, respectively, as

$$W_{\bar{\mathbf{k}}',\mathbf{k},\bar{\mathbf{k}},\mathbf{k}'}^{v',c,v,c'} = \int \psi_{\bar{\mathbf{k}}'}^{v'}(\mathbf{r})^* \psi_{\mathbf{k}}^{c}(\mathbf{r}')^* w(\mathbf{r},\mathbf{r}') \psi_{\bar{\mathbf{k}}}^{v}(\mathbf{r}) \psi_{\mathbf{k}'}^{c'}(\mathbf{r}') d^d\mathbf{r} d^d\mathbf{r}', \qquad (S12)$$

$$V_{\bar{\mathbf{k}}',\mathbf{k},\mathbf{k}',\bar{\mathbf{k}}}^{v',c,c',v} = \int \psi_{\bar{\mathbf{k}}'}^{v'}(\mathbf{r})^* \psi_{\mathbf{k}}^{c}(\mathbf{r}')^* v(\mathbf{r},\mathbf{r}') \psi_{\mathbf{k}'}^{c'}(\mathbf{r}) \psi_{\bar{\mathbf{k}}}^{v}(\mathbf{r}') d^d\mathbf{r} d^d\mathbf{r}', \qquad (S13)$$

with $v(\mathbf{r},\mathbf{r}') = e^2/|\mathbf{r}-\mathbf{r}'|$ being the unscreened Coulomb potential, and $w(\mathbf{r},\mathbf{r}')$, the screened one defined in terms of the inverse dielectric function via $w(\mathbf{r},\mathbf{r}') = \int \epsilon^{-1}(\mathbf{r},\mathbf{r}'') v(\mathbf{r}'',\mathbf{r}') d^d\mathbf{r}''$ (for generality, $d$ denotes the dimension of the system). The $e$-$h$ exchange interactions are determined by the unscreened interaction kernel.

The screened Coulomb potential in a 2D case is the standard Rytova-Keldysh expression[41,42], *i.e.*

$$w(\rho) = -\frac{\pi e^2}{4\pi \chi_{2D}} \left[ H_0\left(\frac{\rho}{2\pi \chi_{2D}}\right) - Y_0\left(\frac{\rho}{2\pi \chi_{2D}}\right) \right], \qquad (S14)$$

where $\rho$ denotes the inter-particle distance, $H_0$ and $Y_0$ are the Struve function and the Bessel function of the second kind, and $\chi_{2D}$ is the 2D polarizability ($\chi_{2D}$ = 11.715 for the MoTe$_2$[28]). It should be noted that the standard Keldysh potential is approximated at the zero-frequency (long-wavelength, or static) limit and thus cannot account for the dynamical screening effects. To simplify the following calculations, both of the interaction kernels and the corresponding Coulomb potentials are transformed to the reciprocal space. Within the zero differential overlap approximation proposed by Berkelbach *et al*[43], interaction kernels $W_{\bar{k}',k,\bar{k},k'}^{v',c,v,c'}$ and $V_{\bar{k}',k,k',\bar{k}}^{v',c,c',v}$ can be rewritten as

$$W_{\bar{k}',k,\bar{k},k'}^{v',c,v,c'} \approx -\frac{1}{\mathcal{A}} \langle \tilde{\psi}_{\bar{k}'}^{v'} | \tilde{\psi}_{\bar{k}}^{v} \rangle \langle \tilde{\psi}_{k}^{c} | \tilde{\psi}_{k'}^{c'} \rangle W_q, \quad (S15)$$

$$V_{\bar{k}',k,k',\bar{k}}^{v',c,c',v} \approx -\frac{1}{\mathcal{A}} \langle \tilde{\psi}_{\bar{k}'}^{v'} | \tilde{\psi}_{k'}^{c'} \rangle \langle \tilde{\psi}_{k}^{c} | \tilde{\psi}_{\bar{k}}^{v} \rangle V_q, \quad (S16)$$

where $\mathcal{A}$ denotes the crystal area, $q$ is the momentum transfer (including the Umklapp process), $W_q$ and $V_q$ can be explicitly expressed as $W_q = -\frac{2\pi e^2}{q(1+2\pi\chi_{2D}q)}$ and $V_q = -2\pi e^2/q$ through the radial Fourier transform in a cylindrical coordinate. Both eq. (S15) & (S16) or the related approximations have been widely used to perform many-body perturbation calculations for excitons[38,43,44] and trions[38,44], which could give reasonably accurate results that agree well with those obtained from the fully first-principle GW-BSE calculations[18,22].

A fully first-principle GW-BSE approach could give more accurate results, where the precise bare Coulomb matrix elements and the dielectric function within the random phase approximation (RPA) can be quantitatively calculated. However, it would be too expensive to perform the four-body calculations in this paper. Therefore, we used a combination of the tight-binding k·p model and the approximate potential expression to reduce the computations. The validity of such a reduction scheme has been checked in Ref. [38] for the two- and three-body perturbation calculations.

The final solution of eq. (S11) is the well-known excitonic Rydberg series. Using a 36× 36×1 Monkhorst-Pack k-grid and a Brillouin zone truncation scheme (the numerical techniques will be described in Methods S14), we obtained the full 2B spectrum, and then applied a scissor operator by aligning the 1*s*-X energy of the ML-MoTe$_2$ to ~ 1.167 eV according to the experimental results. With the same scissor, the energies of the 2*p*-X, 2*s*-X, … plasma (continuum band edge) were shifted to 1.296 eV, 1.325 eV and 1.432 eV, respectively. These 2B states will be used to construct the dipole matrix elements with those of 4B states to obtain the possible spectral features and then to compare with the experimental measurements. It should be noted that here the 1*s*-X binding energy of 0.265 eV for the ML-MoTe$_2$ was evaluated to be smaller than the previous theoretical result of 0.375 eV[28]. We attribute the margin to the relatively sparse k-grids for the Brillouin-zone sampling (~ 36×36×1 in our calculations. But generally, up to ~ 100×100×1 or even ~ 1000×1000×1 in Ref. [38]). In fact, diagonalizing the many-body Hamiltonian matrix with a large number of k-grid points seems relatively easy for the 2B case, but it would become an exceedingly large computational challenge for the 4B case. Fortunately, the binding energies of high-order correlated entities such as trions in TMDC systems are one order of magnitude smaller than that of excitons and are found to be much less dependent on the k-grid density[38]. We have also tested out a rapid convergence with a k-grid size of ~ 36×36×1 (see Methods S14 for the results of the convergence tests), and we found the similar k-grid sizes were be used as well for the calculations of the trions and bi-excitons in Ref. [32,38,45]. For these reasons, we adopted the k-grid of ~ 36×36×1 for all the many-body perturbation calculations in this paper.

### **S12.2. Four-body (4B) perturbation theory**

The 2B-BSE discussed above is used to study the *e-h* 2B excitations from the ground state (vacuum) of those charge-neutral systems. Beyond the conventional 2B formalism, the 3-body Bethe-Salpeter equation (3B-BSE)[18,38,44] has been developed to describe the trions and their fine structures in doped systems. Our results by solving such 3B-BSE are shown in Fig. 5a in the main text. However, dealing with 4B states along the same line of the approach involves much more challenges in both of theoretical description and numerical simulation. We have seen so far only two published papers deriving the 4B perturbation equations, both of which are devoted to the bi-exciton fine structures that are spectrally located between T and X as the results of the *e-h* exchange interactions[32,45]. To account for the new rich features seen both below and above T (between T and X) in our pump-probe experiments, developing a more general 4B perturbation theory, by studying the 4-body Bethe-Salpeter equation (4B-BSE), is quite necessary, which has been less understood ever before to the best of our knowledge.

Similarly to the 2B case (see Methods S12.1), we express the 4B state ($|e_1h_1e_2h_2\rangle$, see eq. (1) in the main text) explicitly by including the real-space amplitudes[46],

$$|e_1h_1e_2h_2\rangle = \frac{1}{2} \sum_{\substack{v_1,c_1,v_2,c_2 \\ \mathbf{q},\mathbf{k}_1,\mathbf{k}_2}} B^{v_1,c_1,v_2,c_2}_{\mathbf{k}_1+\mathbf{q}-\mathbf{Q},\mathbf{k}_1,\mathbf{k}_2-\mathbf{q},\mathbf{k}_2} \det\begin{pmatrix} \psi^{v_1}_{\mathbf{k}_1+\mathbf{q}-\mathbf{Q}}(\mathbf{r}_{h_1}) & \psi^{v_1}_{\mathbf{k}_1+\mathbf{q}-\mathbf{Q}}(\mathbf{r}_{h_2}) \\ \psi^{v_2}_{\mathbf{k}_2-\mathbf{q}}(\mathbf{r}_{h_1}) & \psi^{v_2}_{\mathbf{k}_2-\mathbf{q}}(\mathbf{r}_{h_2}) \end{pmatrix}^* \det\begin{pmatrix} \psi^{c_1}_{\mathbf{k}_1}(\mathbf{r}_{e_1}) & \psi^{c_1}_{\mathbf{k}_1}(\mathbf{r}_{e_2}) \\ \psi^{c_2}_{\mathbf{k}_2}(\mathbf{r}_{e_1}) & \psi^{c_2}_{\mathbf{k}_2}(\mathbf{r}_{e_2}) \end{pmatrix}.$$

(S17)

Here, $\mathbf{r}_{e_1}$, $\mathbf{r}_{h_1}$, $\mathbf{r}_{e_2}$ and $\mathbf{r}_{h_2}$ are the real-space coordinates of electrons $e_1$, $e_2$ and holes $h_1$, $h_2$, $B^{v_1,c_1,v_2,c_2}_{\mathbf{k}_1+\mathbf{q}-\mathbf{Q},\mathbf{k}_1,\mathbf{k}_2-\mathbf{q},\mathbf{k}_2}$ is the amplitude of the component in "$e_1h_1e_2h_2$" many-body Hilbert space corresponding to a doubly excited-state manifold whereby $e_1$ is created at the single-particle state ($c_1$, $\mathbf{k}_1$), $h_1$ is created at ($v_1$, $\mathbf{k}_1 + \mathbf{q} - \mathbf{Q}$), $e_2$ is created at ($c_2$, $\mathbf{k}_2$), and $h_2$ is created at ($v_2$, $\mathbf{k}_2 - \mathbf{q}$), with the total momentum of $\mathbf{Q}$. As can be seen, the microscopic basis is constructed as the direct product of the $e_1e_2$ and $h_1h_2$ determinants. Therefore, eq. (S17) fully exhibits the exchange symmetries between the two electrons or two holes. $B^{v_1,c_1,v_2,c_2}_{\mathbf{k}_1+\mathbf{q}-\mathbf{Q},\mathbf{k}_1,\mathbf{k}_2-\mathbf{q},\mathbf{k}_2}$ are determined by the following 4B-BSE,

$$i\hbar \partial_t B^{v_1,c_1,v_2,c_2}_{\mathbf{k}_1+\mathbf{q}-\mathbf{Q},\mathbf{k}_1,\mathbf{k}_2-\mathbf{q},\mathbf{k}_2}(\mathbf{0};\mathbf{Q}) = \left(\tilde{\varepsilon}^{QP}_{c\mathbf{k}_1} + \tilde{\varepsilon}^{QP}_{c\mathbf{k}_2} - \tilde{\varepsilon}^{QP}_{v\mathbf{k}_1+\mathbf{q}-\mathbf{Q}} - \tilde{\varepsilon}^{QP}_{v\mathbf{k}_2-\mathbf{q}} - i2\gamma\right) B^{v_1,c_1,v_2,c_2}_{\mathbf{k}_1+\mathbf{q}-\mathbf{Q},\mathbf{k}_1,\mathbf{k}_2-\mathbf{q},\mathbf{k}_2}(\mathbf{0};\mathbf{Q})$$

$$+ \sum_{(c'_1,c'_2,\mathbf{q}')} \left\{ W^{c_1,c_2,c'_1,c'_2}_{\mathbf{k}_1,\mathbf{k}_2,\mathbf{k}_1+\mathbf{q}',\mathbf{k}_2-\mathbf{q}'} \left[ B^{v_1,c'_1,v_2,c'_2}_{\mathbf{k}_1+\mathbf{q}-\mathbf{Q},\mathbf{k}_1+\mathbf{q}',\mathbf{k}_2-\mathbf{q},\mathbf{k}_2-\mathbf{q}'}(\mathbf{q}';\mathbf{Q}) - B^{v_2,c'_1,v_1,c'_2}_{\mathbf{k}_2-\mathbf{q},\mathbf{k}_1+\mathbf{q}',\mathbf{k}_1+\mathbf{q}-\mathbf{Q},\mathbf{k}_2-\mathbf{q}'}(\mathbf{q}';\mathbf{Q}) \right] \right.$$

$$- W^{c_1,c_2,c'_2,c'_1}_{\mathbf{k}_1,\mathbf{k}_2,\mathbf{k}_2-\mathbf{q}',\mathbf{k}_1+\mathbf{q}'} \left[ B^{v_1,c'_2,v_2,c'_1}_{\mathbf{k}_1+\mathbf{q}-\mathbf{Q},\mathbf{k}_2-\mathbf{q}',\mathbf{k}_2-\mathbf{q},\mathbf{k}_1+\mathbf{q}'}(\mathbf{k}_1-\mathbf{k}_2+\mathbf{q}';\mathbf{Q}) \right.$$

$$\left. \left. - B^{v_2,c'_2,v_1,c'_1}_{\mathbf{k}_2-\mathbf{q},\mathbf{k}_2-\mathbf{q}',\mathbf{k}_1+\mathbf{q}-\mathbf{Q},\mathbf{k}_1+\mathbf{q}'}(\mathbf{k}_1-\mathbf{k}_2+\mathbf{q}';\mathbf{Q}) \right] \right\}$$

$$- \sum_{(v'_1,c'_2,\mathbf{q}')} \left( W^{v'_1,c_2,v_1,c'_2}_{\mathbf{k}_1+\mathbf{q}+\mathbf{q}'-\mathbf{Q},\mathbf{k}_2,\mathbf{k}_1+\mathbf{q}-\mathbf{Q},\mathbf{k}_2+\mathbf{q}'} - V^{v'_1,c_2,c'_2,v_1}_{\mathbf{k}_1+\mathbf{q}+\mathbf{q}'-\mathbf{Q},\mathbf{k}_2,\mathbf{k}_2+\mathbf{q}',\mathbf{k}_1+\mathbf{q}-\mathbf{Q}} \right) \left[ B^{v'_1,c_1,v_2,c'_2}_{\mathbf{k}_1+\mathbf{q}+\mathbf{q}'-\mathbf{Q},\mathbf{k}_1,\mathbf{k}_2-\mathbf{q},\mathbf{k}_2+\mathbf{q}'}(\mathbf{q}';\mathbf{Q}) \right.$$

$$\left. - B^{v_2,c_1,v'_1,c'_2}_{\mathbf{k}_2-\mathbf{q},\mathbf{k}_1,\mathbf{k}_1+\mathbf{q}+\mathbf{q}'-\mathbf{Q},\mathbf{k}_2+\mathbf{q}'}(\mathbf{q}';\mathbf{Q}) - B^{v'_1,c'_2,v_2,c_1}_{\mathbf{k}_1+\mathbf{q}+\mathbf{q}'-\mathbf{Q},\mathbf{k}_2+\mathbf{q}',\mathbf{k}_2-\mathbf{q},\mathbf{k}_1}(\mathbf{q}';\mathbf{Q}) + B^{v_2,c'_2,v'_1,c_1}_{\mathbf{k}_2-\mathbf{q},\mathbf{k}_2+\mathbf{q}',\mathbf{k}_1+\mathbf{q}+\mathbf{q}'-\mathbf{Q},\mathbf{k}_1}(\mathbf{q}';\mathbf{Q}) \right]$$

$$- \sum_{(v'_1,c'_1,\mathbf{q}')} \left( W^{v'_1,c_1,v_1,c'_1}_{\mathbf{k}_1+\mathbf{q}+\mathbf{q}'-\mathbf{Q},\mathbf{k}_1,\mathbf{k}_1+\mathbf{q}-\mathbf{Q},\mathbf{k}_1+\mathbf{q}'} - V^{v'_1,c_1,c'_1,v_1}_{\mathbf{k}_1+\mathbf{q}+\mathbf{q}'-\mathbf{Q},\mathbf{k}_1,\mathbf{k}_1+\mathbf{q}',\mathbf{k}_1+\mathbf{q}-\mathbf{Q}} \right) \left[ B^{v'_1,c'_1,v_2,c_2}_{\mathbf{k}_1+\mathbf{q}+\mathbf{q}'-\mathbf{Q},\mathbf{k}_1+\mathbf{q}',\mathbf{k}_2-\mathbf{q},\mathbf{k}_2}(\mathbf{q}';\mathbf{Q}) \right.$$

$$\left. - B^{v_2,c'_1,v'_1,c_2}_{\mathbf{k}_2-\mathbf{q},\mathbf{k}_1+\mathbf{q}',\mathbf{k}_1+\mathbf{q}+\mathbf{q}'-\mathbf{Q},\mathbf{k}_2}(\mathbf{q}';\mathbf{Q}) - B^{v'_1,c_2,v_2,c'_1}_{\mathbf{k}_1+\mathbf{q}+\mathbf{q}'-\mathbf{Q},\mathbf{k}_2,\mathbf{k}_2-\mathbf{q},\mathbf{k}_1+\mathbf{q}'}(\mathbf{q}';\mathbf{Q}) + B^{v_2,c_2,v'_1,c'_1}_{\mathbf{k}_2-\mathbf{q},\mathbf{k}_2,\mathbf{k}_1+\mathbf{q}+\mathbf{q}'-\mathbf{Q},\mathbf{k}_1+\mathbf{q}'}(\mathbf{q}';\mathbf{Q}) \right]$$

$$+ \sum_{(v'_1,v'_2,\mathbf{q}')} \left\{ W^{v'_1,v'_2,v_1,v_2}_{\mathbf{k}_1+\mathbf{q}+\mathbf{q}'-\mathbf{Q},\mathbf{k}_2-\mathbf{q}-\mathbf{q}',\mathbf{k}_1+\mathbf{q}-\mathbf{Q},\mathbf{k}_2-\mathbf{q}} \left[ B^{v'_1,c_1,v'_2,c_2}_{\mathbf{k}_1+\mathbf{q}+\mathbf{q}'-\mathbf{Q},\mathbf{k}_1,\mathbf{k}_2-\mathbf{q}-\mathbf{q}',\mathbf{k}_2}(\mathbf{q}';\mathbf{Q}) - B^{v'_1,c_2,v'_2,c_1}_{\mathbf{k}_1+\mathbf{q}+\mathbf{q}'-\mathbf{Q},\mathbf{k}_2,\mathbf{k}_2-\mathbf{q}-\mathbf{q}',\mathbf{k}_1}(\mathbf{q}';\mathbf{Q}) \right] \right.$$

$$- W^{v'_2,v'_1,v_1,v_2}_{\mathbf{k}_2-\mathbf{q}-\mathbf{q}',\mathbf{k}_1+\mathbf{q}+\mathbf{q}'-\mathbf{Q},\mathbf{k}_1+\mathbf{q}-\mathbf{Q},\mathbf{k}_2-\mathbf{q}} \left[ B^{v'_2,c_1,v'_1,c_2}_{\mathbf{k}_2-\mathbf{q}-\mathbf{q}',\mathbf{k}_1,\mathbf{k}_1+\mathbf{q}+\mathbf{q}'-\mathbf{Q},\mathbf{k}_2}(\mathbf{k}_1-\mathbf{k}_2+2\mathbf{q}-\mathbf{q}'-\mathbf{Q};\mathbf{Q}) \right.$$

$$\left. \left. - B^{v'_2,c_2,v'_1,c_1}_{\mathbf{k}_2-\mathbf{q}-\mathbf{q}',\mathbf{k}_2,\mathbf{k}_1+\mathbf{q}+\mathbf{q}'-\mathbf{Q},\mathbf{k}_1}(\mathbf{k}_1-\mathbf{k}_2+2\mathbf{q}-\mathbf{q}'-\mathbf{Q};\mathbf{Q}) \right] \right\}$$

$$- \sum_{(v'_2,c'_1,\mathbf{q}')} \left( W^{v'_2,c_1,v_2,c'_1}_{\mathbf{k}_2-\mathbf{q}+\mathbf{q}',\mathbf{k}_1,\mathbf{k}_2-\mathbf{q},\mathbf{k}_1+\mathbf{q}'} - V^{v'_2,c_1,c'_1,v_2}_{\mathbf{k}_2-\mathbf{q}+\mathbf{q}',\mathbf{k}_1,\mathbf{k}_1+\mathbf{q}',\mathbf{k}_2-\mathbf{q}} \right) \left[ B^{v_1,c'_1,v'_2,c_2}_{\mathbf{k}_1+\mathbf{q}-\mathbf{Q},\mathbf{k}_1+\mathbf{q}',\mathbf{k}_2-\mathbf{q}+\mathbf{q}',\mathbf{k}_2}(\mathbf{q}';\mathbf{Q}) \right.$$

$$\left. - B^{v'_2,c'_1,v_1,c_2}_{\mathbf{k}_2-\mathbf{q}+\mathbf{q}',\mathbf{k}_1+\mathbf{q}',\mathbf{k}_1+\mathbf{q}-\mathbf{Q},\mathbf{k}_2}(\mathbf{q}';\mathbf{Q}) - B^{v_1,c_2,v'_2,c'_1}_{\mathbf{k}_1+\mathbf{q}-\mathbf{Q},\mathbf{k}_2,\mathbf{k}_2-\mathbf{q}+\mathbf{q}',\mathbf{k}_1+\mathbf{q}'}(\mathbf{q}';\mathbf{Q}) + B^{v'_2,c_2,v_1,c'_1}_{\mathbf{k}_2-\mathbf{q}+\mathbf{q}',\mathbf{k}_2,\mathbf{k}_1+\mathbf{q}-\mathbf{Q},\mathbf{k}_1+\mathbf{q}'}(\mathbf{q}';\mathbf{Q}) \right]$$

$$- \sum_{(v'_2,c'_2,\mathbf{q}')} \left( W^{v'_2,c_2,v_2,c'_2}_{\mathbf{k}_2-\mathbf{q}+\mathbf{q}',\mathbf{k}_2,\mathbf{k}_2-\mathbf{q},\mathbf{k}_2+\mathbf{q}'} - V^{v'_2,c_2,c'_2,v_2}_{\mathbf{k}_2-\mathbf{q}+\mathbf{q}',\mathbf{k}_2,\mathbf{k}_2+\mathbf{q}',\mathbf{k}_2} \right) \left[ B^{v_1,c_1,v'_2,c'_2}_{\mathbf{k}_1+\mathbf{q}-\mathbf{Q},\mathbf{k}_1,\mathbf{k}_2-\mathbf{q}+\mathbf{q}',\mathbf{k}_2+\mathbf{q}'}(\mathbf{q}';\mathbf{Q}) \right.$$

$$\left. - B^{v'_2,c_1,v_1,c'_2}_{\mathbf{k}_2-\mathbf{q}+\mathbf{q}',\mathbf{k}_1,\mathbf{k}_1+\mathbf{q}-\mathbf{Q},\mathbf{k}_2+\mathbf{q}'}(\mathbf{q}';\mathbf{Q}) - B^{v_1,c'_2,v'_2,c_1}_{\mathbf{k}_1+\mathbf{q}-\mathbf{Q},\mathbf{k}_2+\mathbf{q}',\mathbf{k}_2-\mathbf{q}+\mathbf{q}',\mathbf{k}_1}(\mathbf{q}';\mathbf{Q}) + B^{v'_2,c'_2,v_1,c_1}_{\mathbf{k}_2-\mathbf{q}+\mathbf{q}',\mathbf{k}_2+\mathbf{q}',\mathbf{k}_1+\mathbf{q}-\mathbf{Q},\mathbf{k}_1}(\mathbf{q}';\mathbf{Q}) \right]$$

.

(S18)

In both the 2B- and 4B-BSEs discussed above, we implicitly assumed the interaction kernels conserve the spins and total momenta for all the many-body states. The total momenta of the 2B and 4B states (defined as the sum of the momenta of all the electrons minus that of all the holes) in eq. (S10), (S11), (S17), & (S18) are limited to $\mathbf{Q}$. To simulate our pump-probe experiments, $\mathbf{Q}$ was restricted to zero due to the negligible momenta carried by the pump and probe photons. To avoid double counting during the calculations, we used a similar counting method for our 4B states to that for the 3B states in the previous studies[18,38,44].

**S12.3. Relationship between the 4B-BSE and the cluster expansion model**

To systematically categorize the interaction kernels and better understand the essence of the 4B perturbation equation, we translate eq. (S18) into the Feynman diagrammatic representation. Equation (S18) is schematically illustrated in Extended Data Fig. 13a. Extended Data Fig. 13b depicts all the *e-h*, *e-e*, and *h-h* screened interaction kernels (the correlation terms in the self-energy). Here, it should be pointed out that the *e-e* and *h-h* exchange interactions are screened, while the *e-h* exchange interactions (the exchange terms in the self-energy) are unscreened. These unscreened interaction kernels are not listed for brevity. In fact, their absence will not affect the later understanding of the 4B perturbation picture. Historically, the 2B-BSE was diagrammed by using the ladder approximation. Here, by converting the self-consistent equation in Extended Data Fig. 13 into the infinite series expansion in Extended Data Fig. 14, the complicated 4B interaction can be also understood as the infinite summation of the quadrupled ladders (virtual scatterings) from order-1, 2, … up to ∞. The above-mentioned interaction kernels in Extended Data Fig. 13b are exactly the 1$^{st}$-order virtual scatterings, which are regarded as the most elementary constituent of the higher-order virtual scatterings shown in Extended Data Fig. 14. These stacking ladders can be divided into several different partial summations. For the diagrams in each of the partial summations, it is exactly the same several propagators that are fully linked. Consequently, each partial summation can be considered equivalent to the corresponding term in the cluster expansion model (see also Fig. 1 in the main text). Finally, the many-body Hamiltonian is constructed by summing up all of them (the clusters of different orders). From the viewpoint of a unitary transformation, the final 4B eigen states obtained by diagonalizing the many-body Hamiltonian above are explicitly the linear combinations (eq. (S17)) of those 4B microscopic basis, but not the strict linear superpositions of the base vectors in the simplified model in Methods S1.

Additionally, there is a fairly good consistency between our theory of the 4B-BSE and the dynamics-controlled truncation scheme within the coherent $\chi^{(3)}$ limit[32,47-49]. So, our theory is more suited for describing the initial generations of the correlated 4B entities around the zero pump-probe delay (t ≈ 0 ps), rather than the incoherent cases where the 2B states are formed by those of hot carriers (*e.g.* PL of the bright-dark bi-exciton in ML-WSe$_2$ or ML-WS$_2$).

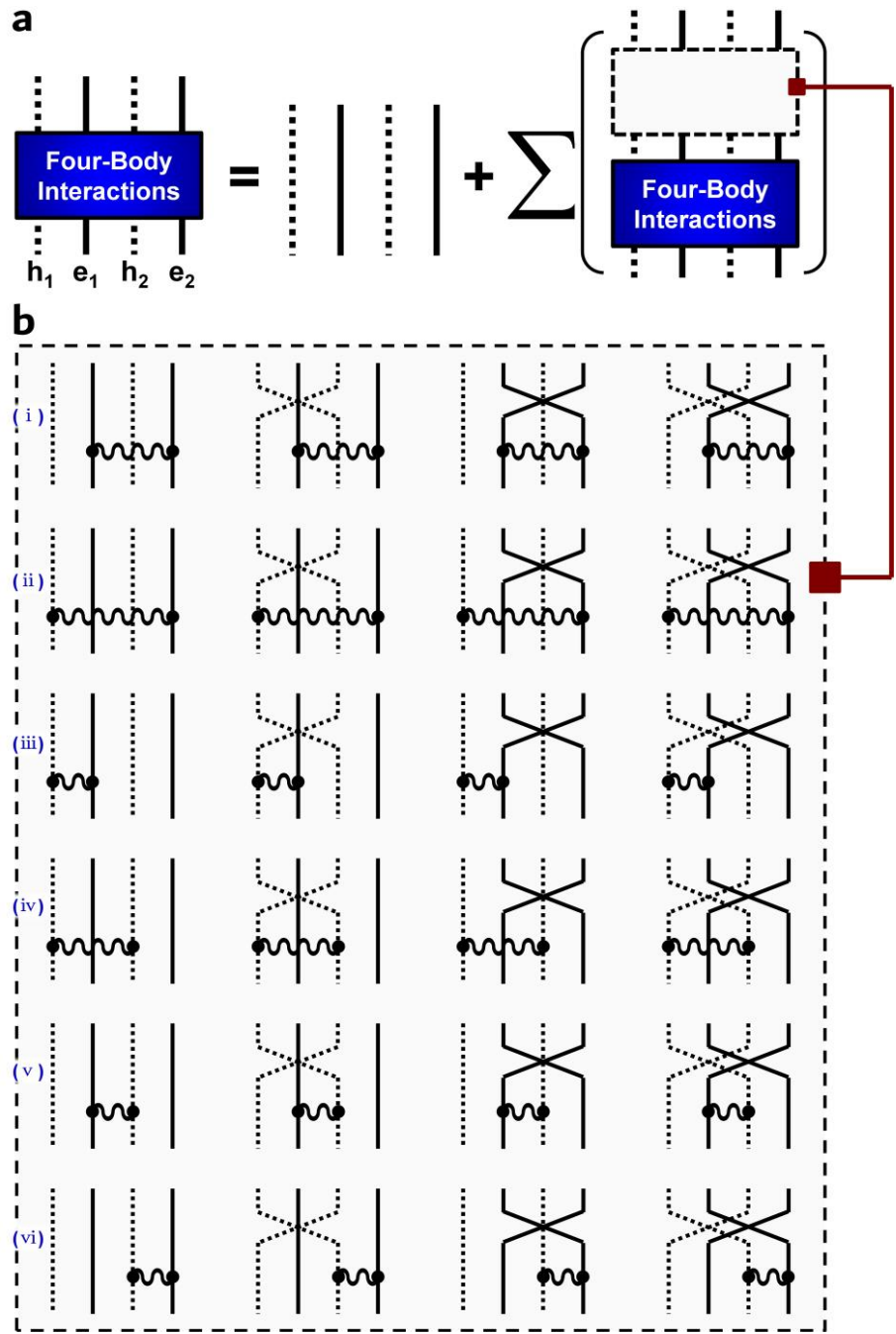

**Extended Data Fig. 13 | a,** Schematic of the self-consistent equation for the correlated 4B system.

**b,** Feynman diagrammatic representations of the screened interaction kernels (labelled as ( i ) - ( vi )) included in the off-diagonal elements of the microscopic many-body Hamiltonian (eq. (S18)).

The unscreened interaction kernels are skipped here for brevity. The dashed-line box in **a** is further detailed by the diagrams in **b**. For all the diagrams in **b**, their signs (positive or negative) derived from the fermionic permutations are not illustrated but given in details in eq. (S18).

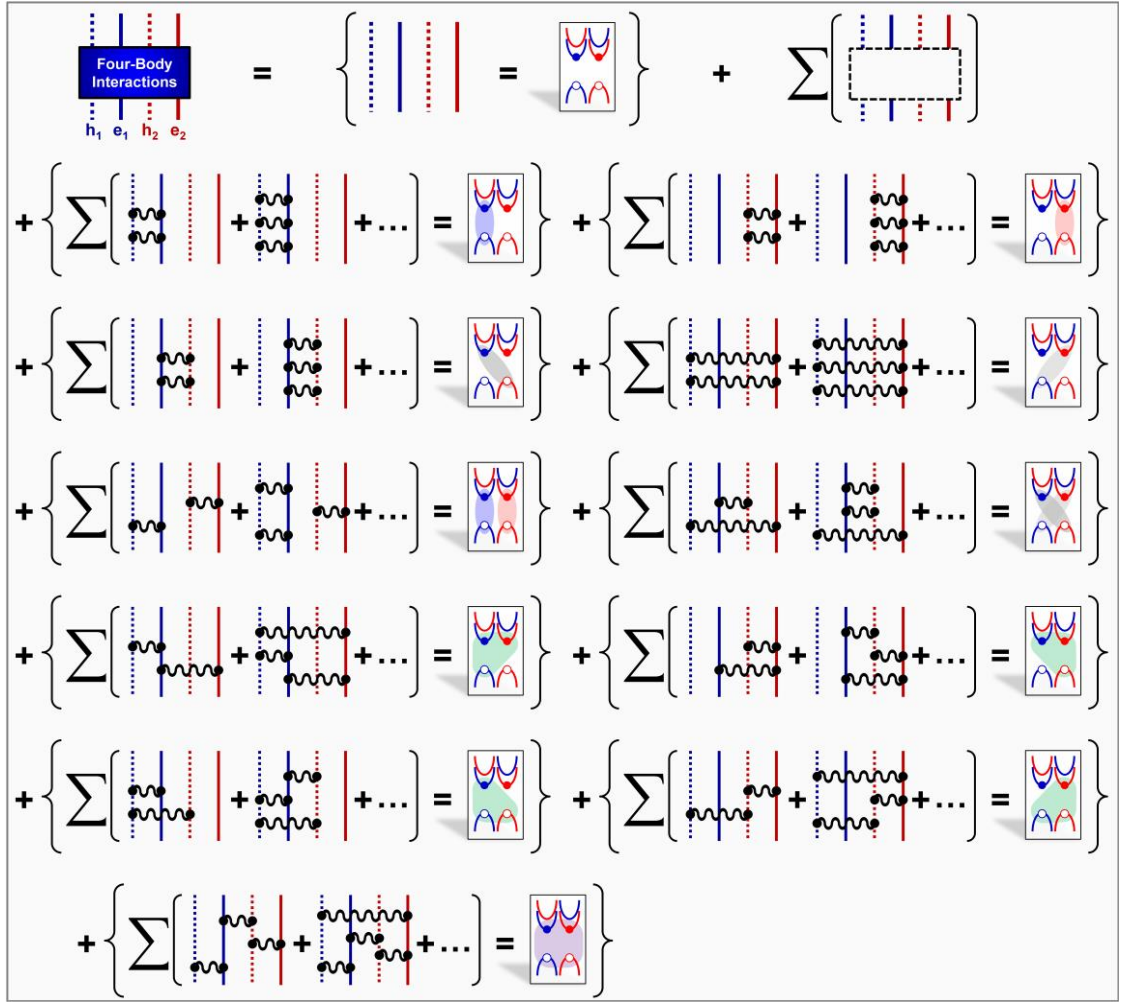

**Extended Data Fig. 14 |** Schematic of the infinite series expansion for the correlated 4B system and its correspondence to the cluster expansion model in Fig. 1 in the main text. We only include the inter-valley cases for brevity, with the K- (spin-up) and K'- (spin-down) valley electrons colored in blue and red, respectively. The summations of the high-order virtual scatterings are classified according to their correspondences to the terms of different orders in the cluster expansion model. Also for the sake of brevity, the diagrams involving the unscreened interactions are not shown.

**S13. Theoretical calculation for the absorption spectra**

The absorption spectra of the 2$e$2$h$ 4B system are determined by all the possible 2B-4B polarizations, which are given by the dipole matrix elements between the various 2B states and the corresponding 4B states. As illustrated in Fig. 2a in the main text, the pump pulse produces a mixture of 2B states (maybe including 1$s$-X, 2$p$-X, 2$s$-X, … plasma) as initial states for the probe process. Absorption of a probe photon leads to a transition from one of those 2B states (the initial states) to one of the 4B states (the final states). Or in other words, the formation of the 4B states intimately relates to those of 2B states. Therefore, first we need to consider among all the 2B states which ones should be included in the theory.

With increase in the pumping density, the 2B states gradually deviate from the pure 1s-exciton gases, and therewith possibly undergoes other processes: *e.g.* the 1s-plasma Mott transition[50-52], the 1s-ns upconversions[24], or the 1s-np intra-excitonic transitions[53]. According to the recent experiments, the time scales are reported to be about 0.1 – 1.0 ps for the ultrafast Mott transition[52], and 0.7 ps to reach the quasi-equilibrium among the 1s-np transitions[53]. The time scales of their occurrences are within the life span (~ 1 ps) of the possible many-body features observed in our experiments. That's say, besides the 1s excitons, these of other *e-h* 2B states (*i.e.* 2p-X, 2s-X, … plasma) should be also considered in the theory.

We note that the processes of the 1s A exciton scattering to the dark exciton or the B exciton would take relatively longer time (estimated to be more than 1 ps for MoS$_2$ in Ref. [54]). Moreover, the CB splitting is much larger for MoTe$_2$ than for MoS$_2$, thus the above scattering time would be longer for MoTe$_2$ than for MoS$_2$. Therefore, the dark exciton and the B exciton with higher single-particle energies are considered not to contribute considerable proportions to the 2B-state mixture in the "soup" around the zero delay. This will be extensively discussed later in Methods S14.

Based on the above understanding, we concern all the single-particle basis only from VBM and CBM (see Methods S14). The 2B states, $|e_3 h_3\rangle$, including the 1s-X, 2p-X, 2s-X, … plasma with the corresponding energies $\varepsilon_{e_3 h_3}$ are solved from the conventional 2B-BSE (eq. (S11)), and the 4B states ($|e_1 h_1 e_2 h_2\rangle_\alpha$) with the $\alpha^{th}$ energies $\varepsilon^{\alpha}_{e_1 h_1 e_2 h_2}$ are calculated from the 4B-BSE (eq. (S18)) (see Methods S14 for the numerical techniques). The polarization operator is written as $\hat{\mathbf{p}} = \sum_{(v,c,\bar{\mathbf{k}},\mathbf{k})} \mathbf{p}^{v,c}_{\bar{\mathbf{k}},\mathbf{k}} \hat{a}_{c\mathbf{k}} \hat{a}^?_{v\bar{\mathbf{k}}}$, where $\mathbf{p}^{v,c}_{\bar{\mathbf{k}},\mathbf{k}}$ is the dipole matrix element as defined in eq. (S8) & (S9). The 2B-4B polarization (eq. (4) in the main text) is then given as

$$\langle e_3 h_3 | \hat{\mathbf{p}} | e_1 h_1 e_2 h_2 \rangle =$$

$$\sum_{(v_3,c_3,\mathbf{k}_3)} \left( A^{v_3,c_3}_{\mathbf{k}_3-\mathbf{Q},\mathbf{k}_3} \right)^* \sum_{\substack{(v_1,c_1,v_2,c_2) \\ \mathbf{q},\mathbf{k}_1,\mathbf{k}_2}} B^{v_1,c_1,v_2,c_2}_{\mathbf{k}_1+\mathbf{q}-\mathbf{Q},\mathbf{k}_1,\mathbf{k}_2-\mathbf{q},\mathbf{k}_2} \left( \Theta_{123}\Theta_{231} \mathbf{p}^{v_1,c_1}_{\mathbf{k}_1-\mathbf{q}+\mathbf{Q},\mathbf{k}_1} \delta_{v_2,v_3} \delta_{\mathbf{q},\mathbf{Q}} \delta_{c_2,c_3} \delta_{\mathbf{k}_2,\mathbf{k}_3} \right.$$

$$+\Theta_{213}\Theta_{312} \mathbf{p}^{v_2,c_1}_{\mathbf{k}_2-\mathbf{q},\mathbf{k}_1} \delta_{v_3,v_1} \delta_{\mathbf{k}_2-\mathbf{q},\mathbf{k}_1} \delta_{c_2,c_3} \delta_{\mathbf{k}_2,\mathbf{k}_3} + \Theta_{123}\Theta_{132} \mathbf{p}^{v_1,c_2}_{\mathbf{k}_1+\mathbf{q}-\mathbf{Q},\mathbf{k}_2} \delta_{v_2,v_3} \delta_{\mathbf{k}_2,\mathbf{k}_1+\mathbf{q}-\mathbf{Q}} \delta_{c_3,c_1} \delta_{\mathbf{k}_3,\mathbf{k}_1},$$

$$\left. +\Theta_{321}\Theta_{213} \mathbf{p}^{v_2,c_2}_{\mathbf{k}_2-\mathbf{q},\mathbf{k}_2} \delta_{v_3,v_1} \delta_{0,\mathbf{q}} \delta_{c_3,c_1} \delta_{\mathbf{k}_3,\mathbf{k}_1} \right)$$

(S19)

where factors $\Theta_{ijk}$ ($\Theta_{123}=\Theta_{312}=\Theta_{231}=1$, or $-1$ otherwise) are derived from the canonical relations of the creation-annihilation operators. The absorption spectrum is finally given by the Fermi's Golden Rule,

$$\varepsilon_2 \propto \frac{2\pi}{\hbar} \sum_{\alpha} \left| \sum_{\mathbf{k}} \langle e_3 h_3 | \mathbf{e} \cdot \hat{\mathbf{p}} | e_1 h_1 e_2 h_2 \rangle_{\alpha} \right|^2 \Gamma\left( \varepsilon^{\alpha}_{e_1 h_1 e_2 h_2} - \varepsilon_{e_3 h_3} - \hbar\omega - i\gamma \right). \quad (S20)$$

Here, **e** denotes the unit vector of the in-plane electric field component for the normal incident field, $\hbar\omega$ stands for the photon energy, and $\Gamma$ represents the broadening function, where the artificial broadening constant, $\gamma$, is also introduced here. In the calculations, we assume the 2B states are only located at the K' valley (*e.g.* locked with σ– circularly polarized photon). In this way, the spectra of the transitions between these 2B states to the inter- or intra-valley 4B states can be obtained by configuring the probe photons (**e**) to be σ+ or σ– circularly polarized, respectively (see Extended Data Fig. 15 for the calculated helicity-resolved spectra).

### S13.1. Calculation for the absorption spectra from different 2B states

Extended Data Fig. 15 shows the calculated absorption spectra for the 2B-4B transitions between the different 2B states (1*s*-X, 2*s*-X, and plasma) and the corresponding inter- (σ– σ+) and intra-valley (σ+ σ+) 4B states. The convergence behaviors of the similar spectra are shown in Methods S14, Extended Data Fig. 19. Extended Data Fig. 15c & 15f also shows the experiment-theory comparison. It is important to point out that the actual absorption spectrum (or the experimental observation) should be determined by an appropriately weighted summation over all the spectra such as those in Extended Data Fig. 15a – 15c (more specifically, all the spectra in Extended Data Fig. 17c) or 15d – 15f. The intuitive illustrations can be seen in Fig. 5g – 5j in the main text and also Extended Data Fig. 18. In other words, it is insufficient to compare the experimental measurement to only one of these theoretical spectra (*e.g.* a direct comparison between the experimental measurement and the theoretical spectra in Extended Data Fig. 15c or 15f is not strictly true).

A 1*s*-X spectral peak will be obtained if the 2B initial state is a 1*s*-X, 2*s*-X, … or plasma and the 4B final state is a non-interacting (1*s*-X)(1*s*-X), (1*s*-X)(2*s*-X), … or (1*s*-X)(plasma), respectively, *i.e.* correspondingly those of △₂ △₂ clusters without wavy lines (see Methods S1). Specifically for Fig. 5d & 5e in the main text, peak 2 (in Fig. 5d) and peak 7 (in Fig. 5e) correspond to a (1*s*-X)(1*s*-X) and (1*s*-X)(plasma) without interactions, respectively. Such a similar sequence of 2B-4B transitions all mark the 1*s*-X line, as shown in Extended Data Fig. 17a & 18a. Besides the 1*s*-X spectral peak, some additional features will be obtained if the final states are related to other clusters such as △₄, *etc*, beyond △₂ △₂.

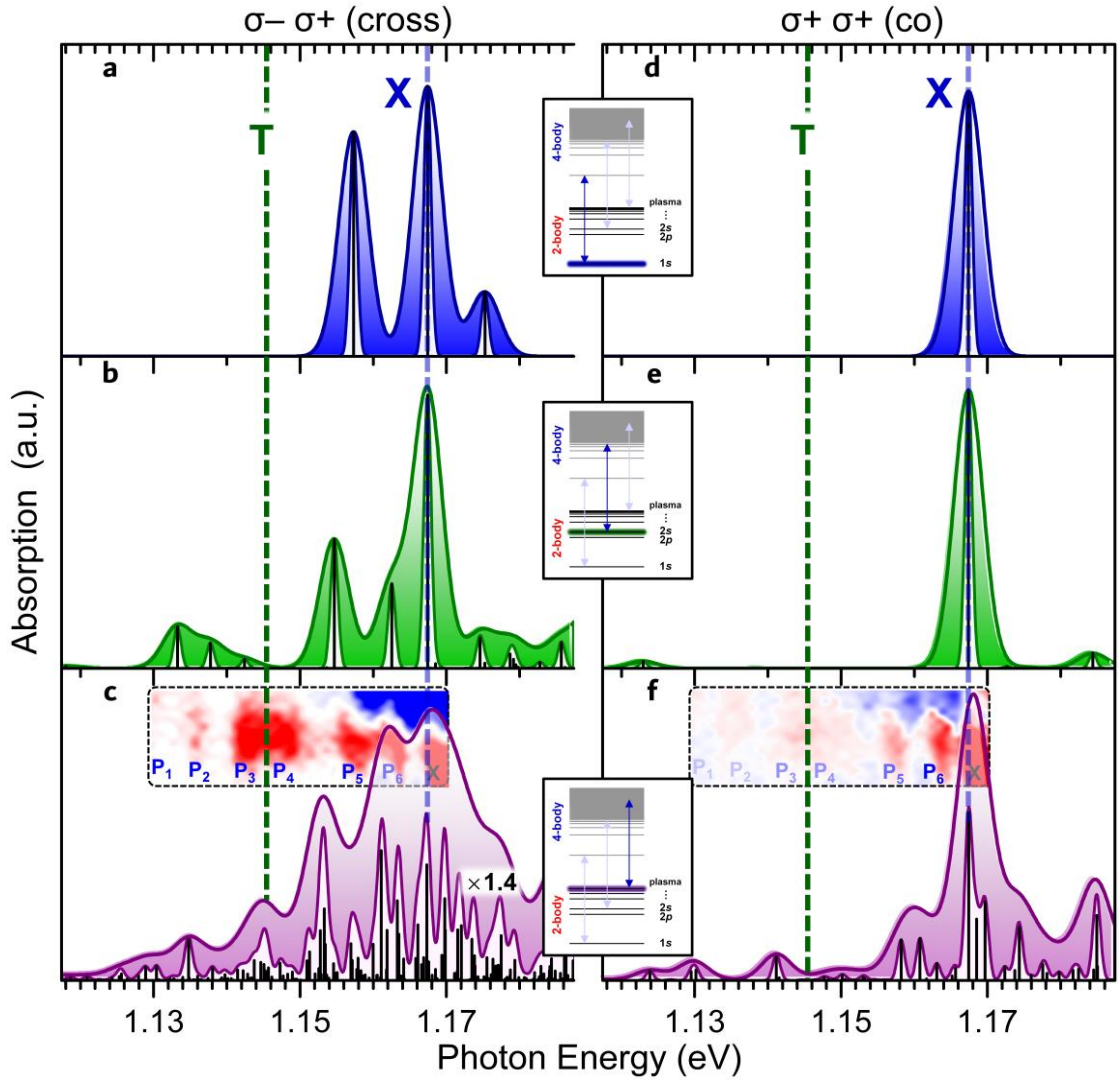

**Extended Data Fig. 15 | a – f,** Calculated absorption spectra for the 2B-4B transitions between the different 2B states (1s-X **(a & d)**, 2s-X **(b & e)**, plasma **(c & f)**) and the inter- (σ− σ+) **(a – c)** and intra- (σ+ σ+) **(d – f)** valley 4B states. The transition spectra with the same 2B states are shown in the same row to compare the cross- (σ− σ+) (left) and co- (σ+ σ+) (right) circularly polarized configurations. To show the broadening effects, two values, 0.5 and 2.0 meV, are used in the Gaussian broadening function (eq. (S20)) for comparison, corresponding to the unfilled and filled spectral profiles, respectively. The same T line with that in Fig. 5a – 5f in the main text, and the X line are marked with the green and blue dashed lines, respectively. The experimental differential absorption shown in Fig. 3d & 3i in the main text is reproduced in **c & f** and horizontally aligned well with the calibrated energy axis.

There are several points that can be directly concluded from Extended Data Fig. 15:

**1)** As can be seen for the same 2B states (the same rows), the spectral features below X are much richer for the (σ− σ+) case (Extended Data Fig. 15a – 15c) than for the (σ+ σ+) case (Extended Data Fig. 15d – 15f). The similar polarization contrast has been observed as well in our experiments for Device #1 (Extended Data Fig. 6a), #2 (Fig. 3d

& 3i in the main text), #4 (Extended Data Fig. 7a & 7b), & #5 (Extended Data Fig. 10c & 10f).

**2)** If the 2B initial state is a 1$s$-X, there is a spectral peak between T and X (Extended Data Fig. 15a, see also Fig. 5d in the main text), corresponding to the inter-valley 4B state. While no peak exists below X for the intra-valley configuration (Extended Data Fig. 15d).

**3)** If the 2B initial state is a 2$s$-X or a plasma and the 4B final states are the inter-valley ones, there are quite a few peaks below X. Besides the ones emerging between T and X, some peaks also newly occur below the T peak (Extended Data Fig. 15b & 15c, see also Fig. 5e in the main text).

**4)** If the 2B initial state is a 2$s$-X and the 4B final states are the intra-valley ones, the spectrum is featureless below X (Extended Data Fig. 15e), similarly to the case of the 1$s$-X (Extended Data Fig. 15d); While if the 2B state is a plasma, quite a few peaks exist below X: some between T and X, and the others below T, corresponding to those of intra-valley 4B states (Extended Data Fig. 15f).

To intuitively display the 2B-4B transitions, the spectra in Extended Data Fig. 15a – 15c are also shown in Extended Data Fig. 16 at the total energy level.

We observed the features that are spectrally below T ($P_1$ – $P_4$) and also between T and X ($P_5$ – $P_6$). According to the theoretical results we described above, the observation of $P_1$ – $P_4$ means that the $e$-$h$ "soup" excited by the pump pulse included those of the highly excited 2B (*e.g.* 2$s$-X and plasma) and 4B states. Or in other words, the $e$-$h$ "soup" included not alone the most ground state of the 2B and 4B manifolds, *e.g.* was not a pure 1$s$-X gas. The occurrences of $P_1$ – $P_4$ reveal the continuous Mott transition at that time the system was undergoing. Combined with the later discussions in Methods S13.2, the polarization contrasts shown in Extended Data Fig. 15c & 15f appear to be sharper for △₄ and △₃ ~ △₁ than for △₂ ~ △₂. It is not difficult to draw another conclusion that the polarization contrasts for all of △₄, △₃ ~ △₁, and △₂ ~ △₂ are sharper than for the △₂△₂ and △₂△₁△₁ continuums, *i.e.* there are almost no such polarization contrasts for those spectral peaks in the very vicinity of X. Therefore, the 4B physics add information and complexity to the highly excited *e-h* system and thus provide new aspects one can further look into.

So far, the coexistence of these *e-h* 2B states and other possible high order excitonic complexes in a continuous Mott transition is less understood. In the *e-h* "soup" (Fig. 2a in the main text), the 4B entities such as a 1*s*-X binding to another 1*s*-X, 2*p*-X, 2*s*-X, ... plasma have not been studied before beyond those of conventional bi-excitons and excited-state bi-excitons. What we have done in the experiments is to find new 4B entities or new 2B-4B transitions by using the 1*s*-X as a probe. We explored the exotic few-body entities beyond the well-defined excitons, trions, and bi-excitons, and thus we newly saw some details in a continuous Mott transition. This will be extensively discussed in the next Section (Methods S13.2).

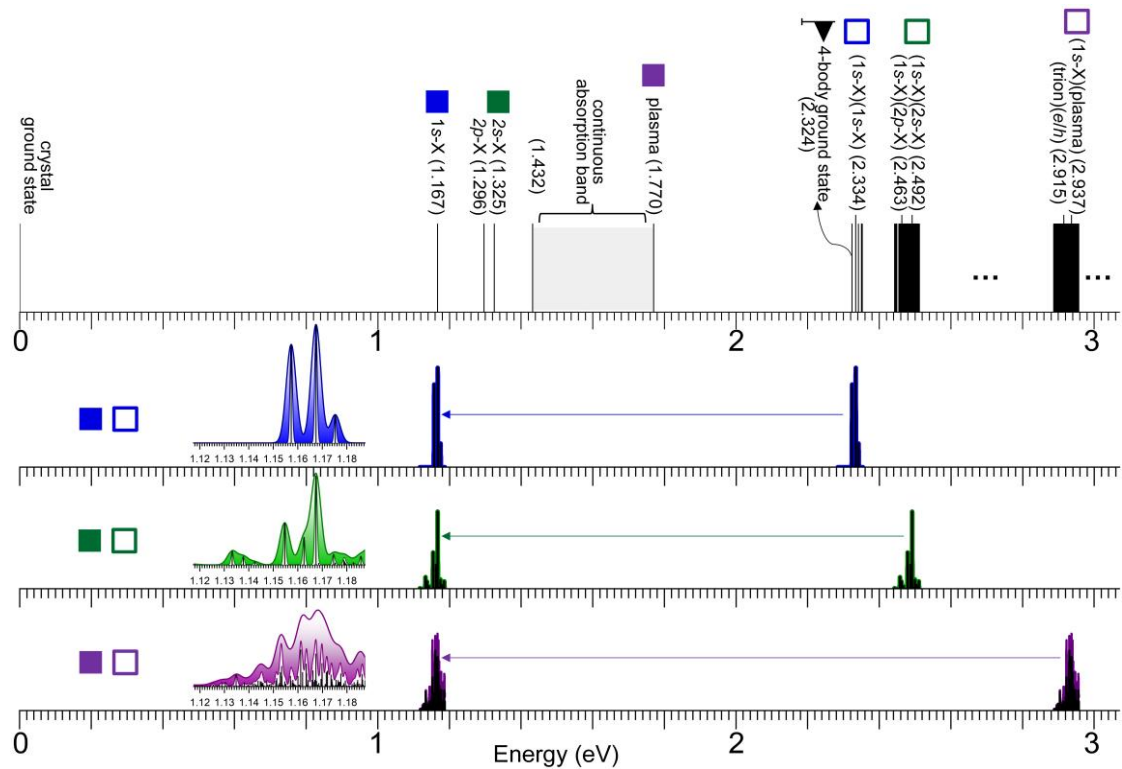

**Extended Data Fig. 16 |** Energy positions of some typical 2B and 4B states calculated from the theory. The 2B-4B transition spectra in Extended Data Fig. 15a, 15b, & 15c are marked at both of the total energy level (right side of ←) and the spectral energy level (left side of ←). Notation ■ and □ mark the 2B initial states and 4B final states, respectively.

## S13.2. Absorption spectra with the Hamiltonian truncated up to different orders

As has been discussed in the main text, the full 4B interaction can be expanded in terms of clusters △△△△, △△△, △△, △△ and △. It is important to show that cluster △ is indispensable for the spectral features observed in our experiments. To show the necessity of cluster △ in producing those low-energy spectral features, we truncated the Hamiltonian up to different orders of clusters and compared the corresponding spectral features for each of the 15 2B states (1$s$-X, 2$p$-X, 2$s$-X, … plasma) solved from eq. (S11). All the results are demonstrated in Extended Data Fig. 17. The convergence behaviors of the similar spectra are shown in Methods S14, Extended Data Fig. 19. The changes from each time one more cluster of higher order introduced (see Methods S1), are demonstrated both in the total energy scales and the optical spectra, as shown in Extended Data Fig. 18 (the 6 examples, selected out of all the 15 cases).

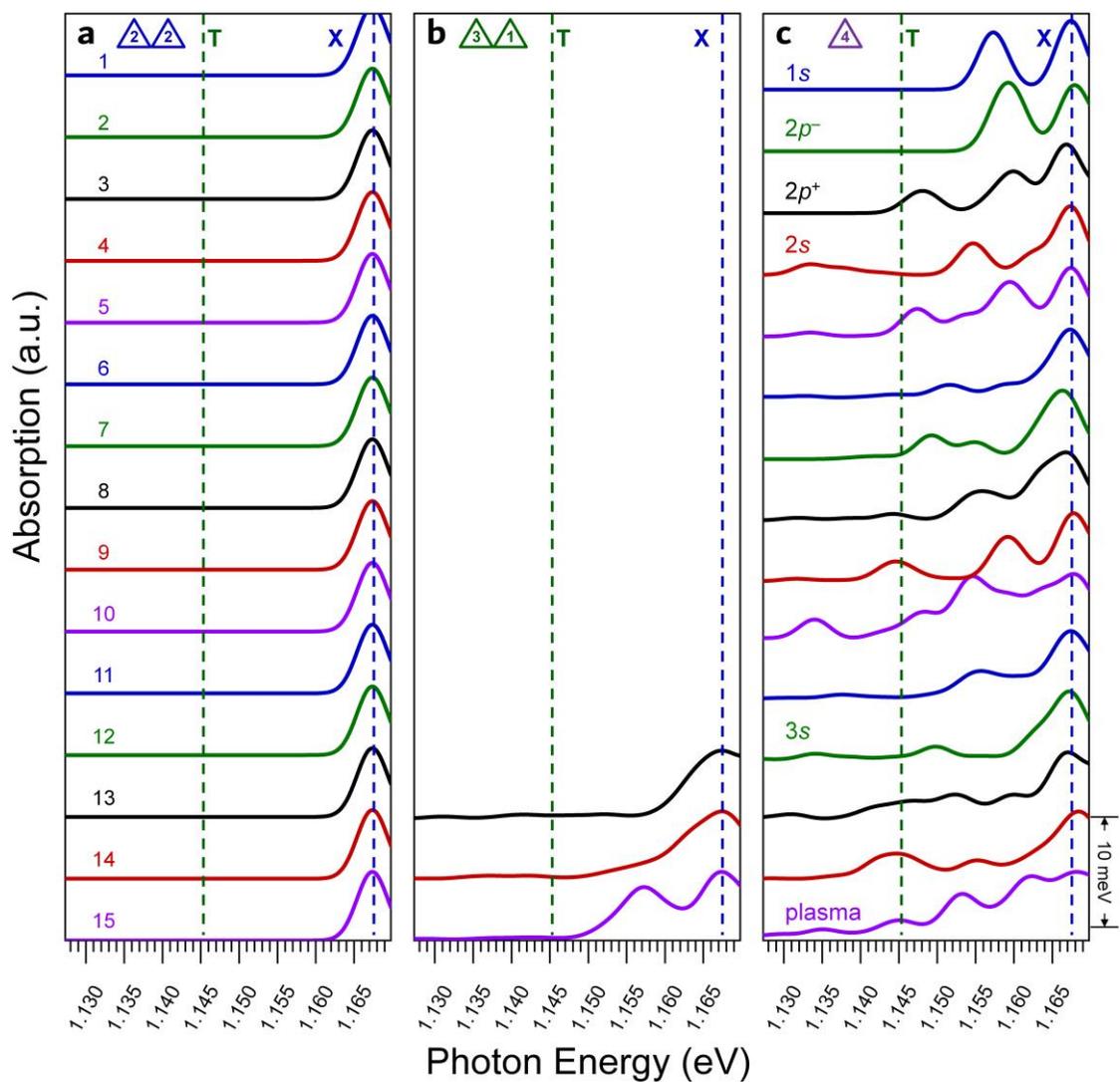

**Extended Data Fig. 17 |** Calculated 2B-4B transition spectra with the Hamiltonian truncated up to different orders of clusters △△ **(a)**, △△ **(b)** and △ **(c)** from the 15 2B states to the corresponding inter-valley 4B states.

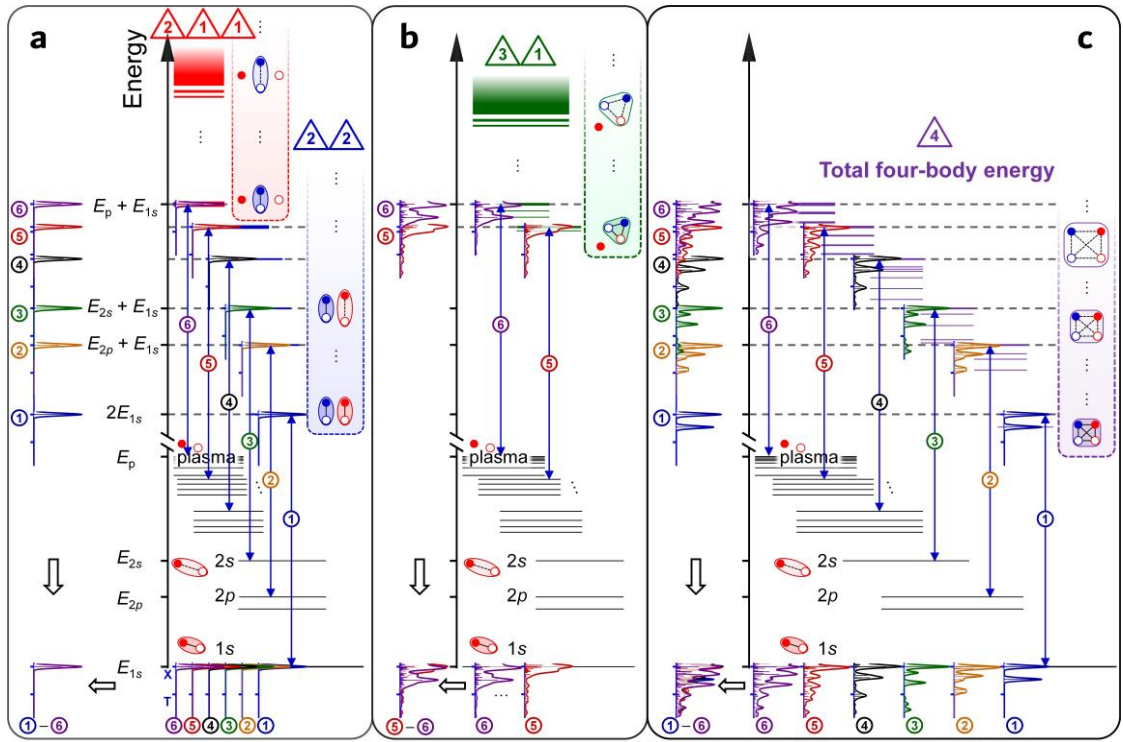

**Extended Data Fig. 18 |** Schematic of the changes that occur both in the total energy spectra and the optical transition spectra, from each time one more cluster of higher order introduced. The full Hamiltonian is truncated up to different orders of clusters △₂△₂ **(a)**, △₃△₁ **(b)** and △₄ **(c)** (see also Fig. 5g – 5j in the main text). As examples, 6 typical cases are selected out of all the 15 2B states in Extended Data Fig. 17, and used to demonstrate the 2B-4B transitions (transitions ① – ⑥), the corresponding total energy spectra (left side of each figure), the optical spectra (bottom side), and the "collapsed" spectra (bottom left corner). The calculations for T⁻h (shown in **b**) and its counterpart, T⁺e, gave the identical results. So we only illustrated the former in **b** for brevity.

The details of our procedure are described as follows:

**Case 1: Truncation up to △₂△₂:** We set to zero the kernels marked by (ⅰ), (ⅱ), (ⅳ), and (ⅴ) in Extended Data Fig. 13, together with all the associated unscreened ones. After diagonalizing the truncated Hamiltonian, we calculated the absorption spectra for the 2B-4B transitions from each of the 2B states, with the results shown in Extended Data Fig. 17a & 18a (see also Fig. 5b in the main text). Within this level, there are no other spectral features below X. We noticed that the same method was adopted in Ref. [32] to extract the total 4B energy of the independent exciton-exciton pair, *i.e.* (1s-X)(1s-X).

**Case 2: Truncation up to** ⟨3⟩⟨1⟩ **:** In this case, we set to zero the kernels (iv) and (v) in Extended Data Fig. 13, together with all the associated unscreened ones. In this way, $h_2$ is intentionally left fully independent from $e_1$ and $h_1$, and also ionized together with $e_2$ due to the 2B wavefunction given the plasma state, although here they are still considered linked with each other. The results are shown in Extended Data Fig. 17b & 18b (see also Fig. 5c in the main text). We see in this special case, the 2B-4B transition spectra from the 12 2B states with lower energies are left blank in Extended Data Fig. 17b. The reason is that clusters ⟨3⟩⟨1⟩ have the total 4B energies much higher than many of ⟨2⟩⟨2⟩ (In 2D TMDCs, the binding energies of X are always one magnitude larger than T). So it is impossible for the 2B states with relatively lower energies (*e.g.* 1*s*-X, 2*p*-X, 2*s*-X) binding to another 1*s*-X to form the high-energy 4B entities such as clusters ⟨3⟩⟨1⟩, unless these 2B states are given around the plasma (*e.g.* 13 – 15 in Extended Data Fig. 17b, see also Extended Data Fig. 18b). For the 3 cases with number 13 – 15, there are new spectral features emerging between T and X, but no obvious features below T.

**Case 3: The full Hamiltonian up to** ⟨4⟩ **:** The results are shown in Extended Data Fig. 17c & 18c (see also Fig. 5d & 5e in the main text). We see clearly in this case there are new spectral features emerging below and near the T resonance. The fully correlated quadruplet term ⟨4⟩, as the highest-order cluster in the 4B cluster expansion, has two prominent effects. First, this term leads to the appearance of new spectral features well below the T resonance, corresponding to the 4B irreducible entities: quadruplons. Second, this term results in the coupling between the clusters of different orders, *i.e.* ⟨3⟩⟨1⟩ and ⟨2⟩⟨2⟩ (without the wavy lines), or in other words, the generation of ⟨3⟩ ~ ⟨1⟩, ⟨2⟩ ~ ⟨2⟩ and other lower-order clusters (with the wavy lines). For reference, both of the two effects described above are well generalized by the simplified model in Methods S1.

Based on above content, we can extend the definition of the 4B "bound" states. For example, the features below X in Extended Data Fig. 15b & 15c correspond to the 4B entities that are energetically lower than the (1s-X)(2s-X) (or △△) and (1s-X)(plasma) (or △△△) continuums, respectively (see transitions ③ & ⑥ in Extended Data Fig. 18, respectively). Thus, it is true that these entities are the bound states when compared with the (1s-X)(2s-X) and (1s-X)(plasma) continuums, respectively. However, they can be also regarded as the highly excited states when compared with the (1s-X)(1s-X) continuum. The 4B irreducible cluster can always lead to new "bound" states at the very different total-energy scales (energetically extending from the most ground state up to the highly excited state of the 4B manifold).

From Extended Data Fig. 17 & 18, the conclusion can be drawn that the 4B irreducible cluster, △₄, or the quadruplon is necessary and sufficient in producing all the experimental spectral features. For each of the same 2B states, the corresponding 4B irreducible entities (cluster △₄) has lower energies and are thus more stable than those of △₂~△₂ (BX) and △₃~△₁ (T$^+$~e and T$^-$~h), indicating the most stable existence of the quadruplon. It is also important to point out that for each of the 2B states with very different energies, the 4B spectrum can be always identified as a sequence of the "bound" states, *i.e.* △₄, △₃~△₁, △₂~△₂, and △₂~△₁~△₁ (for the high-energy excited states), or △₄ and △₂~△₂ (for the low-energy excited states. The description of "bound" states means that the states have lower total 4B energies than the corresponding △₂△₂ or △₂△₁△₁ continuums. We see surprisingly the 4B cluster expansion physics can occur at the total energy scales with large disparities (several hundred meV), but finally were captured by the dressed excitonic spectroscopy with the spectral range of ~ 40 meV.

**S14. Numerical techniques and the convergence tests**

First of all, we point out again the difference between the band structures of Mo- and W-based monolayers and the related consequence. For the zero-momentum excitations, it is the bright 2B states that occupy the energetically lowest bands for the Mo-based monolayers, while is the spin-forbidden dark 2B states for the W-based monolayers. Thus for the W-based monolayers, the well-defined bi-exciton states, such as the bright-dark bi-excitons[6,7,20,21,55] and the brightened dark-dark bi-excitons[19], are energetically lower than the bright-bright ones. Such property is easy to cause the rapid scattering (relaxing from the bright-bright states to the bright-dark or dark-dark ones) and lead to the low-energy spectral features for the W-based materials. While for the Mo-based materials, the bright-dark and dark-dark bi-excitons are energetically higher than the bright-bright ones. So the scattering process described above for the W-based systems becomes quite difficult for the Mo-based systems (see also Methods S4 & S13). We observed the low-energy spectral features ($P_1 - P_4$) for the ML-MoTe$_2$, the possibility can be greatly avoided that they are caused by the bright-dark and dark-dark bi-excitons. And also for the purpose of reducing the computations, we concerned all the single-particle states in the theoretical model to be restricted to CBM and VBM.

As we know, the 1$s$ excitons and the related trions and bi-excitons are wannier-type excitations and well localized in the vicinity of K and K'. Therefore, a Brillouin zone truncation scheme[32,38,45] was applied to further reduce the computational load. The reduced single-particle basis was determined by a circle around the K and K' points with the radius of $k_{max}$. We used the Monkhorst-Pack mesh to perform the discretization for the Brillouin zone. The diagonalization of the extremely large many-body Hamiltonian matrices were performed by utilizing an iterative Krylov space method as complemented in the SLEPc package[56] for the PETSc toolkit[57].

To show the robustness of the results against $k_{max}$, we performed the convergence test of the results with the same k-grid of 33 × 33 × 1 over different $k_{max}$ = 0.199, 0.227, &

0.255 Å⁻¹ (Extended Data Fig. 19a – 19c). The results show reasonably good convergence with a linewidth-broadening larger than 0.5 meV at $k_{max}$ = 0.227 Å⁻¹. Here, we discuss the applicability and the reliability of the **k**-space truncation scheme to describe the higher Rydberg states of excitons (2p-X, 2s-X, … plasma). These highly excited 2B states were calculated in Ref. [43] by using the 2- and 3-band model with the full FBZ included. Despite calculating without the **k**-space truncations, the **k**-space plot of the product $A_{vc}^{X}(\mathbf{k})P^{vc}(k)$ showed that the polarizations for those of 2p-X, 2s-X, and 3d-X are also in the vicinity of K and K' similarly to the case of 1s-X[43]. The spectrum containing such a series of 1s-X, … plasma in Ref. [43] was well reproduced in Ref. [38] by using the **k**-space truncation. Therefore, we believed the reduced model could give the reasonably reliable results for those of higher excitonic Rydberg states and the related 3B and 4B entities. To consider the accurate influences of the side valleys (*e.g.* Q, Γ, M, *etc*), a fully fist-principle calculation should be further performed with the entire FBZ included.

Extended Data Fig. 19d – 19f show the convergence tests of the 2B-4B transition spectra with respect to the k-mesh densities. For brevity, only three cases are demonstrated as representations, *i.e.* for the energetically lowest one (Extended Data Fig. 19d) and for the energetically highest one (Extended Data Fig. 19e & 19f) among the 15 cases in Extended Data Fig. 17. Extended Data Fig. 19g – 19i show the comparisons of the spectra with the Monkhorst-Pack mesh shifted 1/2 grid (K(K') included) or not (K(K') excluded). As can be seen, the results have a reasonable convergence behavior with the k-grid of ~ 36 × 36 × 1, and the calculated spectral features are not sensitive to the different schemes of the Monkhorst-Pack k-grids. The similar k-grid density (*e.g.* ~ 39 × 39 × 1), typically used in Ref. [45], proved sufficient for the many-body perturbation calculation for BX. In fact, such a level of k-grid density has nearly reached the upper limit of our current computation ability.

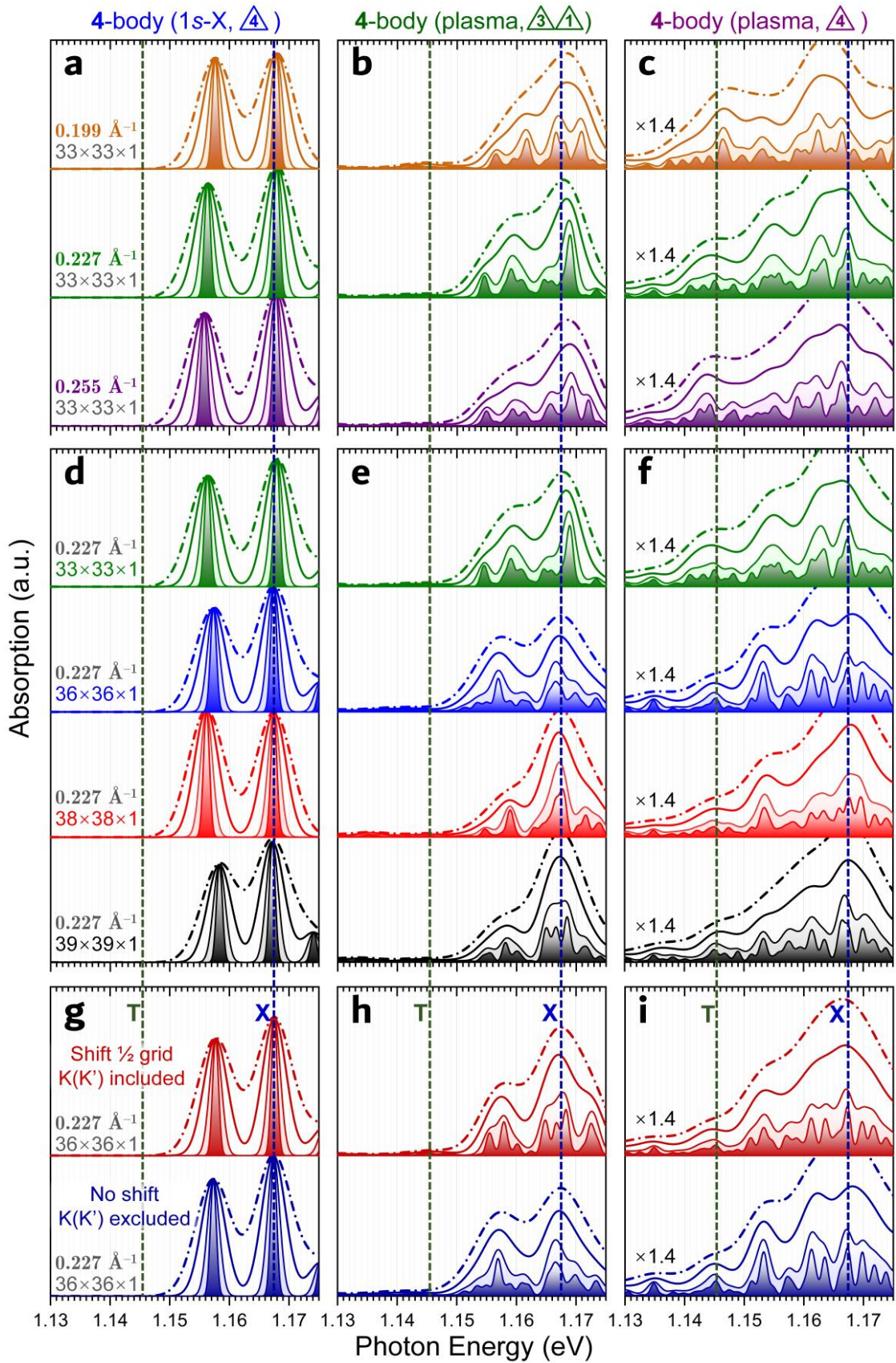

**Extended Data Fig. 19 | a – c,** Convergences of the 2B-4B transition spectra for the ML-MoTe$_2$ testing over various k-space truncation radius. **d – f,** Convergences of the spectra testing over various k-mesh densities. **g – i,** Comparisons of the spectra calculated with the Monkhorst–Pack k-

mesh shifted 1/2 grid or not. In cases of **a, d, & g,** the 2B states are the 1s-X and the 4B states are solved from the full 4B-BSE. In cases of **b, c, e, f, h, & i** the 2B states are identically the plasma, but the 4B states are solved from the 4B-BSE truncated up to △△ (**b, e, & h**) and △ (**c, f, & i**).

The green and blue dashed lines mark the calculated energies of the trion (T) and exciton (X). The Gaussian broadening parameter, $\gamma$ (eq. (3) in the main text), is chosen to be 0.5, 1.0, 2.0, & 3.0 meV for comparisons.

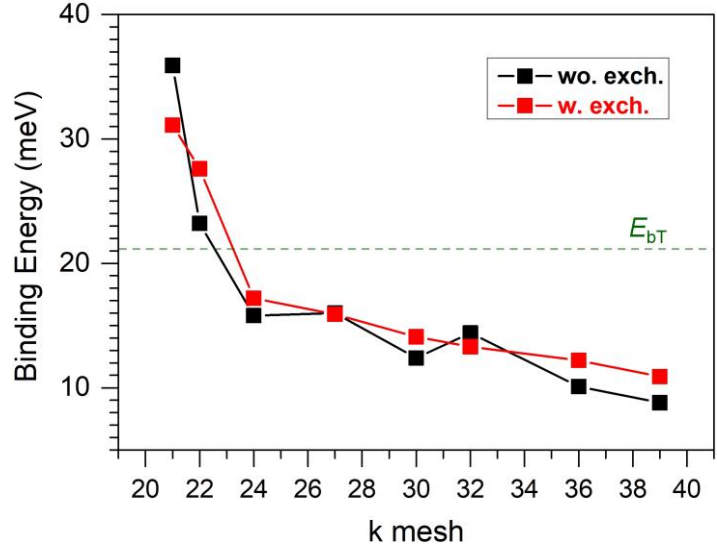

**Extended Data Fig. 20 |** Numerical convergence of the energy difference between peak $1$ and $2$ in Fig. 5d in the main text testing over various k-mesh densities with (red) or without (black) accounting for the *e-h* exchange. For better comparisons, the calculated binding energy of T ($E_{bT}$) is marked with the green dashed line.

Extended Data Fig. 20 shows the convergence test of the energy difference between peak $1$ and $2$ in Fig. 5d in the main text with respect to the k-mesh densities for the two cases, with or without the *e-h* exchange interactions. Both the values converge to 10 – 15 meV (within 2 meV accuracies), thus are smaller than the converged binding energy of T ($E_{bt}$: ~ 21 meV, as marked by the green dashed line). The result shows a good consistency with the previous Monte-Carlo calculation[28]. The convergence trends versus the k-mesh densities can be seen with the similar accuracies in Ref. [32,45] for the many-body perturbation calculation for BX.

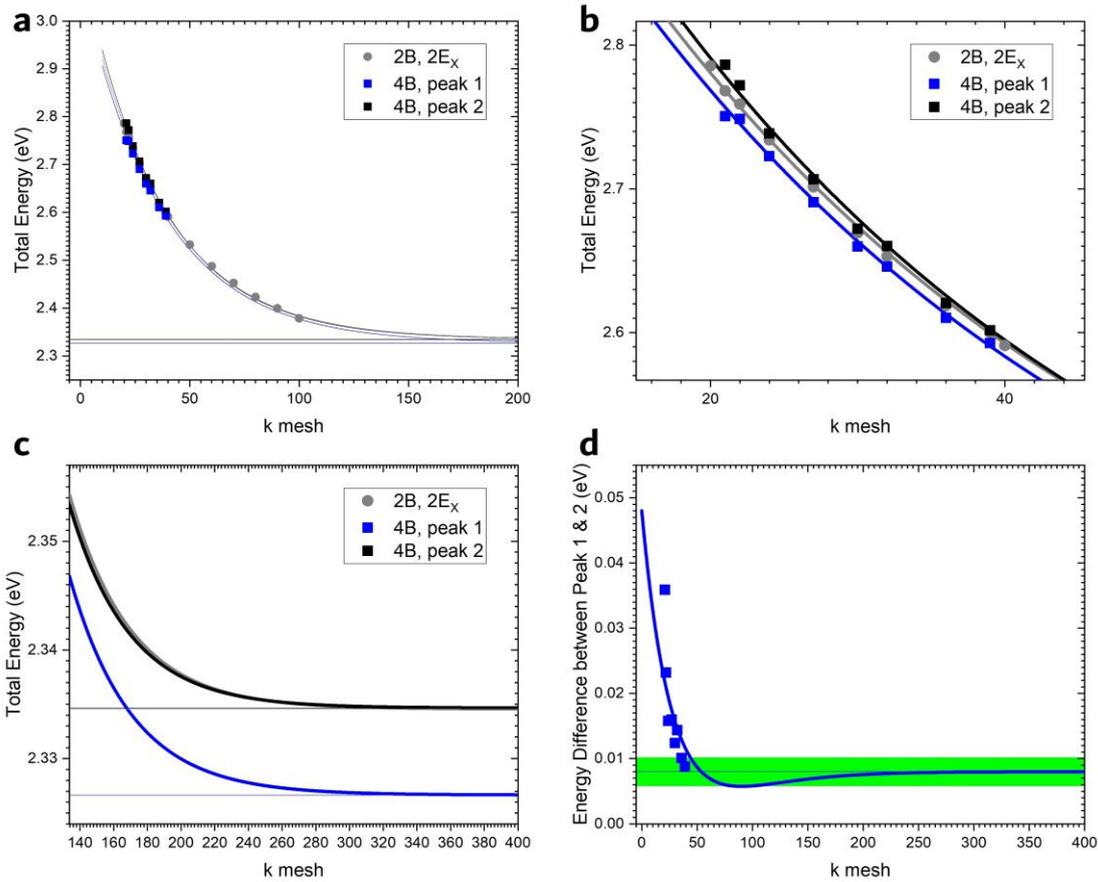

**Extended Data Fig. 21 | a, b, c,** Numerical convergence of the 2B and 4B total energies and corresponding exponential fittings versus k-mesh density, with zoom-in figures shown in **b & c**. The horizontal lines mark the asymptotic line (converged values) of the exponential fittings. **d,** Numerical convergence of the energy difference (same to Extended Data Fig. 20), shown together with the difference between the two fitting lines (blue and black lines in **a – c**). The green band marks the accuracy of ± 2 meV around the converged value.

We performed the calculation of the 2B states using a much grid size, up to 100 × 100 × 1. As shown in Extended Data Fig. 21a, the 2B total energy converges at the grid size of ~200 × 200 × 1. Since performing calculation for the 4B states using such a large grid is not feasible with our computational resources, we assumed that the 4B total energies converge similarly to that of 2B states with the increase of k-grid size. The limited 4B calculation results are then fitted using similar grid-size dependence to that of the 2B states. Such fittings are shown in Extended Data Fig. 21a, with a zoom-in figure shown in Extended Data Fig. 21b. Extended Data Fig. 21c shows the fitting to large grid-size. As one can see there, the 4B total energies converge at quite large grid size, similarly to the 2B case. However, the energy difference between the two total energies converge much more quickly, as shown in Extended Data Fig. 21d. For the grid size used in our paper, 39 × 39 × 1, the energy difference is within 2 meV of the converged value, as shown in Extended Data Fig. 21d. Therefore, we believe that the grid size that we used is sufficient for the energy difference that is the key quantities of the concern of our paper. We note that the grid sizes that we used are at the similar level to those in the literature[32,45]. We also would like to mention that similar conclusion about the quicker convergence of the energy differences than the total energies was also confirmed and discussed in Ref. [38] for the calculation of the 3B states.

**S15. Other possible origins of new spectral features**

**1) Defects and phonon replicas:**

**i)** Spectral features from defects[4,6-8,55,58] and phonon replicas[9-12] were found for other ML-TMDCs in the similar spectral range to our $P_1 - P_6$ as emission peaks in PL spectra, or through theoretical calculation for defect absorption[23]. However, we have not seen report of the observation of exciton phonon replicas in TDAS or TDRS in pump-probe spectroscopy, even though defect-related effects were identified on carrier scattering or cooling time in pump-probe spectroscopy[59-63]. However, it seems that pump-induced peaks from defects or phonon replicas have not been seen in TDAS. In addition, none of the defect- or phonon-assisted processes in those papers[4,6-12,23,55,58-63] could explain the polarization contrast of $P_1 - P_6$ seen in our experiments in, *e.g.* Fig. 3d and Fig. 3i in the main text.

**ii)** Even if we suppose defects or exciton phonon replicas could contribute to GSA peaks in the TDAS, the GSA processes, *e.g.* 0B-2B transitions (the 2B state is possibly trapped by defects or dressed by phonons), would decrease with pump, similar to exciton bleaching, known as Pauli blocking or GSB. While the key features of $P_1 - P_6$ we observed show increasing absorption with pump (for the reverse trends of the absorptions of $P_1 - P_6$ and X with increase in pumping, see Extended Data Fig. 11, Methods S8) in contrast to the typical GSA trend, thus the GSA related to defects or exciton phonon replicas can be excluded.

**iii)** Very few closely-related papers (*e.g.* Ref. [61]) that we searched in literature reported the study of influence of possible defect ESA process but not discussing about possible spectral peaks. In this paper, the ESA effect was evaluated for $MoS_2$ to occur likely at $E_{1s-X} - E_{probe} \approx 1.83 - 1.31 = 0.52$ eV = 520 meV below 1s-X, corresponding to a possible ESA transition from defect exciton state to continuous-absorption edge. If such an ESA occurred to $MoTe_2$, they would be spectrally located at $E_g - E_{defect} \approx 1.80 - 1.14 = 0.66$ eV, or ~660 meV below 1s-X, similarly so much to the case of $MoS_2$. However, we point out that the measured spectral range of interest here is ~40 meV below 1s-X, such that the above ESA effect is not expected in this spectral range. As only our own speculation, possible ESA processes related to defects or exciton phonon replicas could correspond to 2B-4B transitions, where the 2B and 4B states are possibly trapped by defects or dressed by phonons. Such ESA processes and associated spectral peaks have not been studied or even discussed in the literature. Even if we suppose defects could contribute in this way to ESA peaks in the TDAS, since the defect density of ML-TMDCs varies from $5 \times 10^{10}$ cm$^{-2}$ (Ref. [7]) to $10^{11} - 10^{12}$ cm$^{-2}$ (Ref. [23,64]), below the highest pumping level of ours up to $1.6 \times 10^{13}$ cm$^{-2}$, it is reasonable to expect that those ESA peaks would increase first and saturates eventually for samples with low to moderate levels of defects (such as our best samples). However, such deduction about defect ESA trend versus pump density is contradict to the results shown in Fig. 4 in the main text, where most of $P_1 - P_6$ is still growing with increase in pumping at the level of $1.6 \times 10^{13}$ cm$^{-2}$. Thus, the ESA related to defects can be excluded. In addition, our theory does not include any defects while producing good agreements with our experiments without adjusting any material parameters.

**iv)** We performed similar PL (to Ref. [4,6-8,55,58]) and absorption spectra measurements to assess the quality of our ML-$MoTe_2$ samples. Specifically in PL experiments, we used a CW laser and a femtosecond pulsed laser to excite the materials, corresponding to weak- and strong-excitation densities, *i.e.* $10^{8-9}$ and $10^{13}$ (similar to the pump density used in the pump-probe experiments) cm$^{-2}$, respectively. As can be seen in Extended Data Fig. 2i, 2j, 3a, 3b, & 4, we did not observe any emission or absorption peaks of defects or phonon replicas for those pre-screened ML-$MoTe_2$ samples with high quality. We did not either observe emission peaks from defects showing up under the strong-excitation density in Extended Data Fig. 4.

**v)** In connection with **iii)** & **iv)**, since we did not observe phonons assisting those of 0B-2B transitions (GSA) or any excitons dressed by phonons, it would be ever more difficult to observe phonon effects in those of weaker 2B-4B processes (ESA). Thus, the ESA related to exciton phonon replicas can be excluded.

For the above reasons, we exclude defects and phonon replicas as potential origins of the spectral features of $P_1 - P_6$.

**2) Emission signals:**

Since our observed signals show increased absorption ($-\Delta\alpha < 0$) with pump, those emission processes can be excluded. In principle, none of emission signals can be recorded as time-resolved signals by the standard pump-probe setup based on the lock-in technique (see Extended Data Fig. 5, where the possible PL signals do not pass through the delay line and thus the delay time associated with PL cannot be obtained), usually such time-resolved emission processes have to be measured by a time-correlated single photon counting (TCSPC) setup, which was not used in our experiments.

**3) Fluctuations of laser powers:**

As discussed in more detail in Methods S5, the data fluctuations caused by the laser-power jitters were evaluated to be two orders smaller than the actual features.

**4) Non-linear effects:**

**i) Absorption increase caused by, *e.g.* BGR:**

The non-linear effects such as BGR were discussed in the main text (Fig. 3l – 3n) and Methods S6 (Extended Data Fig. 7). BGR always causes a redshift for X with pump and thus leads to an asymmetric line-shape (negative and positive bands) around X in TDAS. As a significant distinction, the signals of BGR near X do not show such an obvious polarization contrast as $P_1 - P_6$ (ESA).

## ii) Second-harmonic generation (SHG):

Possible SHG signals are in the spectral range of ~ 2.3 – 2.4 eV (our laser energy is ~ 1.17 eV), which is far away out of the spectral range of $P_1 - P_6$, *i.e.* ~ 1.1 – 1.2 eV. Our detector was equipped with low-pass grating such that all the high-frequency signals are filtered out.

## iii) Other non-linear mixing effects:

Since our experiments were performed with both the pump and probe beams normal to the sample surfaces and the small thickness of the samples of ~ 60 nm (for the value of thickness in more detail see Methods S2), the phase-matching conditions cannot be met and the non-linear mixing effects can also be excluded. Normally, four-wave mixing or any $\chi^{(3)}$, $\chi^{(5)}$, … contributions typically require much more sophisticated techniques such as 2D coherent Fourier-transform spectroscopy[65] or quantum-optical spectroscopy[66], which cannot be obtained through standard pump-probe spectroscopy.

## 5) Bi-exciton (BX) and bi-exciton fine structure (BXFS):

First, the 4B-BSE truncated up to cluster $\triangle\triangle$ and $\triangle\triangle$ cannot explain the observed peaks of $P_1 - P_6$, especially $P_1 - P_4$ that are below the feature of T.

Second, the BX-related spectral features have been calculated theoretically[28,32-34,45] to be between T and X. Specifically, the BX resonance of 14.4 meV below X for ML-MoTe2 was calculated to be between T and X[28]. Using a simplified configuration-interaction model, BXFS was described in Ref. [32] by a 6×6 Hamiltonian. The model could only yield three peaks for BXFS, all distributing between T and X as well. Therefore, the bi-exciton fine structure does not explain our observed features as rich as six peaks.

## 6) Charged entities, *i.e.* trion, charged bi-exciton (or trion-exction), *etc*:

The optical signals of charged entities were the weakest under the charge-neutral conditions[6,7,20,21]. However, we found $P_1 - P_6$ were the strongest under the charge-neutral condition, whose signals clearly dropped when the extra carriers were injected into the sample. Therefore, we exclude charged entities as potential origins of $P_1 - P_6$.

**7) Charge-neutral *N*-body entities (*N* is an even integer > 4): tri-exciton △~△~△ (*N* = 6), quad-exciton △~△~△~△ (*N* = 8), and dropleton △~△~△~...~△ (with larger *N*):**

The series of tri-excitons, quad-excitons, and much bigger dropletons are only observable in non-classical quantum-optical spectroscopy[66] or multi-dimensional coherent Fourier transform spectroscopy based on non-linear four-wave mixing[65]. In other words, these multi-exciton complexes have no contributions to the TDAS obtained from the classical pump-probe experiments, despite the fact that we cannot deny the possibilities of their existence in our highly excited sample.